\pdfoutput=1

\documentclass[final,12pt]{elsarticle}

\usepackage{amssymb}
\usepackage{comment}
\usepackage{graphicx} 
\usepackage{amsmath}    
\usepackage{amssymb}    
\usepackage{amsthm}
\usepackage{amsfonts}
\usepackage{amssymb}
\usepackage{booktabs}
\usepackage{pdflscape}
\usepackage{lscape}
\usepackage{tikz}
\usetikzlibrary{arrows.meta, positioning}
\usepackage{float}
\usepackage{amsmath}
\usepackage{tikz}
\usetikzlibrary{automata, positioning}
\usepackage{geometry}
\usepackage{graphicx}
\usepackage{float}
\usepackage{algorithm}
\usepackage{algpseudocode}
\usepackage{multirow}
\usepackage{subcaption}
\usepackage{rotating}
\usepackage{cleveref}
\usepackage{mathrsfs}
\usepackage{longtable}

\usepackage{lineno}

\journal{Journal of Computational Physics}

\begin{document}

\begin{frontmatter}




\title{\textbf{SPARK}: \textbf{S}pectral \textbf{P}reconditioned \textbf{A}nalytic \textbf{R}esidual \textbf{K}olmogorov Arnold Networks for incompressible Navier Stokes equations} 

\author[NDSU]{Dathi Rajasekhar}

\author[NDSU]{Thien-Tam Nguyen}
\author[NDSU]{Trung Bao Le (trung.le@ndsu.edu)}
\affiliation[NDSU]{organization={North Dakota State University, Dept. Civil, Construction, and Environmental Engineering},
            addressline={1410 14th Ave N},
            city={Fargo},
            postcode={58102},
            state={ND},
            country={United States}}

\begin{abstract}
We propose Spectral Jet KAN (SJ-KAN) to reduce the cost of differentiation
and the network complexity of Physics-Informed Neural Networks (PINNs) for the
Navier--Stokes (NS) equations. The input coordinates are embedded through
Chebyshev bases whose spatial derivatives are known in closed form, and all
differentiation matrices are computed once for the collocation points before
training. A single linear layer compresses the Chebyshev bases into a few
latent features, and a single nonlinear Chebyshev Kolmogorov--Arnold Network
(CKAN) layer maps these features to the physical variables. Because the basis
derivatives are known in advance and the compression is linear, the spatial
derivatives in the NS equations are carried through the network analytically by
a fixed sequence of matrix products and elementwise operations. Automatic
differentiation (AD) is used only in backpropagation, to compute the gradients
of the loss with respect to the network weights. We examine three variants of
SJ-KAN (JetKAN, JetKAN-F, JetKAN-H), which differ in how the output layer maps
the latent features to the physical variables. All variants are tested on the
Taylor--Green vortex and on the lid driven cavity at Reynolds numbers
$\mathrm{Re} = 100$, $1000$, $5000$ and $10{,}000$. The results are compared
with a KAN that uses AD (PIKAN) and with a standard PINN. The SJ-KAN variants
reach comparable or better accuracy than both, with up to $47$ times fewer
parameters and up to $8.2$ times faster training per epoch. SJ-KAN also solves
the lid driven cavity flow at $\mathrm{Re} = 10{,}000$ without any reference
data and with fewer than $9{,}000$ trainable parameters. These results show
that analytic derivative propagation, a small spectral input layer and a single
KAN layer give a faster and simpler solver than methods based on AD, also for
flows at high Reynolds numbers.

\end{abstract}


\begin{highlights}
\item A novel design to accelerate the convergence of Physics Informed Kolgomorov Arnold Network (PIKAN) for solving incompressible Navier-Stokes equations with the use of Chebyshev bases,  analytic spectral jet, and linear bottleneck 
\item Obtain a similar accuracy as automatic differentiation with 8.2 times lower training cost per epoch
\item Data-free PIKAN solution of lid-driven cavity flow up to $Re = 10,000$
\end{highlights}


\end{frontmatter}


\clearpage

\section{Introduction} \label{sec:introduction}
The incompressible Navier--Stokes (NS) equations govern a wide range of flows in science and engineering. The two-dimensional $NS$ equations read:
\begin{equation}
  \nabla \cdot \boldsymbol{U} = 0,
  \label{eqn:1. continuity}
\end{equation}
\begin{equation}
  \frac{\partial \boldsymbol{U}}{\partial t}
  + (\boldsymbol{U} \cdot \nabla)\boldsymbol{U}
  = -\nabla p + \nu\,\nabla^2 \boldsymbol{U},
  \label{eqn:2_momentum}
\end{equation}
where $\boldsymbol{U} = (u,v)$ is the velocity field, $p$ is the pressure divided by the constant density, and $\nu$ is the kinematic viscosity. Classical numerical methods, including finite differences~\cite{strikwerda2004finite, le2013fluid}, finite elements~\cite{gallagher1975finite}, finite volumes~\cite{versteeg2007introduction}, and spectral methods~\cite{hussaini1989spectral}, discretize the domain and
produce finite-dimensional algebraic systems. These solvers perform well when the governing equations and boundary conditions are fully specified in forward problems \cite{gilmanov2015numerical, borazjani2013parallel, akerkouch2025shape, ahmed2026impacts}. For inverse problems with incomplete boundary data, classical numerical methods are inefficient at recovering the missing information~\cite{karniadakis2021physics} because parameter optimization requires repeated forward solves. 

As an alternative framework, Physics-informed neural networks (PINNs)~\cite{lagaris1998artificial, raissi2019physics} approximate the solution directly, penalizing the governing equation's residual at collocation points sampled from the computational domain. This formulation handles inverse problems naturally, since unknown parameters enter the loss function as trainable variables and are learned jointly with the network weights~\cite{raissi2020hidden, cai2021espresso}. PINNs have since been applied to the incompressible Navier--Stokes equations at low-to-moderate Reynolds
numbers~\cite{jin2021nsfnets, cuomo2022scientific} successfully. Two challenges, however, remain with this PINN framework: 1) the computation of spatial derivatives; and 2) network complexity, as discussed below. 

First, the conventional PINN implementations evaluate the required spatial derivatives using nested reverse-mode automatic differentiation (AD)~\cite{baydin2018automatic}.For second-order coupled systems like the Navier--Stokes equations, the AD cost scales with both network depth and derivative order~\cite{toscano2025pinns}, as shown in Equation \ref{eqn:2_momentum}. This cost grows with the network depth and the number of collocation points. To alleviate this issue, recent works embed spectral methods directly into the PINN framework to reduce or eliminate the differentiation cost ~\cite{yu2025sinn}.  Yu et al.~\cite{yu2025sinn} proposed the Spectral-Informed Neural Network (SINN), which operates entirely in the Fourier spectral domain and replaces high-order spatial derivatives with algebraic multiplications in frequency space, substantially lowering memory use and training time. In addition, Huang et al.~\cite{huang2025chebyshev} replaced AD with Chebyshev differentiation matrices (CD-PINN), cutting computation time by up to $86\%$. Yin et al.~\cite{yin2024chebyshev} constructed neurons from Chebyshev basis functions (CSNN) that satisfy boundary conditions by construction, removing the boundary-penalty term from the loss.However, the computed derivatives are still evaluated through $AD$. Chen et al.~\cite{chen2025cspinn} combined Chebyshev spectral differentiation with Sobolev regularization (CS-PINN), addressing both derivative cost and solution regularity. Xiong et al.~\cite{xiong2025svsnn} decomposed multivariable functions into products of univariate spectral representations with learnable frequencies (SV-SNN), computing spatial derivatives analytically while relying on $AD$ for temporal derivatives~\cite{xiong2025svsnn}. Among these spectral formulations, those that replace $AD$ with spectral differentiation retain fixed activation functions in the underlying network~\cite{yu2025sinn, huang2025chebyshev, chen2025cspinn}. In contrast, those that introduce learnable spectral representations still rely on $AD$ for at least part of the differentiation evaluation~\cite{yin2024chebyshev, xiong2025svsnn}. 

Second, the standard PINN architecture is built on a multilayer perceptron (MLP) with fixed activation functions that tend to learn low-frequency components first~\cite{rahaman2019spectral}. This property, known as spectral bias, makes it difficult to resolve the sharp gradients present in fluid flows~\cite{tancik2020fourier, shukla2024comprehensive}. Efforts to address these issues include Fourier-feature embeddings~\cite{tancik2020fourier} and adaptive loss-weighting strategies~\cite{wang2021understanding}. These works have partially alleviated spectral bias but do not, by themselves, remove the need for $AD$-based derivative evaluation. Studies characterizing training failures in PINNs~\cite{krishnapriyan2021characterizing} have improved convergence in specific settings but again do not change how the residual derivatives are evaluated. Recently, Kolmogorov--Arnold Networks 
(KANs)~\cite{liu2024kan} have been developed to offer a structurally different alternative to MLPs. Instead of learning weights that multiply fixed activation functions at each node, a KAN places a learnable univariate function on every edge, a design motivated by the Kolmogorov--Arnold representation theorem~\cite{kolmogorov1957representation}. Training a KAN with a physics-informed loss yields a Physics-Informed KAN (PIKAN)~\cite{wang2025kinn}. For example, separable PIKANs (SPIKANs)~\cite{jacob2025spikans} reduce cost further by assigning each spatial dimension its own KAN. However, this formulation requires a factorizable grid of collocation points~\cite{jacob2025spikans}. Feature-enriched KANs (FEKANs)~\cite{menon2025fekan} accelerate convergence by augmenting the input with fixed nonlinear transformations, without adding trainable parameters. 

KAN-based architectures can match or exceed MLP accuracy but exhibit distinct gradient pathologies during training. In the original KAN architecture, the B-spline parameterization was computationally expensive and sensitive to the number of collocation points~\cite{shukla2024comprehensive}. Replacing the B-spline basis with low-order orthogonal polynomials such as Chebyshev, Legendre~\cite{shukla2024comprehensive} and Jacobi~\cite{aghaei2024fkan} functions substantially improved both accuracy and training speed on the reported benchmarks. Chebyshev-basis PIKANs in particular proved faster and more accurate than B-spline PIKANs across a range of benchmark problems ~\cite{shukla2024comprehensive}, including the incompressible Navier--Stokes equations~\cite{guo2025chebpikan, cui2025pikan}. Despite these advances~\cite{mostajeran2025scaledcpikans}, polynomial-based PIKANs evaluate a learnable polynomial and thus increase the number of floating-point operations per edge~\cite{shukla2024comprehensive, guo2025chebpikan}. Furthermore, differentiating through these polynomial evaluations makes $AD$ even more expensive per parameter than in an equally sized MLP ~\cite{shukla2024comprehensive, cui2025pikan}. Currently, the theoretical understanding of PIKAN training dynamics and their extension to coupled multi-output systems has received increasing attention~\cite{shukla2024comprehensive}. To date, systematic studies of how architectural choices, such as network depth, polynomial degree, and basis type, affect multi-output accuracy and training stability on canonical fluid benchmarks remain limited. 

The lid-driven cavity is the classical benchmark for validating PINN-based approaches for Navier-Stokes equations, as the flow transitions to turbulence at sufficiently high Reynolds numbers around $Re=10{,}000$ ~\cite{ghia1982highre} to $Re=21{,}000$ ~\cite{erturk2005numerical}. Current PINN-based approaches perform best in the low-Reynolds-number regime, and higher-Reynolds-number flows remain an open problem~\cite{cai2021review}. Wang et al.~\cite{wang2023multiplicity} studied vanilla PINNs on the lid-driven cavity at $Re=2{,}000$, $3{,}000$, and $5{,}000$ and found that, without regularization, the network can converge to an unstable, physically unrealizable solution. Xiong et al.~\cite{xiong2025jpikan} reported a Jacobi-polynomial PIKAN (J-PIKAN) that can resolve the flow accurately up to $Re=4{,}000$. To our knowledge, this is the highest Reynolds number previously reported for a PIKAN solver on the lid-driven cavity flow problem.

The present work (the Spectral--Jet KAN, or SJ--KAN) aims to address both challenges within a single framework by combining spectral representation and Kolmogorov--Arnold Network architectures. We validate the proposed architecture on canonical flow problems such as the Taylor--Green vortex and the lid-driven cavity flow at various Reynolds numbers. We finally discuss the performance of this framework with respect to its architecture (spectral order and latent dimensions), as well as potential future improvements.

\section{Preliminaries}
\label{sec:2. KAN}
The Kolmogorov--Arnold representation theorem
\cite{kolmogorov1957representation, braun2009constructive}
states that any continuous multivariate function can be written as a
finite composition of continuous univariate functions and addition.
Specifically, for any continuous function
\(f : [0,1]^d \rightarrow \mathbb{R}\), there exist continuous univariate
functions \(\phi_{q,\kappa} : [0,1] \rightarrow \mathbb{R}\) and
\(\Phi_q : \mathbb{R} \rightarrow \mathbb{R}\) such that, for the input
vector \(\mathbf{x} = (x_1, \dots, x_d)\),
\begin{equation}
  f(\mathbf{x}) = f(x_1, \dots, x_d)
  = \sum_{q=1}^{2d+1} \Phi_q\!\left( \sum_{\kappa=1}^{d} \phi_{q,\kappa}(x_\kappa) \right).
\label{eq:3. KA theorm}
\end{equation}

\noindent The original constructive proof \cite{kolmogorov1957representation} yields a two-layer network structure of width $2d+1$. Collecting the $d$ input coordinates into the vector $\mathbf{x} = (x_1, x_2, \dots, x_d) \in \mathbb{R}^d$, the representation \Cref{eq:3. KA theorm} is written compactly as
\begin{equation}
    f(\mathbf{x}) \approx \mathrm{KAN}(\mathbf{x}) = \Phi_{\text{out}} \circ \Phi_{\text{in}} \circ \mathbf{x}
\label{eq:4. KA compact}
\end{equation}

Here $\mathbf{x}$ is the complete input vector rather than a single variable, so the composition acts on all $d$ coordinates at once and \Cref{eq:4. KA compact} is a restatement of \Cref{eq:3. KA theorm} in operator form. The two function matrices collect the univariate functions already introduced in \Cref{eq:3. KA theorm}: $\Phi_{\mathrm{in}}$ gathers the inner functions $\phi_{q,\kappa}$ and $\Phi_{\mathrm{out}}$ gathers the outer functions $\Phi_q$,

$$
\Phi_{\text{in}} =
\begin{pmatrix}
\phi_{1,1}(\cdot) & \cdots & \phi_{1,d}(\cdot) \\
\vdots & \ddots & \vdots \\
\phi_{2d+1,1}(\cdot) & \cdots & \phi_{2d+1,d}(\cdot)
\end{pmatrix}, \quad
\Phi_{\text{out}} =
\begin{pmatrix}
\Phi_{1}(\cdot) & \cdots & \Phi_{2d+1}(\cdot)
\end{pmatrix}
$$
Kolmogorov Arnold Network (KAN) \cite{liu2024kan} can be extended to a larger number of layers as:

\begin{equation}
    f_{\alpha_L}(\mathbf{x}) \approx \sum_{\alpha_{L-1}=1}^{d_{L-1}} \phi_{L-1,\, \alpha_L,\, \alpha_{L-1}} \left( \sum_{\alpha_{L-2}=1}^{d_{L-2}} \cdots \sum_{\alpha_2=1}^{d_2} \phi_{2,\, \alpha_3,\, \alpha_2} \left( \sum_{\alpha_1=1}^{d_1} \phi_{1,\, \alpha_2,\, \alpha_1} \left( \sum_{\alpha_0=1}^{d_0} \phi_{0,\, \alpha_1,\, \alpha_0}(x_{\alpha_0}) \right) \right) \cdots \right) \quad 
\label{eq: 5. KAN}
\end{equation}

\noindent The right-hand side of \Cref{eq: 5. KAN} defines the $L$-layer Kolmogorov-Arnold Network, denoted $\mathrm{KAN}(\mathbf{x})$. Here $L$ is the number of layers, $\{d_l\}_{l=0}^{L}$ are the numbers of nodes in each layer, with $d_0 = d$ inputs and $d_L$ outputs, and $\alpha_l \in \{1, \dots, d_l\}$ denotes the index of a node in layer $l$. The function $\phi_{l,\, \alpha_{l+1},\, \alpha_l}$ is the univariate activation function placed on the edge connecting node $\alpha_l$ of layer $l$ to node $\alpha_{l+1}$ of layer $l+1$, for $l = 0, \dots, L-1$, and $f_{\alpha_L}$ is the $\alpha_L$-th output. The choice of functional form for $\phi$ distinguishes the different KAN architectures; in the present work $\phi$ is a Chebyshev expansion. Generalizing this to a multi-layer KAN, we define the $l$-th KAN layer $\Phi_l$ as a function matrix of shape $(d_{l+1}, d_l)$:

$$
\Phi_l =
\begin{pmatrix}
\phi_{l,1,1}(\cdot) & \cdots & \phi_{l,1,d_l}(\cdot) \\
\vdots & \ddots & \vdots \\
\phi_{l,d_{l+1},1}(\cdot) & \cdots & \phi_{l,d_{l+1},d_l}(\cdot)
\end{pmatrix}
$$

Here, the entry in row $\alpha_{l+1}$ and column $\alpha_l$ of $\Phi_l$ is the function $\phi_{l,\, \alpha_{l+1},\, \alpha_l}(\cdot)$ connecting node $\alpha_l$ of layer $l$ to node $\alpha_{l+1}$ of layer $l+1$. A KAN with $L$ layers is then a composition of these layers:

$$
\mathrm{KAN}(\mathbf{x}) = \Phi_{L-1} \circ \cdots \circ \Phi_1 \circ \Phi_0 \circ \mathbf{x}
$$

\subsection{Chebyshev KAN (CKAN)}
\label{subsec:2.1. CKAN}

An alternative implementation of KANs utilizes Chebyshev polynomials \cite{dragomir2003survey, cesarano2019orthogonality} as the basis for the univariate activation functions $\phi$ placed on the edges of each layer, as introduced in \Cref{sec:2. KAN}. Chebyshev polynomials of the first kind, denoted $T_n(x)$ with $n$ the polynomial degree, form a sequence of orthogonal polynomials defined recursively by
\begin{equation}
T_0(x) = 1, \qquad T_1(x) = x, \qquad T_{n+1}(x) = 2x\, T_n(x) - T_{n-1}(x)
\end{equation}
where $x \in [-1,1]$ and $T_n(x) \in [-1,1]$ for every degree $n$. In Chebyshev-based KANs (CKANs) \cite{ss2024chebyshev}, the univariate functions $\phi(x)$ are represented as a linear combination of these Chebyshev polynomials:
\begin{equation}
\phi(x) = \sum_{n=0}^{P} c_n\, T_n(x)
\label{eq:7. ckan univariate}
\end{equation}
where $P$ is the highest degree retained and the $c_n$ are the trainable coefficients of the expansion. This formulation reduces the number of trainable parameters in a CKAN with $N_l$ ($=L$) layers and $N_n$ neurons per layer (i.e.\ $d_l = N_n$ for all hidden layers), using Chebyshev polynomials up to degree $P$, to approximately $\mathcal{O}(N_l N_n^2 P)$. To ensure the stability of training, CKANs typically restrict the input to each layer to the range $[-1,1]$, since $T_n$ is defined only on this interval. This is achieved by applying the hyperbolic tangent ($\tanh$) activation function after each KAN layer. Consequently, the forward pass of an $L$-layer CKAN for an input vector $\mathbf{x}$ can be described as:
\begin{equation}
\mathrm{CKAN}(\mathbf{x}) = \Phi_{L-1} \circ \tanh \circ \cdots \circ \Phi_1 \circ \tanh \circ \Phi_0 \circ \tanh \circ \mathbf{x}
\end{equation}
where each $\Phi_l$ is the function matrix of shape $(d_{l+1}, d_l)$ defined in \Cref{sec:2. KAN}, whose entries are the univariate functions of \Cref{eq:7. ckan univariate}. Specifically, the activation function for the initial input and subsequent layers incorporates this $\tanh$ scaling. Writing $x_l$ for the scalar input to the function at layer $l$, the first layer takes the network input directly:
\begin{equation}
\phi_0(x_0) = \omega\, \beta\!\left[\tanh(x_0)\right] + \sum_{n=0}^{P} c_n\, T_n\!\left[\tanh(x_0)\right]
\label{eq: 9. ckan first layer}
\end{equation}
And for subsequent layers $(l \geq 1)$, where the input is the output of the preceding layer:
\begin{equation}
\phi_l(x_l) = \omega\, \beta\!\left[\tanh\!\left(\phi_{l-1}(x_{l-1})\right)\right]
+ \sum_{n=0}^{P} c_n\, T_n\!\left[\tanh\!\left(\phi_{l-1}(x_{l-1})\right)\right]
\label{eq: 10. ckan l layer}
\end{equation}
In \Cref{eq: 9. ckan first layer,eq: 10. ckan l layer}, the summation is the Chebyshev polynomial expansion of \Cref{eq:7. ckan univariate} evaluated at the $\tanh$-scaled input, $\beta$ is a fixed non-polynomial base function applied to the same scaled input, which supplies a smooth residual component alongside the polynomial terms, and $\omega$ is a trainable weight controlling the contribution of that base term. The $\tanh$ therefore acts only to restrict the domain, while the learnable behaviour of the edge is carried by the coefficients $c_n$ and the weight $\omega$. The use of Chebyshev polynomials as an orthogonal basis provides a strong foundation for function approximation within the KAN framework \cite{ss2024chebyshev}.

ChebyKAN~\cite{ss2024chebyshev} replaces the B-spline basis of the original KAN~\cite{liu2024kan} with a truncated Chebyshev series $\phi(x) = \sum_{n=0}^{P} c_n T_n(x)$, reducing the parameter count per edge to $P+1$ coefficients. However, for the Navier--Stokes equations at moderate to high Reynolds numbers, achieving sufficient accuracy requires increasing 
$P$ or adding more layers, which grows the parameter count as 
$\mathcal{O}(n_\mathrm{in}\,n_\mathrm{out}\,P)$ per layer, where $n_\mathrm{in} = d_l$ and $n_\mathrm{out} = d_{l+1}$ are the numbers of input and output nodes of layer $l$. Since the activations operate directly on the raw coordinates $(x, y)$, the network carries the full burden of resolving the spatial structure during training, without exploiting the fact that Chebyshev polynomials admit closed-form derivatives. Furthermore, computing second-order residual terms such as 
$\partial^2 u/\partial x^2$ relies on automatic differentiation, which requires propagating a second-order computational graph at every iteration, and this cost scales with network depth and width and dominates the per-epoch wall time for large collocation sets. These three issues, namely parameter scaling with polynomial degree, lack of spectral pre-processing of the inputs, 
and the overhead of automatic differentiation for second-order derivatives, motivate the architecture proposed in this work.

\section{Spectral jet accelerator for KAN to solve Navier-Stokes equations}
Consider the Navier--Stokes equations \Cref{eqn:1. continuity} - \ref{eqn:2_momentum} in a domain $\Omega$ with boundary conditions on $\partial\Omega$ and initial conditions at $t = 0$. In Cartesian coordinates, the two-dimensional incompressible Navier--Stokes equations read
\begin{align}
\frac{\partial u}{\partial x} + \frac{\partial v}{\partial y} &= 0, \label{eq:ldc_continuity} \\
\frac{\partial u}{ \partial t} + u\frac{\partial u}{\partial x} + v\frac{\partial u}{\partial y} &= -\frac{\partial p}{\partial x} + \frac{1}{Re}\left( \frac{\partial^2 u}{\partial x^2} + \frac{\partial^2 u}{\partial y^2} \right), \label{eq:ldc_mom_u} \\
\frac{\partial v}{ \partial t} + u\frac{\partial v}{\partial x} + v\frac{\partial v}{\partial y} &= -\frac{\partial p}{\partial y} + \frac{1}{Re}\left( \frac{\partial^2 v}{\partial x^2} + \frac{\partial^2 v}{\partial y^2} \right), \label{eq:ldc_mom_v}
\end{align}
where $Re$ is the Reynolds number, the ratio of inertial to viscous
forces.

In the general PIKAN formulation, the governing equations are written compactly as
\begin{align}
\frac{\partial \mathbf{U}}{\partial t} + \mathcal{N}[\mathbf{U}] &= 0, 
    \quad \mathbf{x} \in \Omega, \quad t \in [0, t_f], \label{eqn:pde}\\
\mathbf{U}(\mathbf{x}, t) &= \mathbf{U}_b(\mathbf{x}, t), 
    \quad \mathbf{x} \in \partial\Omega, \quad t \in [0, t_f], \label{eqn:bc}\\
\mathbf{U}(\mathbf{x}, 0) &= \mathbf{U}_0(\mathbf{x}), 
    \quad \mathbf{x} \in \Omega, \label{eqn:ic}
\end{align}
where $\mathbf{x} = (x, y)$ is the position vector, $\mathbf{U}(\mathbf{x}, t) = (u, v, p)^{\top}$ is the unknown solution vector, $t_f$ is the final time, and $\mathcal{N}[\cdot]$ is the nonlinear differential operator containing the advection, pressure gradient, and viscous diffusion terms. The functions $\mathbf{U}_b(\mathbf{x}, t)$ and $\mathbf{U}_0(\mathbf{x})$ are the prescribed boundary data on $\partial\Omega$ and the prescribed initial state at $t = 0$, respectively. We approximate the solution using a KAN, denoted $\hat{\mathbf{U}}(\mathbf{x}, t)$. The training objective is to minimize a composite loss function that accounts for the PDE residual, the boundary conditions, and the initial condition:
\begin{equation}
\mathcal{L} = \lambda_R \mathcal{L}_R 
            + \lambda_{BC} \mathcal{L}_{BC} 
            + \lambda_{IC} \mathcal{L}_{IC},
\label{eqn:total_loss}
\end{equation}
where $\lambda_R$, $\lambda_{BC}$, and $\lambda_{IC}$ are the weights 
assigned to each loss term. Using the mean squared error, these terms are 
defined as
\begin{align}
\mathcal{L}_R &= \frac{1}{N_R} \sum_{m=1}^{N_R} 
    \left| \frac{\partial \hat{\mathbf{U}}}{\partial t}(\mathbf{x}_m, t_m) 
    + \mathcal{N}[\hat{\mathbf{U}}](\mathbf{x}_m, t_m) \right|^2, 
    \label{eqn:pde_loss}\\
\mathcal{L}_{BC} &= \frac{1}{N_{BC}} \sum_{m=1}^{N_{BC}} 
    \left| \hat{\mathbf{U}}(\mathbf{x}_m, t_m) 
    - \mathbf{U}_b(\mathbf{x}_m, t_m) \right|^2, 
    \quad (\mathbf{x}_m, t_m) \in \mathcal{S}_{BC},
    \label{eqn:bc_loss}\\
\mathcal{L}_{IC} &= \frac{1}{N_{IC}} \sum_{m=1}^{N_{IC}} 
    \left| \hat{\mathbf{U}}(\mathbf{x}_m, 0) 
    - \mathbf{U}_0(\mathbf{x}_m) \right|^2, 
    \quad \mathbf{x}_m \in \mathcal{S}_{IC},
    \label{eqn:ic_loss}
\end{align}
where $m$ is the collocation-point index, and $N_R$, $N_{BC}$, and $N_{IC}$ are the number of collocation points sampled within the domain $\Omega$, on the boundary $\partial\Omega$, and at the initial time $t = 0$, respectively. $\mathcal{S}_{BC}$ and $\mathcal{S}_{IC}$ denote the sets of sampled boundary and initial condition points.

The total loss $\mathcal{L}$ in \Cref{eqn:total_loss} is a weighted sum of three terms: $\mathcal{L}_{R}$ penalises the PDE residual evaluated at a set of interior collocation points, $\mathcal{L}_{BC}$ enforces the boundary conditions on $\partial\Omega$, and $\mathcal{L}_{IC}$ enforces the initial 
conditions at $t = 0$. The scalar weights $\lambda_{R}$, $\lambda_{BC}$, and 
$\lambda_{IC}$ control the relative contribution of each term and are treated 
as hyperparameters. Minimising $\mathcal{L}$ with respect to the network 
parameters yields an approximate solution $\hat{\mathbf{U}}$ that satisfies the 
governing equations in a least-squares sense. Evaluating $\mathcal{L}_{R}$ 
requires second-order derivatives of the network output, which in a standard 
PIKAN are obtained through automatic differentiation (AD). Because AD builds 
a second-order computational graph at every iteration, its cost grows with both 
the network depth and the number of collocation points, and it typically 
dominates the training wall time. \Cref{subsec: 3.1 sjkan} introduces the 
spectral-jet formulation that computes these derivatives analytically, avoiding 
the repeated graph construction required by AD.

\subsection{Spectral-Jet KAN (SJ--KAN)}
\label{subsec: 3.1 sjkan}

The SJ-KAN architecture constructs the solution using a tensor-product Chebyshev basis, projects the basis evaluations through a linear bottleneck, and applies a single KAN activation layer~\cite{liu2024kan,guo2025chebpikan}. The derivatives of the Chebyshev polynomials with respect to the spatial coordinates are known in closed form, so we can precompute the differentiation matrices once before training. The PDE residual derivatives are then obtained by propagating these matrices through the network layers analytically. We refer to the resulting tuple of function values and their required derivatives at each collocation point as the \emph{spectral jet}.\\
 
\begin{sidewaysfigure}
    \centering
    \includegraphics[width=\linewidth]{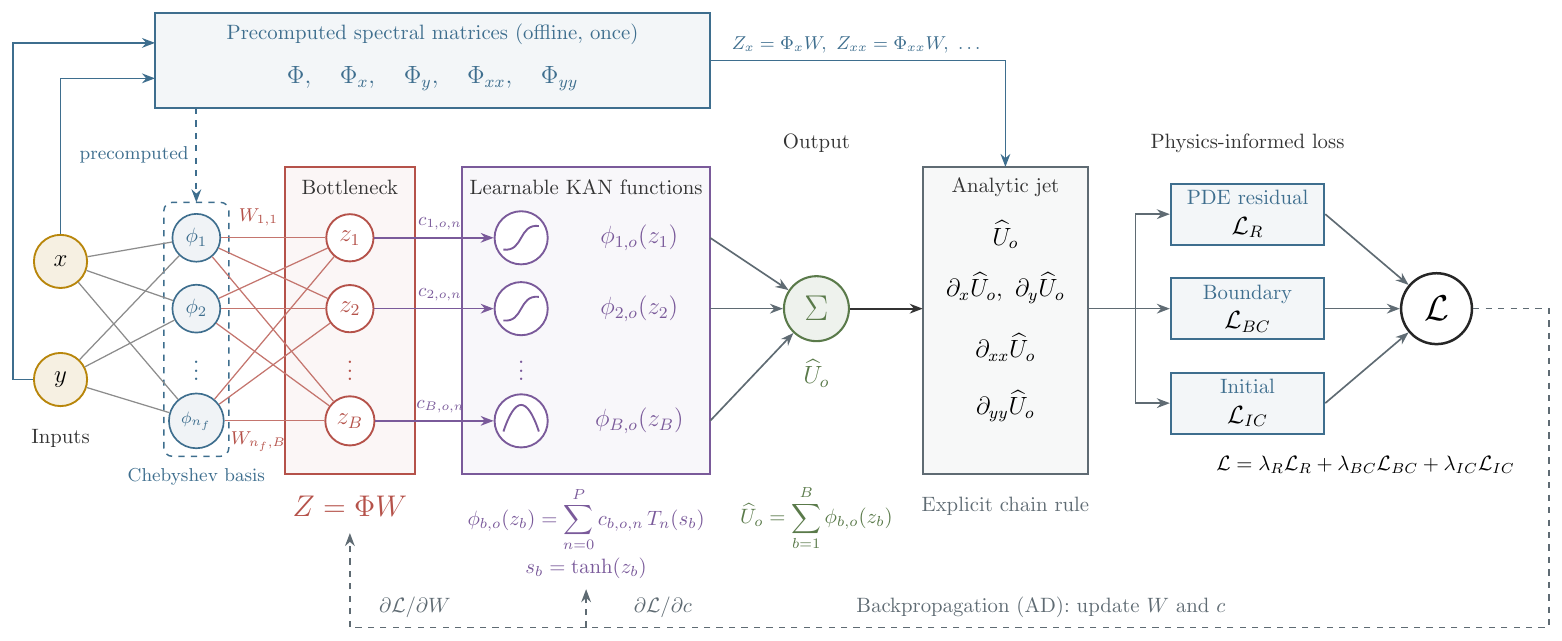}
    \caption{SPARK architecture. The inputs $(x,y)$ are mapped to the tensor-product Chebyshev basis functions $\phi_{ij}=T_i(\xi_x)\,T_j(\xi_y)$ through fixed edges with no trainable parameters (gray). Since this embedding is fixed, its values and derivatives are computed once and stored in $\Phi$, $\Phi_x$, $\Phi_y$, $\Phi_{xx}$, and $\Phi_{yy}$. The trainable weights $W$ project the basis onto $B$ latent channels, $Z=\Phi W$, and each channel passes through a learnable Chebyshev edge function $\phi_{b,o}$ with coefficients $c_{b,o,n}$. The output $\widehat{U}_o$ and its derivatives are obtained from the analytic jet without automatic differentiation, and backpropagation updates only $W$ and $c_{b,o,n}$.}
    \label{fig:spark_architecture}
\end{sidewaysfigure}

The construction proceeds in four stages as shown in \Cref{fig:spark_architecture}. 
\Cref{subsubsec:3.1.1 Coordinate normalization and feature matrix}--\ref{subsubsec:3.1.4 KAN activation layer and analytic jet} 
detail the procedure, and \Cref{subsubsec: 3.1.5 Model variants} 
presents the three model variants, which differ only in how the latent 
features are mapped to the physical output variables. Step-by-step 
derivations are given in the \ref{appendix}.


\subsubsection{Coordinate normalization and feature matrix}
\label{subsubsec:3.1.1 Coordinate normalization and feature matrix}

Chebyshev polynomials $T_n$ are defined on $[-1,1]$, so each physical 
coordinate is first mapped to that interval. For a two-dimensional domain 
$(x,y)\in[0,H_x]\times[0,H_y]$, the normalizations are
\begin{equation}
  \xi_x = \frac{2x}{H_x} - 1, \qquad \xi_y = \frac{2y}{H_y} - 1,
  \label{eq:18. normalization}
\end{equation}
with scale factors
\begin{equation}
  h_x = \frac{H_x}{2}, \qquad h_y = \frac{H_y}{2},
  \label{eq:19. scale factors}
\end{equation}
which convert derivatives between normalized and physical coordinates via the 
chain rule: $\partial/\partial x = (1/h_x)\,\partial/\partial\xi_x$. Using the first kind Chebyshev polynomials $T_n$, which satisfy 
$|T_n(\xi)|\le 1$ on $[-1,1]$, a two-dimensional basis function of 
degree $(i,j)$ is defined as the product
\begin{equation}
  \phi_{ij}(x,y) = T_i(\xi_x)\,T_j(\xi_y),
  \qquad i = 0,\ldots,K_x-1,\quad j = 0,\ldots,K_y-1,
  \label{eq:20. tensor product}
\end{equation}
where $i$ and $j$ are the polynomial degrees in the $x$- and $y$-directions, and $K_x$ and $K_y$ are the numbers of Chebyshev modes in each direction, giving $n_f = K_x K_y$ basis functions in total. Given $N$ collocation points $\{(x_m,y_m)\}_{m=1}^{N}$, all $n_f$ basis 
functions are evaluated at every point and stored in the \emph{feature matrix}
\begin{equation}
  \boldsymbol{\Phi} \in \mathbb{R}^{N \times n_f},
  \qquad \boldsymbol{\Phi}_{m,(i,j)} = \phi_{ij}(x_m,y_m).
  \label{eq:21. featurematrix}
\end{equation}
Row $m$ of $\boldsymbol{\Phi}$ is the evaluation of the complete basis at the 
$m$-th collocation point. This matrix is computed once before training and 
remains fixed throughout.

\subsubsection{Precomputed differentiation matrices}
\label{subsubsec:3.1.2 Precomputed differentiation matrices}

A key property of Chebyshev polynomials is that the derivative of any 
Chebyshev series is itself a Chebyshev series, and the coefficients of the 
derivative are obtained by a single matrix multiplication. If 
$f(\xi)=\sum_{n=0}^{K-1}a_n T_n(\xi)$, then 
$f'(\xi)=\sum_{n=0}^{K-1}a'_n T_n(\xi)$ with 
$\mathbf{a}' = \mathbf{D}_1\mathbf{a}$, where $K$ is the number of modes of 
the one-dimensional series ($K = K_x$ along $x$ and $K = K_y$ along $y$; in 
this work $K_x = K_y = K$), and the $K\times K$ Chebyshev 
differentiation matrix has entries
\begin{equation}
  (\mathbf{D}_1)_{kn} =
  \begin{cases}
    2n/\gamma_k, & k < n \text{ and } (n+k) \text{ odd},\\
    0, & \text{otherwise},
  \end{cases}
  \qquad \gamma_0 = 2,\quad \gamma_{k\ge1} = 1.
  \label{eq:22. precomputed matrices}
\end{equation}
The second-derivative matrix is $\mathbf{D}_2 = \mathbf{D}_1\mathbf{D}_1$. 
Both $\mathbf{D}_1$ and $\mathbf{D}_2$ depend only on $K$ and are computed 
once before training. Because the basis function $\phi_{ij} = T_i(\xi_x)\,T_j(\xi_y)$ introduced 
in \Cref{eq:20. tensor product} is a tensor product, differentiating with respect to $x$ 
acts only on the $T_i(\xi_x)$ factor. Let $(\mathbf{D}_1)_x$ and 
$(\mathbf{D}_2)_x$ denote the first- and second-derivative matrices of size 
$K_x \times K_x$ applied along the $x$-direction of the tensor-product basis, 
and $(\mathbf{D}_1)_y$, $(\mathbf{D}_2)_y$ the corresponding 
$K_y \times K_y$ matrices along the $y$-direction. Combining these with the 
scale factors $h_x$, $h_y$ from \Cref{eq:19. scale factors} gives the derivative feature 
matrices
\begin{equation}
  \boldsymbol{\Phi}_x = \frac{1}{h_x}\,(\mathbf{D}_1)_x\,\boldsymbol{\Phi},
  \qquad
  \boldsymbol{\Phi}_{xx} = \frac{1}{h_x^2}\,(\mathbf{D}_2)_x\,\boldsymbol{\Phi},
  \qquad
  \boldsymbol{\Phi}_y = \frac{1}{h_y}\,(\mathbf{D}_1)_y\,\boldsymbol{\Phi},
  \qquad
  \boldsymbol{\Phi}_{yy} = \frac{1}{h_y^2}\,(\mathbf{D}_2)_y\,\boldsymbol{\Phi},
  \label{eqn:23 . precomputed matrices}
\end{equation}
each in $\mathbb{R}^{N\times n_f}$. All five matrices 
$\boldsymbol{\Phi},\,\boldsymbol{\Phi}_x,\,\boldsymbol{\Phi}_{xx},\,
\boldsymbol{\Phi}_y,\,\boldsymbol{\Phi}_{yy}$ are computed once and stored 
as fixed buffers; no differentiation is performed during training.


\subsubsection{Bottleneck projection}
\label{subsubsec:3.1.3 Bottleneck projection}

The $n_f$ spectral features are projected onto $B$ latent channels, 
where $B \ll n_f$, using a trainable weight matrix 
$\mathbf{W} \in \mathbb{R}^{n_f \times B}$,
\begin{equation}
  \mathbf{Z} = \boldsymbol{\Phi}\,\mathbf{W} \in \mathbb{R}^{N \times B}.
  \label{eqn:24 . bottleneck}
\end{equation}
This reduces the input dimension of the KAN layer from $n_f$ to $B$, so 
the number of trainable Chebyshev coefficients scales with $B$ instead of 
$n_f$. The bottleneck width $B$ is a hyperparameter that controls the 
trade-off between accuracy and parameter count. The $b$-th column of 
$\mathbf{Z}$, written $\mathbf{z}_b \in \mathbb{R}^{N}$ for 
$b = 1,\ldots,B$, is the $b$-th latent channel evaluated at all $N$ 
collocation points. Since $\mathbf{W}$ does not depend on $(x,y)$, differentiating \Cref{eqn:24 . bottleneck} gives

\begin{equation}
  \mathbf{Z}_x = \boldsymbol{\Phi}_x\,\mathbf{W},
  \quad
  \mathbf{Z}_{xx} = \boldsymbol{\Phi}_{xx}\,\mathbf{W},
  \quad
  \mathbf{Z}_y = \boldsymbol{\Phi}_y\,\mathbf{W},
  \quad
  \mathbf{Z}_{yy} = \boldsymbol{\Phi}_{yy}\,\mathbf{W},
  \label{eq:25. Zderiv}
\end{equation}

where $\boldsymbol{\Phi}_x, \boldsymbol{\Phi}_{xx}, \boldsymbol{\Phi}_y, 
\boldsymbol{\Phi}_{yy}$ are the precomputed matrices 
from \Cref{eqn:23 . precomputed matrices}. For instance, $(\mathbf{Z}_x)_{m,b} = 
\partial z_{m,b} / \partial x$ at the $m$-th collocation point. No automatic 
differentiation is involved. The detailed derivation is given in \ref{sec:C. Linearity of the Bottleneck Derivatives}.


\subsubsection{KAN activation layer and analytic jet}
\label{subsubsec:3.1.4 KAN activation layer and analytic jet}

Each latent channel $\mathbf{z}_b \in \mathbb{R}^N$ from \Cref{eqn:24 . bottleneck} is passed through a learnable univariate activation $\phi_{b,o}$, defined as a 
Chebyshev series of degree $P$ in the same form as the CKAN activation 
from \Cref{eq:7. ckan univariate},
\begin{equation}
  \phi_{b,o}(z_b) = \sum_{n=0}^{P} c_{b,o,n}\,T_n(s_b),
  \qquad s_b = \tanh(z_b),
  \quad b = 1,\ldots,B, \quad o = 1,\ldots,O,
  \label{eq:26. kan layer}
\end{equation}
where $z_b$ and $s_b$ denote values at a single collocation point (the index $m$ is omitted), $c_{b,o,n}$ are the trainable Chebyshev coefficients, $P$ is the 
polynomial degree as defined in \Cref{subsec:2.1. CKAN}, $B$ is the bottleneck 
width from \Cref{eqn:24 . bottleneck}, and $O$ is the number of output channels, 
which equals $3$ for the Navier--Stokes fields $(u, v, p)$. The $\tanh$ 
nonlinearity maps $z_b$ to $s_b \in (-1,1)$, the domain of the Chebyshev 
polynomials $T_n$. The predicted output for channel $o$ at all $N$ collocation 
points is
\begin{equation}
  \hat{\mathbf{U}}_o = \sum_{b=1}^{B}\phi_{b,o}(\mathbf{z}_b) \in \mathbb{R}^N,
  \label{eq:27. velocity}
\end{equation}
and the full output is $\hat{\mathbf{U}} = [\hat{\mathbf{U}}_1,\ldots,\hat{\mathbf{U}}_O] 
\in \mathbb{R}^{N \times O}$.

\paragraph{Derivatives of the KAN activation.}
To propagate derivatives through~\Cref{eq:26. kan layer}, define the Chebyshev 
derivative sums for each channel pair $(b,o)$,
\begin{equation}
  g_{b,o}^{(1)} = \sum_{n=0}^{P}c_{b,o,n}\,T_n'(s_b),
  \qquad
  g_{b,o}^{(2)} = \sum_{n=0}^{P}c_{b,o,n}\,T_n''(s_b),
  \label{eq:gcoeffs}
\end{equation}
where $T_n'(s)$ and $T_n''(s)$ are the first and second derivatives of 
the $n$-th Chebyshev polynomial evaluated at $s_b = \tanh(z_b)$. Let 
$\mathbf{T}(s_b) = [T_0(s_b),\ldots,T_P(s_b)]^{\top} \in \mathbb{R}^{P+1}$ 
denote the vector of Chebyshev polynomial values at $s_b$. The derivative 
values $[T_0'(s_b),\ldots,T_P'(s_b)]$ are obtained as 
$\mathbf{D}_1^{\top}\,\mathbf{T}(s_b)$, and the second derivatives as 
$\mathbf{D}_2^{\top}\,\mathbf{T}(s_b)$, using the differentiation matrices 
$\mathbf{D}_1$ and $\mathbf{D}_2$ from~\Cref{eq:22. precomputed matrices} with $K = P+1$. Let $\sigma_b = \tanh'(z_b) = 1 - s_b^2$. Applying the chain rule to 
$\phi_{b,o}(z_b)$, where the inner function is $s_b = \tanh(z_b)$ and the 
outer function is the Chebyshev series in~\Cref{eq:26. kan layer}, gives
\begin{equation}
  \frac{\mathrm{d}\phi_{b,o}}{\mathrm{d}z_b}
    = g_{b,o}^{(1)}\,\sigma_b,
  \qquad
  \frac{\mathrm{d}^2\phi_{b,o}}{\mathrm{d}z_b^2}
    = g_{b,o}^{(2)}\,\sigma_b^2 
      - 2\,s_b\,\sigma_b\,g_{b,o}^{(1)}.
  \label{eq:29. dphidzb}
\end{equation}
The full step-by-step derivation expanding every product-rule and chain-rule 
term is given in \ref{sec:D. Derivation of the Analytic Jet}.

\paragraph{Propagating to physical coordinates (the spectral jet).}
Since each $z_b$ depends on $(x,y)$ through the bottleneck 
projection~\Cref{eqn:24 . bottleneck} and its 
derivatives~\Cref{eq:25. Zderiv}, the chain rule gives the first and second 
derivatives of $\hat{U}_o$ from~\Cref{eq:27. velocity} with respect to the 
physical coordinates,
\begin{align}
  \frac{\partial\hat{U}_o}{\partial x}
    &= \sum_{b=1}^{B}
         \frac{\mathrm{d}\phi_{b,o}}{\mathrm{d}z_b}
         \,
         \frac{\partial z_b}{\partial x},
  \label{eq:30. dudx}\\[6pt]
  \frac{\partial^2\hat{U}_o}{\partial x^2}
    &= \sum_{b=1}^{B}
       \left[
         \frac{\mathrm{d}^2\phi_{b,o}}{\mathrm{d}z_b^2}
         \!\left(\frac{\partial z_b}{\partial x}\right)^{\!2}
         +
         \frac{\mathrm{d}\phi_{b,o}}{\mathrm{d}z_b}
         \,
         \frac{\partial^2 z_b}{\partial x^2}
       \right],
  \label{eq:31. d2udx2}
\end{align}
and analogously for the $y$-direction. Here the sum runs over all $B$ 
bottleneck channels, the terms 
$\partial z_b/\partial x$ and $\partial^2 z_b/\partial x^2$ are read 
from the columns of $\mathbf{Z}_x$ and $\mathbf{Z}_{xx}$ 
in~\eqref{eq:25. Zderiv}, and the activation derivatives 
$\mathrm{d}\phi_{b,o}/\mathrm{d}z_b$ and 
$\mathrm{d}^2\phi_{b,o}/\mathrm{d}z_b^2$ are given 
by~\Cref{eq:29. dphidzb}. Collecting the output and all its required spatial derivatives defines the 
\emph{spectral jet},
\begin{equation}
  \mathrm{Jet}(\hat{\mathbf{U}}) =
  \Bigl(\hat{\mathbf{U}},\;
        \tfrac{\partial\hat{\mathbf{U}}}{\partial x},\;
        \tfrac{\partial^2\hat{\mathbf{U}}}{\partial x^2},\;
        \tfrac{\partial\hat{\mathbf{U}}}{\partial y},\;
        \tfrac{\partial^2\hat{\mathbf{U}}}{\partial y^2}\Bigr),
  \label{eq:32. jet}
\end{equation}
where each component is in $\mathbb{R}^{N \times O}$. The viscous and 
pressure terms in the Navier--Stokes residual, including the Laplacian 
$\Delta\hat{U}_o = \partial^2\hat{U}_o/\partial x^2 + 
\partial^2\hat{U}_o/\partial y^2$, are assembled directly from these 
components. Every term in~\Cref{eq:32. jet} is computed from Chebyshev 
evaluations at $s_b$ and matrix multiplications with the precomputed 
matrices; automatic differentiation is used only in the backward pass to 
compute the gradients of the loss with respect to the trainable parameters 
$\mathbf{W}$ and $c_{b,o,n}$, not for the PDE derivatives themselves.


\subsubsection{Model variants}
\label{subsubsec: 3.1.5 Model variants}

We consider three variants of SJ-KAN, named JetKAN, JetKAN-F, and
JetKAN-H. All three use the spectral basis, the bottleneck projection, and
the analytic jet described in
\Cref{subsubsec:3.1.1 Coordinate normalization and feature matrix,subsubsec:3.1.4 KAN activation layer and analytic jet}.
They differ only in how the output is formed from these components.
JetKAN and JetKAN-F are two implementations of the same model, while
JetKAN-H adds a linear output path. Each variant is described below, and
Table~\ref{tab:arch-summary} lists them together with the two baselines.


\paragraph{JetKAN (einsum variant).}
This is the baseline SJ-KAN implementation. The Chebyshev coefficients from 
\Cref{eq:26. kan layer} are stored as a single tensor 
$\mathbf{c} \in \mathbb{R}^{B \times O \times (P+1)}$. The output 
$\hat{\mathbf{U}} \in \mathbb{R}^{N \times O}$ from~\Cref{eq:27. velocity} is 
computed by contracting over the bottleneck index $b$ and the polynomial 
index $n$ simultaneously,
\begin{equation}
  \hat{U}_{m,o} = \sum_{b=1}^{B} \sum_{n=0}^{P} 
    T_n(s_{m,b})\, c_{b,o,n},
  \qquad m = 1,\ldots,N, \quad o = 1,\ldots,O,
  \label{eq:33. jetkan}
\end{equation}
where $s_{m,b} = \tanh(z_{m,b})$ and $z_{m,b} = (\mathbf{Z})_{m,b}$ is the 
$b$-th latent channel at the $m$-th collocation point. The derivatives in the 
spectral jet~\Cref{eq:32. jet} are assembled with the same double contraction, 
replacing $T_n(s_{m,b})$ with the corresponding derivative terms 
from~\Cref{eq:29. dphidzb}--\eqref{eq:31. d2udx2}.
The tensor $\mathbf{c}$ can be written as $B$ matrices of size 
$O \times (P+1)$, one for each bottleneck channel,
\begin{equation}
  \mathbf{c} = \left\{
  \begin{bmatrix}
    c_{1,1,0} & c_{1,1,1} & \cdots & c_{1,1,P} \\
    c_{1,2,0} & c_{1,2,1} & \cdots & c_{1,2,P} \\
    \vdots    & \vdots    &        & \vdots    \\
    c_{1,O,0} & c_{1,O,1} & \cdots & c_{1,O,P}
  \end{bmatrix},
  \;\ldots,\;
  \begin{bmatrix}
    c_{B,1,0} & c_{B,1,1} & \cdots & c_{B,1,P} \\
    c_{B,2,0} & c_{B,2,1} & \cdots & c_{B,2,P} \\
    \vdots    & \vdots    &        & \vdots    \\
    c_{B,O,0} & c_{B,O,1} & \cdots & c_{B,O,P}
  \end{bmatrix}
  \right\}
  \in \mathbb{R}^{B \times O \times (P+1)}.
  \label{eq:c-tensor}
\end{equation}

\paragraph{JetKAN-F (fused variant).}
In JetKAN, the double sum in~\Cref{eq:33. jetkan} contracts over the 
bottleneck index $b$ and the polynomial index $n$ separately, which 
requires storing the intermediate Chebyshev evaluations 
$T_n(s_{m,b})$ for all $N \times B \times (P+1)$ entries before 
summing. JetKAN-F avoids this intermediate storage by reshaping the 
coefficient tensor $\mathbf{c} \in \mathbb{R}^{B \times O \times (P+1)}$ 
into a two-dimensional matrix 
$\mathbf{c}_f \in \mathbb{R}^{B(P+1) \times O}$ and similarly 
concatenating the Chebyshev evaluations along the $b$ and $n$ indices, 
so that~\Cref{eq:33. jetkan} reduces to a single matrix--matrix product,
\begin{equation}
  \hat{U}_{m,o} = \sum_{b=1}^{B} \sum_{n=0}^{P} 
    T_n(s_{m,b})\, c_{b,o,n},
  \qquad m = 1,\ldots,N, \quad o = 1,\ldots,O,
  \label{eq:34. jetkanf}
\end{equation}
where $\mathbf{T}_s \in \mathbb{R}^{N \times B(P+1)}$ collects all 
Chebyshev values $T_n(s_{m,b})$ in the same flattened order as 
$\mathbf{c}_f$. The output is mathematically identical 
to~\Cref{eq:33. jetkan}; the only difference is computational. The 
matrix--matrix product in~\Cref{eq:34. jetkanf} maps directly to a 
single GPU kernel call, whereas the einsum formulation 
in~\Cref{eq:33. jetkan} may require multiple passes depending on the 
framework backend. This fusion reduces peak memory usage by eliminating 
the $N \times B \times (P+1)$ intermediate tensor and runs faster 
on GPU hardware, particularly when $B$ and $P$ are large. 
The same reshaping applies to the derivative terms 
in~\Cref{eq:29. dphidzb}--\eqref{eq:31. d2udx2}, so all components of the 
spectral jet in~\Cref{eq:32. jet} are also computed as matrix--matrix 
products.
The matrix $\mathbf{c}_f$ holds the same coefficients as $\mathbf{c}$ 
in~\Cref{eq:c-tensor}, arranged as one matrix of size $B(P+1) \times O$,
\begin{equation}
  \mathbf{c}_f =
  \left[
  \begin{array}{cccc}
    c_{1,1,0} & c_{1,2,0} & \cdots & c_{1,O,0} \\
    c_{1,1,1} & c_{1,2,1} & \cdots & c_{1,O,1} \\
    \vdots    & \vdots    &        & \vdots    \\
    c_{1,1,P} & c_{1,2,P} & \cdots & c_{1,O,P} \\ \hline
    \vdots    & \vdots    &        & \vdots    \\ \hline
    c_{B,1,0} & c_{B,2,0} & \cdots & c_{B,O,0} \\
    c_{B,1,1} & c_{B,2,1} & \cdots & c_{B,O,1} \\
    \vdots    & \vdots    &        & \vdots    \\
    c_{B,1,P} & c_{B,2,P} & \cdots & c_{B,O,P}
  \end{array}
  \right]
  \in \mathbb{R}^{B(P+1) \times O}.
  \label{eq:cf-matrix}
\end{equation}


\paragraph{JetKAN-H (hybrid variant).}
JetKAN-H adds a trainable linear map 
$\mathbf{W}_{\mathrm{lin}} \in \mathbb{R}^{n_f \times O}$ that projects the 
full feature matrix $\boldsymbol{\Phi}$ from~\Cref{eq:21. featurematrix} 
directly to the $O$ output channels. The output of this linear 
map is added to the KAN output from~\Cref{eq:33. jetkan}, giving 
the total prediction,
\begin{equation}
  \hat{U}_{m,o} = (\boldsymbol{\Phi}\,\mathbf{W}_{\mathrm{lin}})_{m,o}
    \;+\; \sum_{b=1}^{B} \sum_{n=0}^{P} 
    T_n(s_{m,b})\, c_{b,o,n},
  \qquad m = 1,\ldots,N, \quad o = 1,\ldots,O,
  \label{eq:hybrid}
\end{equation}
where the first term is the linear spectral expansion and the second 
term is the JetKAN output from~\Cref{eq:33. jetkan}. The spatial derivatives follow the same structure. For example, the 
first derivative with respect to $x$ is
\begin{equation}
  \frac{\partial \hat{U}_{m,o}}{\partial x}
    = (\boldsymbol{\Phi}_x\,\mathbf{W}_{\mathrm{lin}})_{m,o}
    \;+\; \sum_{b=1}^{B}
      \frac{\mathrm{d}\phi_{b,o}}{\mathrm{d}z_b}
      \,\frac{\partial z_b}{\partial x}\bigg|_m,
  \label{eq:hybridderiv}
\end{equation}
and similarly for $\partial^2/\partial x^2$, $\partial/\partial y$, 
and $\partial^2/\partial y^2$. The linear head contributes 
$\boldsymbol{\Phi}_x\,\mathbf{W}_{\mathrm{lin}}$, 
$\boldsymbol{\Phi}_{xx}\,\mathbf{W}_{\mathrm{lin}}$, 
$\boldsymbol{\Phi}_y\,\mathbf{W}_{\mathrm{lin}}$, and 
$\boldsymbol{\Phi}_{yy}\,\mathbf{W}_{\mathrm{lin}}$ using the precomputed 
derivative matrices from~\Cref{eqn:23 . precomputed matrices}. The KAN branch 
contributes the jet components 
from~\Cref{eq:30. dudx}--\eqref{eq:31. d2udx2}. Both contributions are 
summed to form the spectral jet~\Cref{eq:32. jet}.

\section{Pseudo-time stepping}
As $Re$ increases, the velocity gradients near the walls become steeper to enforce the no-slip boundary condition. The residual loss depends strongly on the convergence of the nonlinear convective term. To stabilise the early phase of training, the implicit pseudo-time stepping ~\cite{zhang2025pseudo, xiong2025jpikan} is used. The
steady momentum equations are replaced by the pseudo-time residuals: \\
\begin{align}
\varepsilon_{\tau,u}^{(r+1)}
  &= \frac{\hat u^{(r+1)} - \hat u^{(r)}}{\delta\tau}
    + \mathcal{R}_u\bigl[\hat u^{(r+1)},\hat v^{(r+1)},\hat p^{(r+1)}\bigr],
    \notag\\[4pt]
\varepsilon_{\tau,v}^{(r+1)}
  &= \frac{\hat v^{(r+1)} - \hat v^{(r)}}{\delta\tau}
    + \mathcal{R}_v\bigl[\hat u^{(r+1)},\hat v^{(r+1)},\hat p^{(r+1)}\bigr],
\label{eq:pseudo_time}
\end{align}
where $r$ is the pseudo-step index, $\delta\tau$ is the pseudo-time step, and $\hat u^{(r)}$ and $\hat v^{(r)}$ are the network
predictions from the previous pseudo-step, which are held fixed
during the update. The steady residuals are
\begin{align}
\mathcal{R}_u &= \hat u\,\hat u_x + \hat v\,\hat u_y + \hat p_x
  - \frac{1}{Re}\bigl(\hat u_{xx} + \hat u_{yy}\bigr), \notag\\
\mathcal{R}_v &= \hat u\,\hat v_x + \hat v\,\hat v_y + \hat p_y
  - \frac{1}{Re}\bigl(\hat v_{xx} + \hat v_{yy}\bigr), \notag\\
\mathcal{R}_c &= \hat u_x + \hat v_y,
\label{eq:ldc_residuals}
\end{align}
with all derivatives taken from the spectral jet in
Eq.~\eqref{eq:32. jet}. Continuity has no pseudo-time term. The training loss is
\begin{equation}
\mathcal{L}_\tau
  = \frac{\lambda_R}{N_R}\sum_{m=1}^{N_R}
    \Bigl[\varepsilon_{\tau,u}(\mathbf{x}_m)^{2}
         +\varepsilon_{\tau,v}(\mathbf{x}_m)^{2}
         +\mathcal{R}_c(\mathbf{x}_m)^{2}\Bigr]
  + \lambda_{BC}\,\mathcal{L}_{BC},
\label{eq:pseudo_time_loss}
\end{equation}
which is Eq.~\eqref{eqn:total_loss} with the steady momentum residuals replaced by the pseudo-time residuals and without an initial-condition term. The pseudo-time term penalizes large changes of the velocity between two pseudo-steps and so prevents large jumps of the solution early in training. As training converges, $\hat u^{(r+1)}$ approaches $\hat u^{(r)}$, the pseudo-time term vanishes, and $\mathcal{L}_\tau$ reduces to the steady residual loss. 

For $Re\geq1000$, the quasi-Newton L-BFGS method ~\cite{liu1989limited} is used for the optimizer in each epoch~\cite{rathore2024challenges}. At $Re=5000$ and $Re=10{,}000$, the residual stays large near the walls and in the corners while it is small in the interior. For these two cases,  two additinal techniques are used: i) balanced residual decay rate (BRDR) weighting~\cite{chen2025self}, which adjusts the weights of the momentum and continuity residuals during training so that no single term controls the gradient; and ii) Residual-based adaptive refinement (RAR)~\cite{lu2021deepxde}, which adds collocation points where the residual is largest, which places more points in the wall layers and the corner regions.

\section{Error Metrics}  \label{sec:error_metrics}
Two error metrics used to evaluate the accuracy of the predicted solution fields $\hat{u}$, $\hat{v}$, $\hat{p}$ against reference values $u^{\mathrm{ref}}$, $v^{\mathrm{ref}}$, $p^{\mathrm{ref}}$ at the $N$ collocation points $m = 1,\ldots,N$. Each predicted field is one output channel of the network: $\hat{\mathbf{U}}_o \in \mathbb{R}^{N}$ from~\Cref{eq:27. velocity}, with $o = 1, 2, 3$ corresponding to $\hat{u}$, $\hat{v}$, and $\hat{p}$, respectively. Its entries are $\hat{U}_{m,o}$, and $\mathbf{U}^{\mathrm{ref}}_o \in \mathbb{R}^{N}$ with entries $U^{\mathrm{ref}}_{m,o}$ denotes the corresponding reference field. The mean squared error measures the average pointwise squared difference between the prediction and the reference. For a single output field $o$, it is defined as:

\begin{equation}
  \mathrm{MSE}_o
  = \frac{1}{N}\sum_{m=1}^{N}
    \bigl(\hat{U}_{m,o} - U_{m,o}^{\mathrm{ref}}\bigr)^2.
  \label{eq:mse}
\end{equation}

When all three solution fields are evaluated jointly, the MSE averages over both the $N$ collocation points and the three fields,
\begin{equation}
  \mathrm{MSE}
  = \frac{1}{3N}\sum_{m=1}^{N}
    \Bigl[
      \bigl(\hat{u}_m - u_m^{\mathrm{ref}}\bigr)^2
    + \bigl(\hat{v}_m - v_m^{\mathrm{ref}}\bigr)^2
    + \bigl(\hat{p}_m - p_m^{\mathrm{ref}}\bigr)^2
    \Bigr],
  \label{eq:mse_uvp}
\end{equation}
where $\hat{u}_m$, $\hat{v}_m$, $\hat{p}_m$ are the predicted horizontal velocity, vertical velocity, and pressure at point $m$, and $u_m^{\mathrm{ref}}$, $v_m^{\mathrm{ref}}$,
$p_m^{\mathrm{ref}}$ are the corresponding reference values. The relative $L^2$ error measures the norm of the pointwise error relative to the norm of the reference field. For a single output field $o$, it is defined as:
\begin{equation}
  \mathrm{Rel}\,L^2_o
  = \frac{\|\hat{\mathbf{U}}_o - \mathbf{U}^{\mathrm{ref}}_o\|_2}
         {\|\mathbf{U}^{\mathrm{ref}}_o\|_2}
  = \frac{\displaystyle\sqrt{\sum_{m=1}^{N}
            \bigl(\hat{U}_{m,o} - U_{m,o}^{\mathrm{ref}}\bigr)^2}}
         {\displaystyle\sqrt{\sum_{m=1}^{N}
            \bigl(U_{m,o}^{\mathrm{ref}}\bigr)^2}},
  \label{eq:rel_l2}
\end{equation}
where the sums run over all $N$ collocation points. For $o = 1, 2, 3$, 
these metrics are denoted $\mathrm{MSE}(u)$, $\mathrm{MSE}(v)$, $\mathrm{MSE}(p)$ and 
$\mathrm{Rel}\,L^2(u)$, $\mathrm{Rel}\,L^2(v)$, $\mathrm{Rel}\,L^2(p)$, respectively. 

\section{Results and discussion}
\subsection{Numerical Experiments}
\label{sec:results}
The performance comparison between the proposed \textit{SJ-PIKAN} variants (\textit{JetKAN}, \textit{JetKAN-F}, \textit{JetKAN-H}) against the autograd baselines (\textit{CKAN} and \textit{PINN}) are carried out using the architectures listed in Table \ref{tab:arch-summary}. Two classical benchmarks for the two-dimensional incompressible Navier--Stokes equations are used to demonstrate the performance: (i)~the Taylor--Green vortex, and (ii)~the lid-driven cavity flow are examined as shown below. All computations were performed on the Thunder Prime cluster of the Center for Computationally Assisted Science and Technology (CCAST) at North Dakota State University. Each case was run on a single NVIDIA A100 GPU. The models were implemented in PyTorch 2.8.0 with CUDA 12.6.

\subsubsection{Case (a): Taylor--Green Vortex}
\label{subsec:tgv}

The classical Taylor--Green vortex~\cite{taylor1937mechanism} is defined on a
square domain $\Omega = [0, 2\pi]^2$. The time-dependent velocity
and pressure fields are given by:
\begin{align}
  u(x, y, t) &= \cos(x)\sin(y)\,e^{-2\nu t},
    \label{eq:tgv_u} \\
  v(x, y, t) &= - \sin(x)\cos(y)\,e^{-2\nu t},
    \label{eq:tgv_v} \\
  p(x, y, t) &= -\tfrac{1}{4}
    \bigl(\cos(2x) + \cos(2y)\bigr)\,e^{-4\nu t},
    \label{eq:tgv_p}
\end{align}
where $\nu = 1/Re$ is the kinematic viscosity. 

The steady-state solution of Taylor-Green vortex ($t = 0$) is thus:\\
\begin{align}
  u(x, y) &= \cos(x)\sin(y), \\
  v(x, y) &= - \sin(x)\cos(y), \\
  p(x, y) &= -\tfrac{1}{4}
    \bigl(\cos(2x) + \cos(2y)\bigr).
\end{align}

These fields serve as the reference solution to compute the error in ~\Cref{eq:rel_l2}. For all cases in this work, the Reynolds number is set to be $Re = 100$. For this problem, the analytical solution is included as an additional exact-solution loss term alongside the PDE and boundary losses.

\begin{figure}[htbp]
    \centering
    \includegraphics[width=0.8\linewidth]{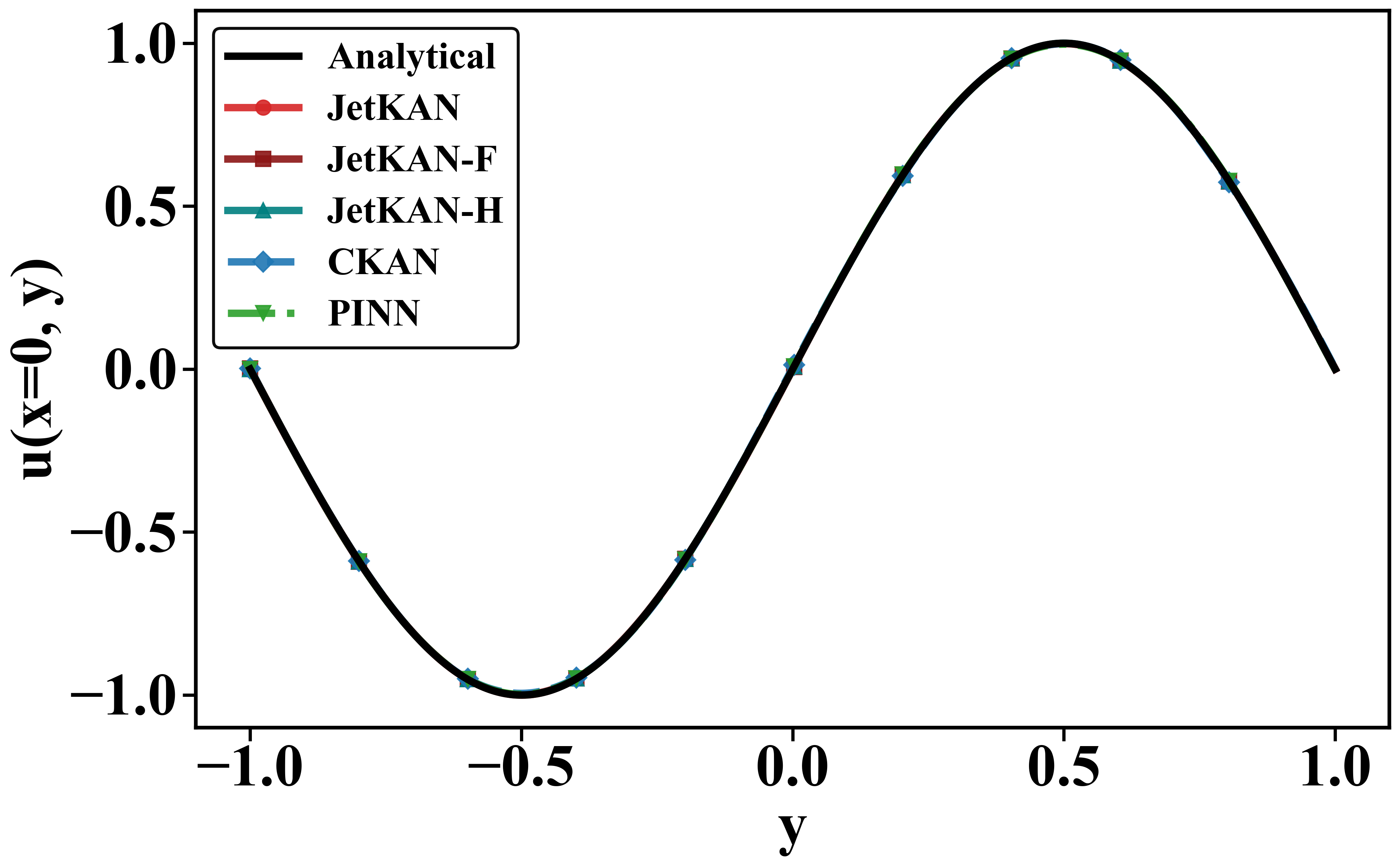}
    \par\noindent\hspace*{0.1\linewidth}\textbf{(a)}
    \vspace{0.5em}

    \includegraphics[width=0.8\linewidth]{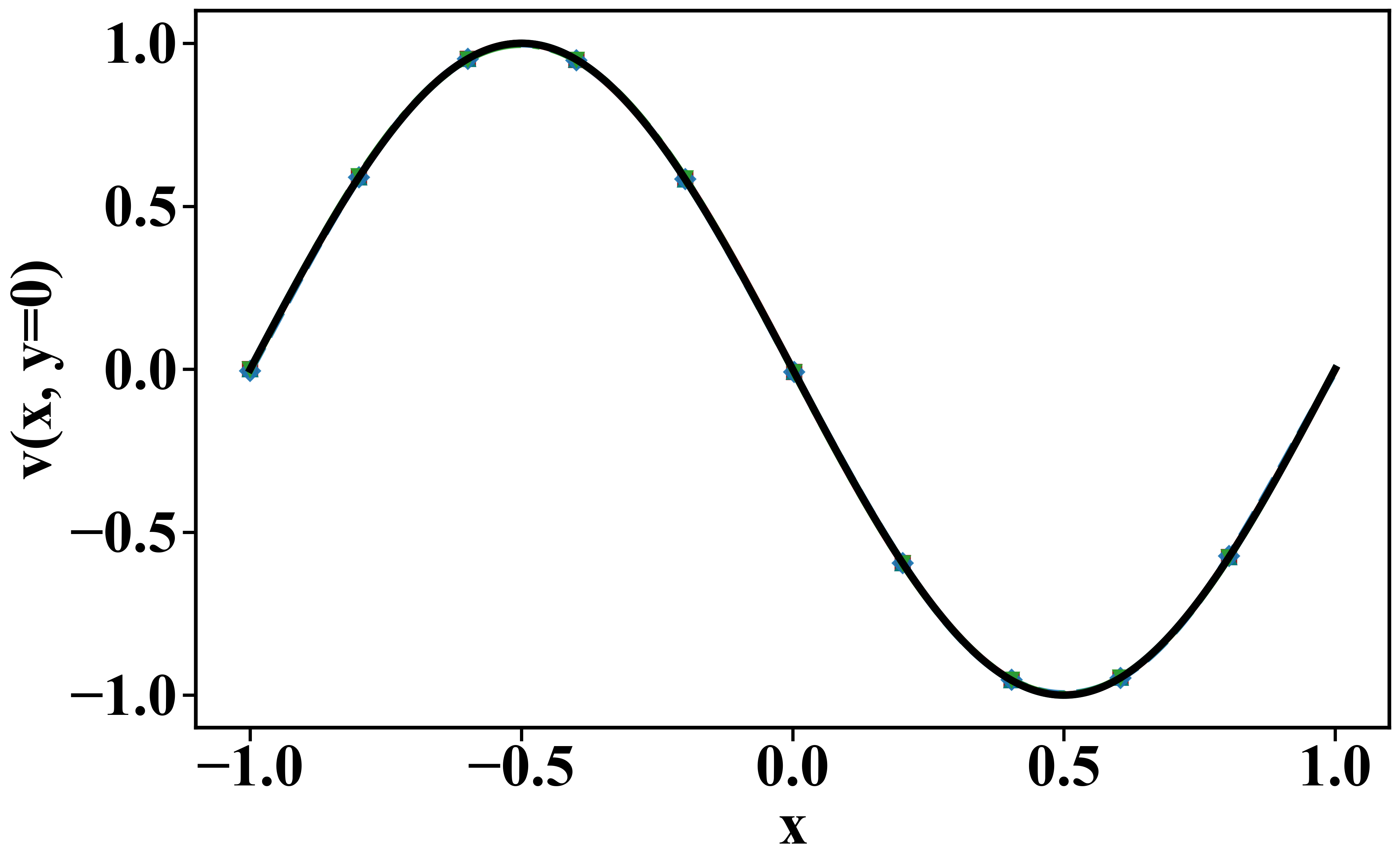}
    \par\noindent\hspace*{0.1\linewidth}\textbf{(b)}
    \vspace{0.5em}

    \caption{Centerline velocity profile validation for the 2D Taylor--Green vortex at $Re = 100$, comparing the SJ-PIKAN family (JetKAN, JetKAN-F, JetKAN-H) and the autograd baselines (CKAN, PINN) against the analytical solution (solid black). (a)~Horizontal velocity component $u$ along the vertical centerline $u(x=0, y)$ at $t = 0$. All five models reproduce the sinusoidal profile, including the zero-crossings and peak magnitudes, to within visual accuracy. (b)~Vertical velocity component $v$ along the horizontal centerline $v(x, y=0)$ at $t = 0$. Consistent agreement with the analytical solution is observed across all models, confirming that each architecture correctly captures the primary vortical structure and local velocity gradients governing viscous dissipation at this Reynolds number.}
    \label{fig:centerline}
\end{figure}

The comparison between the proposed \textit{SJ-PIKAN} variants (\textit{JetKAN}, \textit{JetKAN-F}, \textit{JetKAN-H}) against the autograd baselines (\textit{CKAN} and \textit{PINN}) are shown in Fig.~\ref{fig:centerline}. In Fig.~\ref{fig:centerline}, panel~(a) shows the $u$-velocity along the vertical centerline $x = \pi$, and panel~(b) shows the $v$-velocity along the horizontal centerline $y = \pi$. The three \textit{SJ-PIKAN} variants agree with the analytical solutions of $u$ and $v$ across the full domain. Despite their convergence performance, both \textit{CKAN} and \textit{PINN} agree well with the analytical solution along the centerline and resolve the primary flow structure as shown in Table~\ref{tab:tgv_results}. In short, all five models agree well with the analytical solution for this Taylor--Green vortex at $Re = 100$.

\begin{figure}[htbp]
    \centering
    \includegraphics[width=0.8\linewidth]{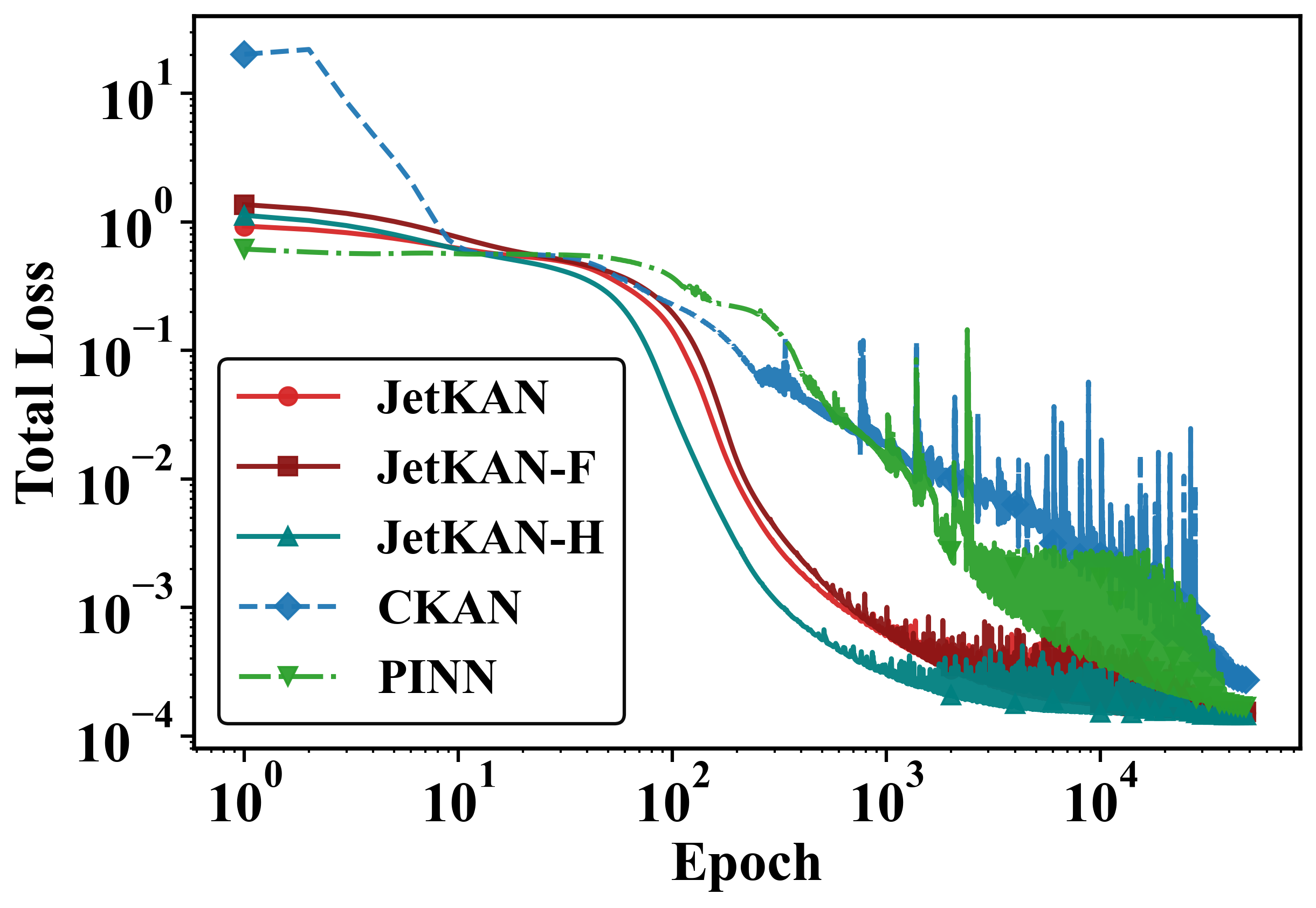}
    \par\noindent\hspace*{0.1\linewidth}\textbf{(a)}
    \vspace{0.5em}

    \includegraphics[width=0.8\linewidth]{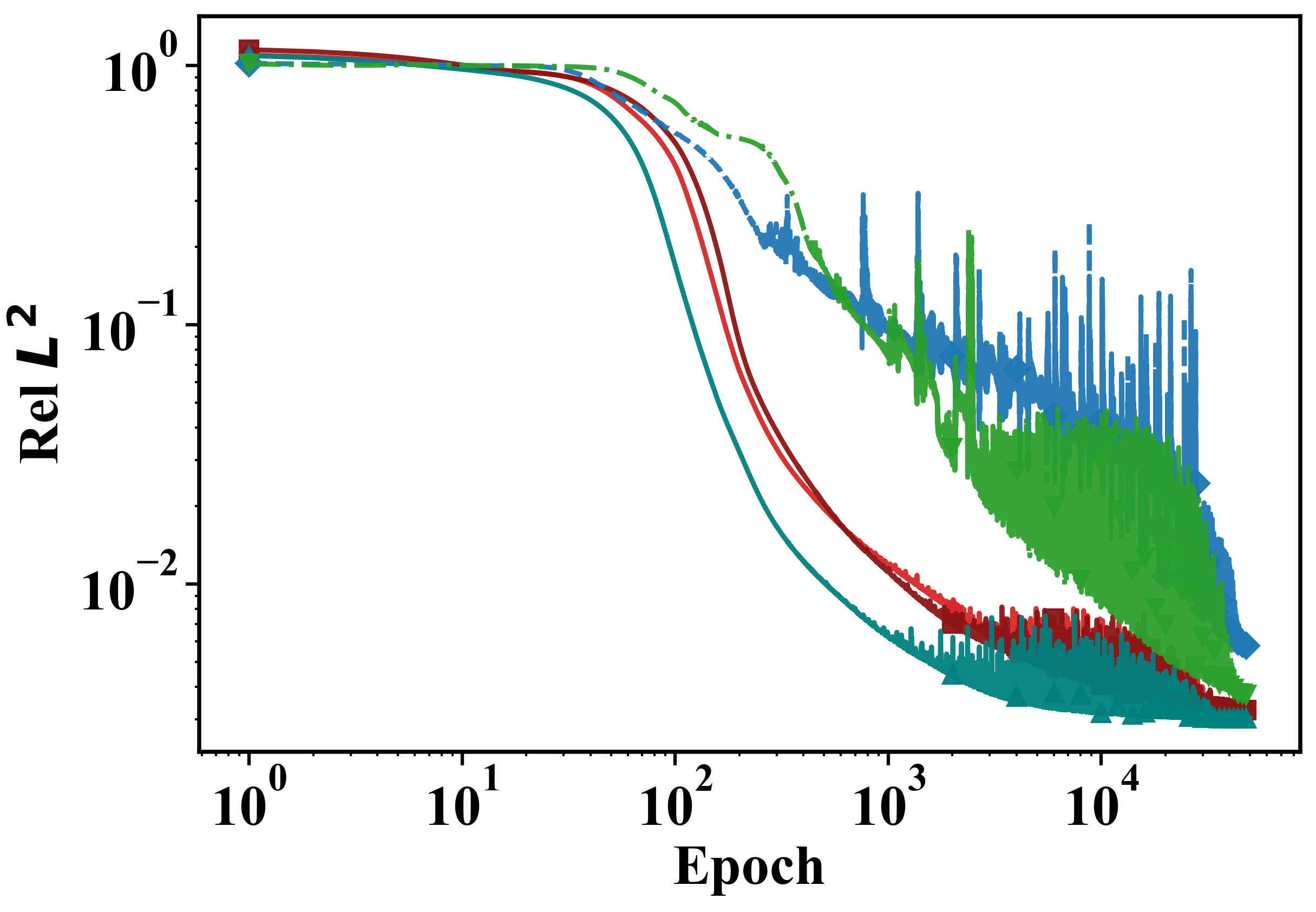}
    \par\noindent\hspace*{0.1\linewidth}\textbf{(b)}
    \vspace{0.5em}

    \caption{Convergence history of the SJ-PIKAN family (JetKAN, JetKAN-F,
    JetKAN-H) and the autograd baselines (CKAN, PINN) on the 2D
    Taylor--Green vortex at $Re = 100$. All models are trained for
    50{,}000 epochs with Adam at a learning rate of $10^{-3}$.
    \textbf{(a)}~Evolution of the total loss on a logarithmic scale.
    The SJ-PIKAN variants converge steadily to $\sim\!10^{-4}$, while
    CKAN and PINN exhibit prolonged plateaus with large persistent
    oscillations, reflecting the difficulty of simultaneously satisfying
    the nonlinear momentum residual and the incompressibility constraint.
    \textbf{(b)}~Evolution of the $\mathrm{Rel}~L^{2}$ error against the
    analytical Taylor--Green solution. The SJ-PIKAN family converges to
    $\mathrm{Rel}~L^{2}$ errors in the range $10^{-3}$, with
    JetKAN-H achieving the lowest final value; CKAN and PINN show
    slower and less stable descent throughout the training window.}
    \label{fig:tgv_loss_l2}
\end{figure}

The training loss histories in Fig.~\ref{fig:tgv_loss_l2}(a) and Table~\ref{tab:tgv_results} show that $AD-based$ methods (\textit{CKAN} and \textit{PINN}) show rapid initial descent followed by a prolonged plateau with persistent oscillations that continue to the end of the training window.These oscillations are a known behavior of the Adam optimizer at a fixed learning rate. As training goes on, the loss surface becomes sharper until the step size is too large for it, so the optimizer overshoots and the loss jumps; this dynamical instability is described by Cohen et al. \cite{cohen2022adaptive}. Consequently, \textit{PINN} reaches an MSE of $7.77 \times 10^{-6}$ with $\mathrm{Rel}\;L^2$ errors of $3.53 \times 10^{-3}$, $3.55 \times 10^{-3}$, and $5.27 \times 10^{-3}$ for $u$, $v$, and $p$. CKAN stagnates at an MSE of $1.84 \times 10^{-5}$ with $\mathrm{Rel}\;L^2$ errors of
$4.95 \times 10^{-3}$, $4.88 \times 10^{-3}$, and $6.81 \times 10^{-3}$ for $u$, $v$, and $p$, which are approximately $1.6$--$1.7\times$ larger than the SJ-PIKAN values. This higher error comes at a computational cost of $144.8$~ms/epoch, which is $18.6\times$ more expensive than JetKAN-F at $7.8$~ms/epoch. 

The $\mathrm{Rel}\;L^2$ error histories in Fig.~\ref{fig:tgv_loss_l2}(b) and the error metrics in Table~\ref{tab:tgv_results} show that the \textit{SJ-PIKAN} variants navigate the coupled nonlinear landscape more effectively than the $AD-based$ methods. \textit{JetKAN} and \textit{JetKAN-F} descend steeply in the initial epochs and continue to decrease steadily throughout the full $50{,}000$ epoch training window, reaching final total losses of $5.69 \times 10^{-6}$ and $5.96 \times 10^{-6}$, respectively. The \textit{JetKAN-H} follows a nearly identical trajectory and achieves the lowest final loss of $5.17 \times 10^{-6}$ among all five models.

The three SJ-PIKAN variants achieve final MSE values in the range $5.17 \times 10^{-6}$ to $5.96 \times 10^{-6}$, with JetKAN-H at the lower end and JetKAN-F at the upper end, though the spread across the family is small. The per-field $\mathrm{Rel}\;L^2$ errors are similarly close across the three variants. JetKAN-H achieves $2.86 \times 10^{-3}$,
$2.83 \times 10^{-3}$, and $4.18 \times 10^{-3}$ for $u$, $v$, and $p$, respectively; JetKAN yields $2.99 \times 10^{-3}$, $3.05 \times 10^{-3}$, and $4.23 \times 10^{-3}$; and JetKAN-F yields $3.10 \times 10^{-3}$, $3.07 \times 10^{-3}$, and
$4.33 \times 10^{-3}$. 

The \textit{SJ-PIKAN} variants achieve comparable or lower errors than \textit{PINN} and \textit{CKAN} with roughly $20\times$ fewer parameters and up to $2.0\times$ shorter wall-clock time per epoch, and they outperform \textit{CKAN} in both accuracy and speed by a wide margin. The \textit{MSE} values of \textit{SJ-PIKAN} is approximately $3.1 \times$ lower than one of \textit{CKAN} method. Moreover, the training time for $CKAN$ is $144.8$~ms/epoch, which is $18.6\times$ more expensive than \textit{JetKAN-F} ($7.8$~ms/epoch). Given similar \textit{MSE} values, \textit{PINN} requires $12{,}867$ parameters and $15.7$~ms/epoch, compared to only $656$ parameters and $7.8$~ms/epoch for \textit{JetKAN-F}. In short, \textit{JetKAN-F} is 18.6 times faster per epoch than \textit{CKAN} and 2.0 times faster than \textit{PINN} on the Taylor--Green vortex against the \textit{AD}-based baselines.

\subsubsection{Case (b): Lid-driven cavity flow}
The lid-driven cavity is defined in a square cavity $\Omega=[0,1]^2$ with the top wall moving tangentially with unit velocity. The boundary conditions are
\begin{align}
u(x, 1) &= 1, \quad v(x, 1) = 0 \quad &\text{(top boundary, moving lid)}, \\
u(x, 0) &= 0, \quad v(x, 0) = 0 \quad &\text{(bottom boundary, no-slip)}, \\
u(0, y) &= 0, \quad v(0, y) = 0 \quad &\text{(left boundary, no-slip)}, \\
u(1, y) &= 0, \quad v(1, y) = 0 \quad &\text{(right boundary, no-slip)}.
\end{align}

The reference solution is taken from a Computational Fluid Dynamics simulation ($128 \times 128$ grid) using the finite volume solver CURVIB~\citep{Ge2007}. The CURVIB solver has been extensively validated for a range of incompressible flow problems~\citep{Le2010, Le2012, Le2019}. Different values of Reynolds number at $Re = 100$, 1000, 5000 and 10{,}000 are used as shown in Table~\ref{tab:ldc_all_re}. 

\begin{figure}[htbp]
    \centering
    \includegraphics[width=1.0\linewidth]{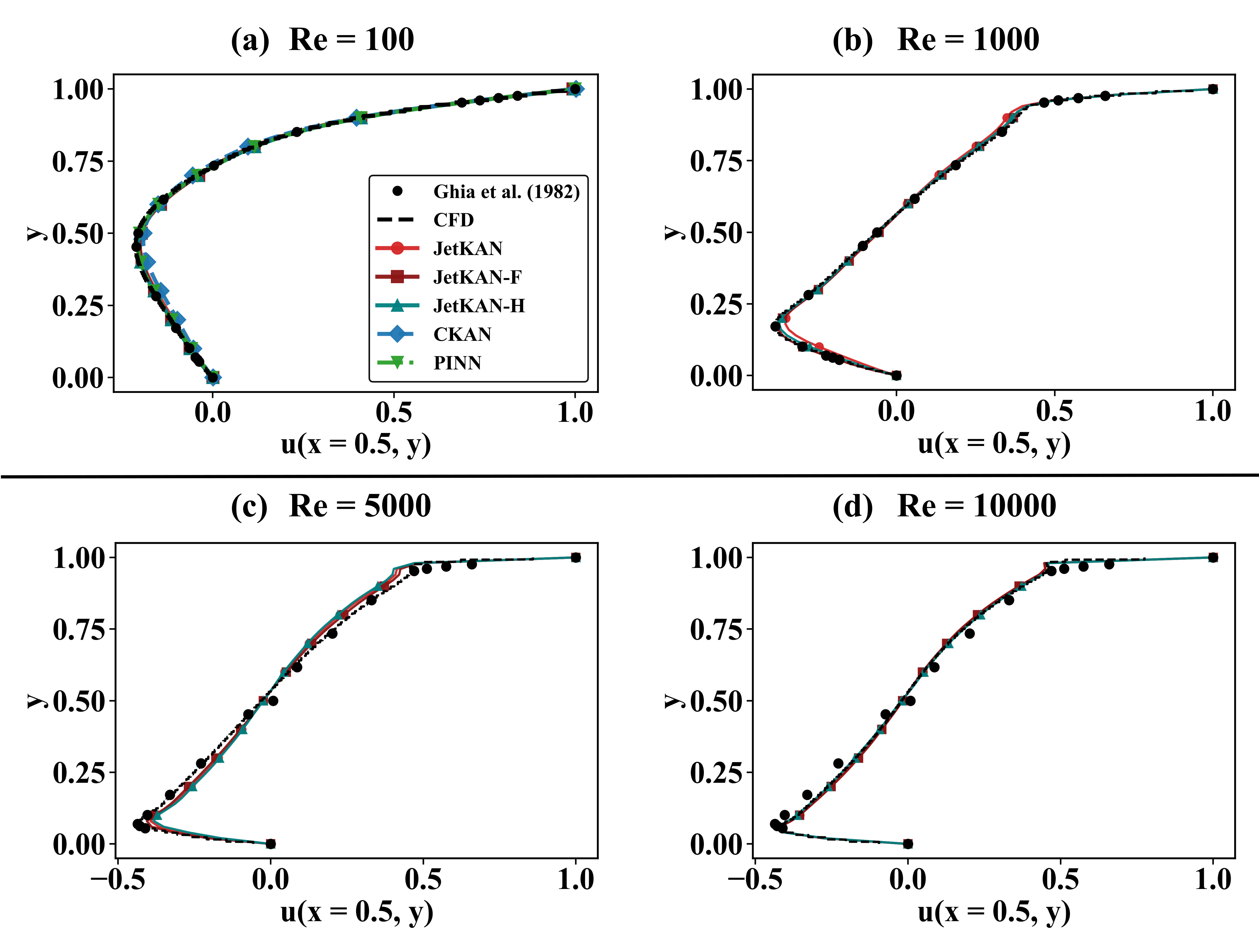}
    \caption{Centerline velocity $u(x=0.5,y)$ for the 2D lid-driven cavity
flow at (a) $Re=100$, (b) $Re=1000$, (c) $Re=5000$, and
(d) $Re=10{,}000$. The SJ-KAN variants (JetKAN, JetKAN-F, JetKAN-H)
are compared with the in-house finite-volume solution (CFD, dashed
line) and the reference data of Ghia et al.~\cite{ghia1982highre}
(filled circles). The autograd baselines CKAN and PINN are shown only
at $Re=100$, since CKAN is not run at higher Reynolds numbers because
of its cost and PINN did not converge to the correct solution at
$Re=1000$. The SJ-KAN variants agree with the reference profiles at
all four Reynolds numbers.}
    \label{fig:u_profiles}
\end{figure}

\begin{figure}[htbp]
    \centering
    \includegraphics[width=1.0\linewidth]{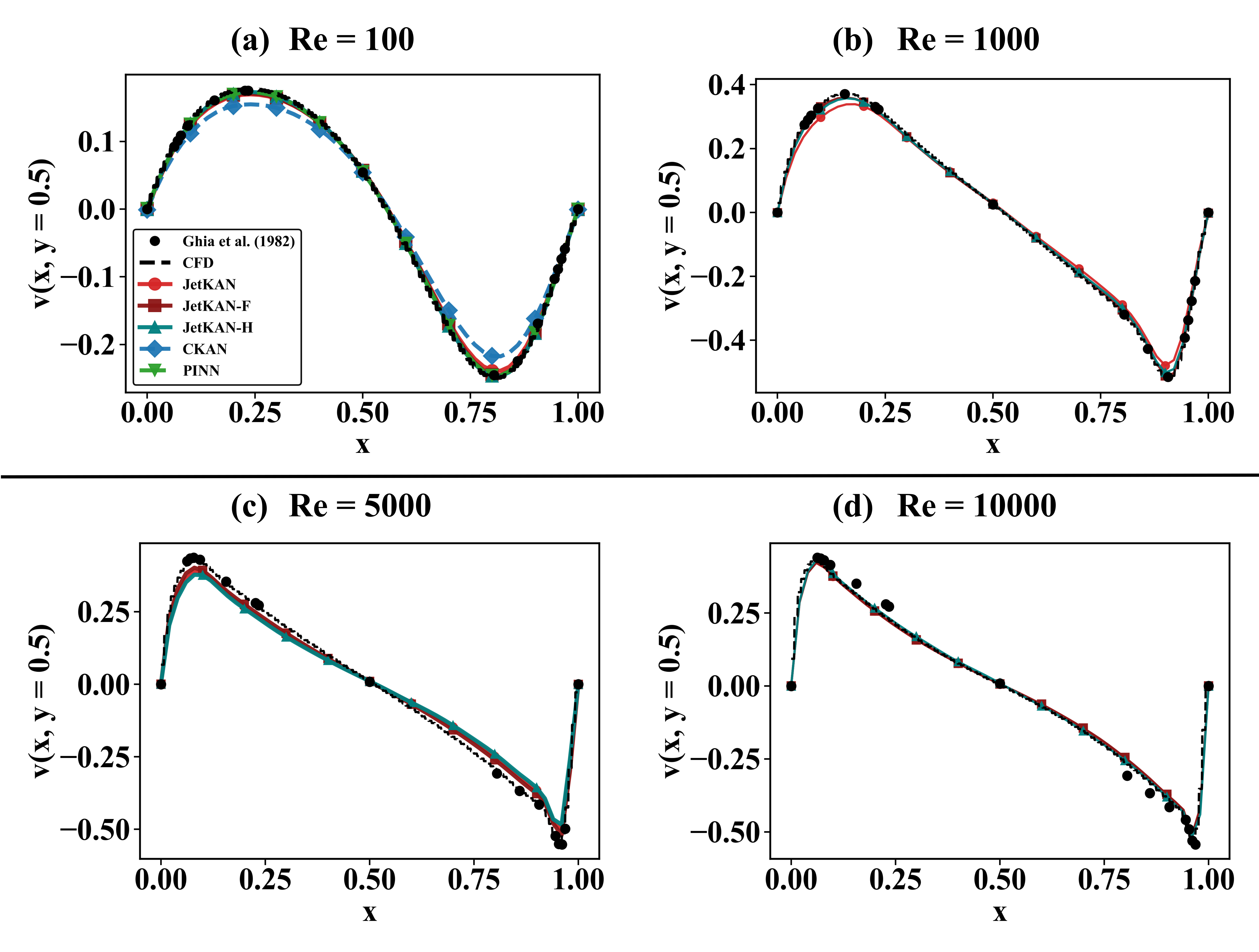}
    \caption{Centerline velocity $v(x,y=0.5)$ for the 2D lid-driven cavity
flow at (a) $Re=100$, (b) $Re=1000$, (c) $Re=5000$, and
(d) $Re=10{,}000$. The SJ-KAN variants (JetKAN, JetKAN-F, JetKAN-H)
are compared with the in-house finite-volume solution (CFD, dashed
line) and the reference data of Ghia et al.~\cite{ghia1982highre}
(filled circles). The autograd baselines CKAN and PINN are shown only
at $Re=100$, since CKAN is not run at higher Reynolds numbers because
of its cost and PINN did not converge to the correct solution at
$Re=1000$. The SJ-KAN variants agree with the reference profiles at
all four Reynolds numbers.}
    \label{fig:v_profiles}
\end{figure}

\begin{figure}[htbp]
    \centering
    \includegraphics[width=1.0\linewidth]{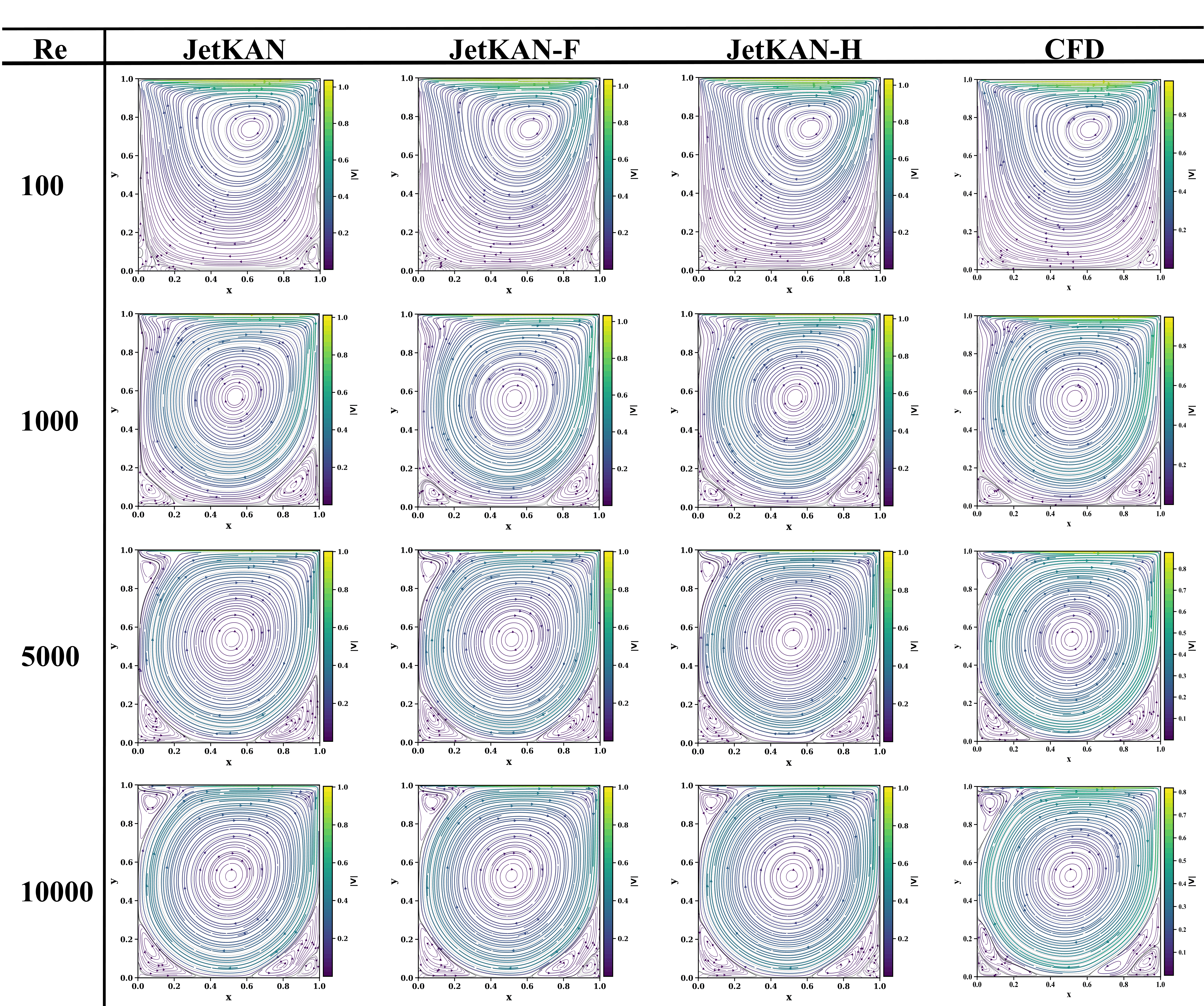}
    \caption{Streamlines of the predicted velocity field for the 2D lid-driven
cavity flow at $Re = 100$, $1000$, $5000$, and $10{,}000$ (rows), comparing
JetKAN, JetKAN-F, and JetKAN-H against the in-house CFD solution (columns).
Streamlines are colored by the local velocity magnitude $|\boldsymbol{u}|$,
with each panel sharing the same colorbar range from $0$ (dark blue) to $1$
(yellow).
At all four Reynolds numbers, JetKAN, JetKAN-F, and JetKAN-H reproduce the
correct cavity topology: a single primary vortex spanning the cavity with
small secondary vortices near the bottom corners that grow in size as $Re$
increases from $100$ to $10{,}000$, and a primary vortex core that shifts
from the upper-right region toward the geometric centre with increasing $Re$.
The streamline patterns of all three SJ--KAN variants are in close visual
agreement with the CFD reference across all Reynolds numbers.}
    \label{fig:lid_streamlines}
\end{figure}

First, all methods \textit{CKAN}, \textit{PINN}, \textit{JetKAN}, \textit{JetKAN-F}, and \textit{JetKAN-H} are tested with $Re=100$. All five models are trained for 100{,}000 epochs with the Adam optimizer. The obtained velocity profiles are compared with the reference CFD solution and the data of Ghia et al.~\cite{ghia1982highre} as seen in Figs.~\ref{fig:u_profiles}, ~\ref{fig:v_profiles}, and Fig.~\ref{fig:lid_streamlines}. 
The predicted profiles of $u(0.5,y)$ and $v(x,0.5)$ in Figs.~\ref{fig:u_profiles}(a) and~\ref{fig:v_profiles}(a) lie on the reference curves, including the minimum of $u$ near $y=0.45$ and the two extrema of $v$ near $x=0.25$ and $x=0.8$. The streamlines in the first row of Fig.~\ref{fig:lid_streamlines} show one primary vortex centred in the
upper right part of the cavity and two small vortices in the bottom
corners, as in the CFD reference solution. In short, all methods are able to reproduce the reference solution well.

There are significant differences in training performance of the models as shown in ~\Cref{tab:ldc_all_re}. At this Reynolds number, \textit{JetKAN-F} uses only $K_x=K_y=8$ Chebyshev modes per direction and a bottleneck of $B=8$ channels, which gives 656 trainable parameters only. \textit{JetKAN-F} reaches $\mathrm{MSE}(u)=9.61\times10^{-5}$, $\mathrm{MSE}(v)=8.70\times10^{-5}$, Rel\,$L^2(u)=4.59\times10^{-2}$ and Rel\,$L^2(v)=6.15\times10^{-2}$
at 8.2\,ms per epoch (Table~\ref{tab:ldc_all_re}). \textit{JetKAN} and \textit{JetKAN-H} give errors of the same order as \textit{JetKAN-F}. \textit{JetKAN-H} is slightly more accurate, with Rel\,$L^2(u)=3.47\times10^{-2}$, but it uses 848 parameters and is slower. \textit{JetKAN-F} is the fastest of the three, and the gap grows with $Re$. At $Re=10{,}000$ it takes 287.4\,ms per epoch, compared with 910.8\,ms for \textit{JetKAN} and 1093.3\,ms for \textit{JetKAN-H} (Table~\ref{tab:ldc_all_re}). We therefore focus on \textit{JetKAN-F} in the rest of this section. CKAN has the highest cost per epoch of all models at this Reynolds number. CKAN, with 11{,}880 parameters, gives the largest errors, Rel\,$L^2(u)=6.54\times10^{-2}$ and Rel\,$L^2(v)=1.26\times10^{-1}$,
at 267.6\,ms per epoch, which is 33 times the cost of JetKAN-F, and it underpredicts both extrema of $v$ in Fig.~\ref{fig:v_profiles}(a). On the contrary, PINN, with 30{,}903 parameters, gives Rel\,$L^2(u)=2.96\times10^{-2}$ and Rel\,$L^2(v)=4.33\times10^{-2}$ at 17.5\,ms per epoch. These errors are 1.4 to 1.5 times lower than those of \textit{JetKAN-F}, but they are obtained with 47 times more parameters and about twice the cost per epoch. In brief, \textit{SJ-PIKAN} (particularly \textit{JetKAN-F}) clearly outperforms \textit{PINN} and \textit{CKAN} at this Reynolds number.

As Reynolds number increases sequentially to $1000$, $5000$, and $10{,}000$, the velocity gradients increase near the walls. \textit{CKAN} and \textit{PINN} are shown only at $Re = 100$, \textit{CKAN} was not run at higher $Re$ because of its high cost per epoch, and \textit{PINN} failed at $Re = 1000$, even with L-BFGS and pseudo-time stepping. Therefore, it is necessary to increase the order of the polynomial basis  to represent such a steep change. The three \textit{SJ-KAN} variants are able to perform the simulations simply by increasing the values of $K_x$, $K_y$, and $B$ as shown below.  

At $Re=1000$, \textit{JetKAN-F} uses $K_x=K_y=8$ and $B=12$ as shown in ~\Cref{tab:ldc_all_re}. It reaches $\mathrm{MSE}(u)=5.38\times10^{-5}$, $\mathrm{MSE}(v)=4.22\times10^{-5}$, Rel\,$L^2(u)=3.41\times10^{-2}$ and Rel\,$L^2(v)=3.20\times10^{-2}$ at 150.0\,ms per epoch (Table~\ref{tab:ldc_all_re}). The major improvement comes from the change of the optimizer (L-BFGS) with the pseudo-time stepping, and from the increase of $B$. In Fig.~\ref{fig:u_profiles}(b) the $u$
profile follows the reference, including the thin layers below the lid and above the bottom wall. In Fig.~\ref{fig:v_profiles}(b) the $v$ profile follows the nearly linear variation across the core of the cavity and the sharp extrema near the side walls. The largest
deviation is a small underprediction of the positive peak of $v$ near $x=0.15$. The streamlines in the second row of Fig.~\ref{fig:lid_streamlines} show the primary vortex moved toward the center of the cavity and the two bottom corner vortices, in agreement with the reference solution. The cost per epoch rises from 8.2\,ms to 150.0\,ms mainly because each L-BFGS iteration evaluates the loss and its gradient several times
during the line search. \textit{JetKAN} gives Rel\,$L^2(u)=7.71\times10^{-2}$
and Rel\,$L^2(v)=9.35\times10^{-2}$, and JetKAN-H gives
Rel\,$L^2(u)=3.47\times10^{-2}$ and Rel\,$L^2(v)=4.16\times10^{-2}$.

At $Re=5000$, JetKAN-F uses $K_x=K_y=12$ and $B=24$, which gives 3{,}888 parameters, and is trained for 20{,}000 L-BFGS iterations. It reaches $\mathrm{MSE}(u)=2.45\times10^{-4}$,
$\mathrm{MSE}(v)=2.73\times10^{-4}$, Rel\,$L^2(u)=7.45\times10^{-2}$ and Rel\,$L^2(v)=7.98\times10^{-2}$ at 189.0\,ms per epoch. The $u$ profile in
Fig.~\ref{fig:u_profiles}(c) shows the nearly linear core and the thin layers next to the lid and the bottom wall. The $v$ profile in Fig.~\ref{fig:v_profiles}(c) shows the narrow peaks near the side walls. JetKAN-F slightly under-predicts the positive peak of $v$ near
$x=0.1$, which is the region of largest error. The streamlines in the third row of Fig.~\ref{fig:lid_streamlines} show the primary vortex centred close to the geometric centre of the cavity, larger vortices in both bottom corners, and an additional vortex in the upper left corner, as in the finite-volume solution. JetKAN gives
Rel\,$L^2(u)=8.70\times10^{-2}$ and JetKAN-H gives Rel\,$L^2(u)=1.81\times10^{-1}$.

At $Re=10{,}000$, SJ-KAN extends this range to $Re=10{,}000$ and resolves the primary vortex, the two bottom corner vortices and the upper left vortex of the finite-volume solution at $Re=5000$ and $Re=10{,}000$ (Fig.~\ref{fig:lid_streamlines}). With 8{,}768 trainable parameters, JetKAN-F reproduces the centreline profiles and the vortex structure of the cavity flow up to $Re=10{,}000$. \textit{JetKAN-F} uses $K_x=K_y=16$ and $B=32$, which gives 8{,}768 parameters, and is trained for 40{,}000 L-BFGS iterations. It reaches $\mathrm{MSE}(u)=2.30\times10^{-4}$, $\mathrm{MSE}(v)=2.78\times10^{-4}$, Rel\,$L^2(u)=7.66\times10^{-2}$ and Rel\,$L^2(v)=8.47\times10^{-2}$
at 287.4\,ms per epoch. The $u$ and $v$ profiles in Figs.~\ref{fig:u_profiles}(d) and~\ref{fig:v_profiles}(d) show the same thin wall layers and narrow side wall peaks as at $Re=5000$. The prediction follows the finite-volume solution closely, and the small
offsets from the points of Ghia et al.~\cite{ghia1982highre} near $x=0.2$ and $x=0.85$ are also present in the finite-volume solution. The streamlines in the last row of Fig.~\ref{fig:lid_streamlines} show the same vortex structure as at $Re=5000$, with the primary vortex near the centre, two bottom corner vortices and the upper left
vortex, in agreement with the finite-volume solution. JetKAN gives
Rel\,$L^2(u)=6.36\times10^{-2}$ and JetKAN-H gives
Rel\,$L^2(u)=7.85\times10^{-2}$ (Table~\ref{tab:ldc_all_re}). Note that the cost per epoch grows from 189.0 to 287.4\,ms because the basis grows from 144 to 256 functions and residual-based adaptive refinement increases the number of collocation points. 

\subsection{Ablation study} 
\label{sec:ablation}

This section presents a parametric study of the SJ-KAN architecture.  We vary one parameter ($K$, $B$, $N$, etc.) at a time and report the results in five parts: (i) differentiation computation (analytic jet vs.\ autograd); (ii) spectral representation; (iii) collocation points; (iv) network depth; and (v) reproducibility. 

\subsubsection{The cost of differentiation}
\label{subsec: jet_vs_ad}
The jet removes automatic differentiation from the PDE residual completely. Every derivative in the Navier--Stokes residual, including the second derivatives in the viscous terms, is built from the precomputed matrices by matrix products and element-wise operations. Automatic differentiation is used only in the backward pass to compute the gradient of the loss with respect to the trainable parameters.

\begin{figure}[htbp]
    \centering
    \includegraphics[width=0.8\linewidth]{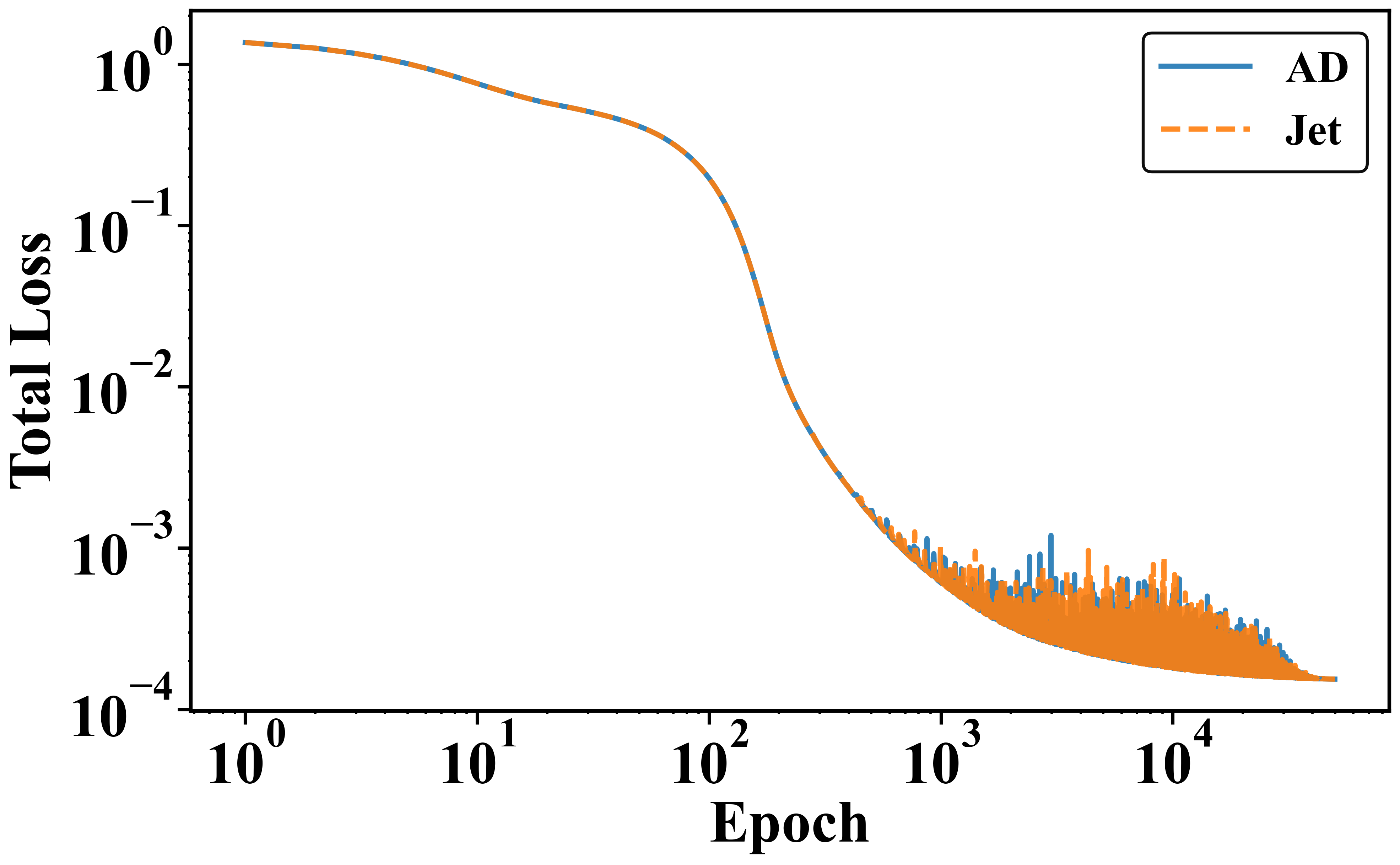}
    \par\noindent\hspace*{0.1\linewidth}\textbf{(a)}
    \vspace{0.5em}

    \includegraphics[width=0.8\linewidth]{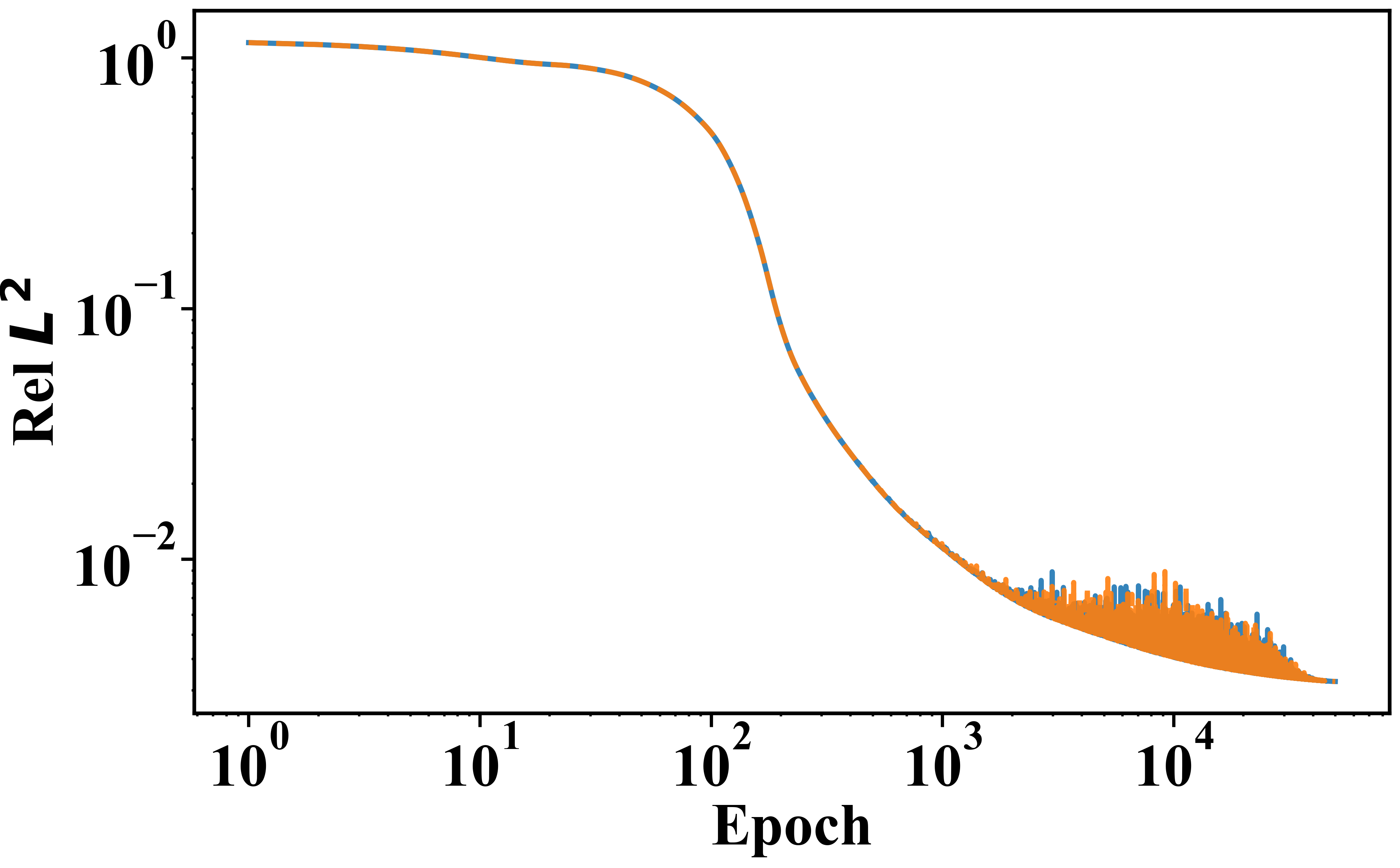}
    \par\noindent\hspace*{0.1\linewidth}\textbf{(b)}
    \vspace{0.5em}

    \caption{Analytic jet versus autograd (AD) ablation on the 2D Taylor--Green vortex ($Re = 100$) for the \textit{JetKAN-F} architecture, with both models sharing 656 parameters, $K_x = K_y = 8$, $B = 8$, $P = 5$, a $51 \times 51$ collocation grid, and Adam at $10^{-3}$ for 50{,}000 epochs; the only difference is the derivative computation strategy. \textbf{(a)} Total loss versus epoch. The jet and AD versions follow the same trajectory and both converge to $\sim 10^{-4}$. \textbf{(b)} Rel\,$L^2$ error versus epoch. The two curves overlap over the full training window, and both settle near $\sim 3 \times 10^{-3}$.}
    \label{fig:jet_ad_loss_l2}
\end{figure}

We compare analytic jet propagation with reverse-mode automatic differentiation (\textit{AD}) on the same \textit{JetKAN-F} architecture. All experiments use \textit{JetKAN-F} on the steady Taylor--Green vortex at $Re=100$ with a $51\times51$ collocation grid, $K_x=K_y=8$, $B=8$, $P=5$, seed 42, and 50{,}000 Adam epochs at a
learning rate of $10^{-3}$. Both models share the same 656 parameters, $K_x = K_y = 8$, $B = 8$, $P = 5$, a $51 \times 51$ grid, initialization, and 50{,}000 epochs with Adam at $\text{lr} = 10^{-3}$. The only difference is how the PDE residual derivatives are computed: analytically, by propagating precomputed Chebyshev basis derivatives through the network with the chain rule (jet), or with PyTorch autograd ($AD$). The results are reported in Table~\ref{tab:jet_ad_table} and the training histories are shown in Figs.~\ref{fig:jet_ad_loss_l2}(a) and (b).

It is observed from Figs.~\ref{fig:jet_ad_loss_l2}(a) and (b) that the total loss and the $\mathrm{Rel}~L^{2}$ error histories of the jet and \textit{AD} versions overlap over the full 50{,}000 epochs. Both versions compute the same derivatives of the same network, so the optimizer receives the same gradients, up to floating-point round-off, at every step. As reported in Table~\ref{tab:jet_ad_table}, the final PDE loss is $1.29 \times 10^{-4}$ and the BC loss is $1.87 \times 10^{-5}$ for both versions. The \textit{MSE} is $5.95 \times 10^{-6}$ for the jet and $5.94 \times 10^{-6}$ for \textit{AD}, and the $\mathrm{Rel}~L^{2}$ errors are $3.10 \times 10^{-3}$, $3.06 \times 10^{-3}$, and $4.33 \times 10^{-3}$ for $u$, $v$, and $p$ in both cases. The AD-to-jet ratio for every accuracy metric is $1.00$, confirming that the jet derivatives agree with the autograd derivatives up to round-off error.

The difference between the two versions is the computational cost. The jet version requires $7.4$~ms/epoch, while \textit{AD} requires $60.7$~ms/epoch, a speedup of $8.2\times$ as shown in \Cref{tab:jet_ad_table}. The total training time is $371.3$~s for the jet and $3{,}035.4$~s for \textit{AD}. In the \textit{AD} version, the Chebyshev basis is evaluated from the collocation coordinates at every epoch. The derivatives $u_x$, $u_y$, $u_{xx}$, $u_{yy}$, $v_x$, $v_y$, $v_{xx}$, $v_{yy}$, $p_x$, and $p_y$ are obtained through ten separate autograd calls. The second derivatives require differentiating through the graph of the first derivatives, and all of these graphs must be kept in memory for the final backward pass. In the jet version, the basis functions and their derivatives are computed once before training and stored as fixed matrices. At each epoch, these matrices are multiplied by the bottleneck weights and passed through the single \textit{KAN} layer with the chain rule, so all derivatives are obtained in one forward pass with a small number of matrix multiplications. The jet therefore achieves the same accuracy as \textit{AD} at a fraction of the cost, allowing longer training or finer collocation grids within the same computational budget.

\textit{SJ-KAN} uses a Chebyshev basis instead, which handles non-periodic domains such as the lid-driven cavity with no-slip walls as well as the periodic Taylor--Green vortex. Huang et al.~\cite{huang2025chebyshev} replaced automatic differentiation with Chebyshev differentiation matrices for the PDE residual, reporting a dramatic reduction in computation time on higher-order nonlinear PDEs. Still, the trainable component remains a standard \textit{MLP}. SJ-KAN uses the same type of differentiation matrix but applies it to the basis before training and carries the derivatives through the bottleneck and the KAN layer in closed form. On the same JetKAN-F network, this reduces the cost per epoch from 60.7 to 7.4\, ms, an 88\,\% reduction, with no change in accuracy (Table~\ref{tab:jet_ad_table}). This is close to the upper end of the range reported for improvement in previous works ~\cite{yin2024chebyshev}. The jet therefore reduces training cost while maintaining the same accuracy.

Against the AD-based baselines, JetKAN-F is 18.6 times faster per epoch than CKAN on the Taylor--Green vortex and 32.6 times faster on the cavity at $Re=100$ (Tables~\ref{tab:tgv_results} and~\ref{tab:ldc_all_re}). Part of this gap comes from the deeper CKAN architecture, and the jet alone accounts for the factor of 8.2 measured on the same network.

\textit{Accuracy and cost of the analytic jet.} On the
Taylor--Green vortex with the same 656-parameter JetKAN-F network,
the analytic jet and reverse-mode automatic differentiation converge
to the same solution. Both reach an MSE of $5.95\times10^{-6}$ and
Rel\,$L^2$ errors of $3.10\times10^{-3}$, $3.06\times10^{-3}$ and
$4.33\times10^{-3}$ for $u$, $v$ and $p$, which shows that the jet
derivatives agree with the automatic-differentiation derivatives to
round-off. The difference is the cost. The jet reduces the time per
epoch from 60.7 to 7.4\, ms, a factor of 8.2, and the total training
time from 3{,}035.4 to 371.3\,s. Against the CKAN baseline, which
also differs in architecture, JetKAN-F is 18.6 times faster per
epoch on the Taylor--Green vortex with a 3.1 times lower MSE, and
32.6 times faster per epoch on the lid-driven cavity at $Re=100$.
The per-epoch cost of JetKAN-F also stays at 8.5 to 8.7\, ms when the
collocation grid is refined from 441 to 40{,}401 points, since at
these grid sizes the GPU time is set mainly by the fixed number of
matrix operations, not by their size.

\subsubsection{Spectral resolution ($K$) and bottleneck width ($B$)} \label{sec:multi_bk}
Recall that the basis functions $\phi_{ij}(x,y)$ generate $n_f = K^2$ tensor-product Chebyshev features, which are collected into the feature matrix $\boldsymbol{\Phi} \in \mathbb{R}^{N \times n_f}$. The bottleneck projection $\mathbf{W} \in \mathbb{R}^{n_f \times B}$ compresses these features into $B$ latent variables $\mathbf{Z} = \boldsymbol{\Phi}\,\mathbf{W}$ before they enter the KAN layer. The polynomial order $K$ controls how many spatial modes are available to approximate the solution, while $B$ controls how many independent combinations of those modes the KAN layer can act on. Both must be large enough for the network to represent the three coupled fields $(u, v, p)$.

\begin{figure}
    \centering
    \captionsetup[subfigure]{labelfont={bf,Large}, labelformat=parens}

    \begin{subfigure}{\linewidth}
        \centering
        \includegraphics[width=1.0\linewidth]{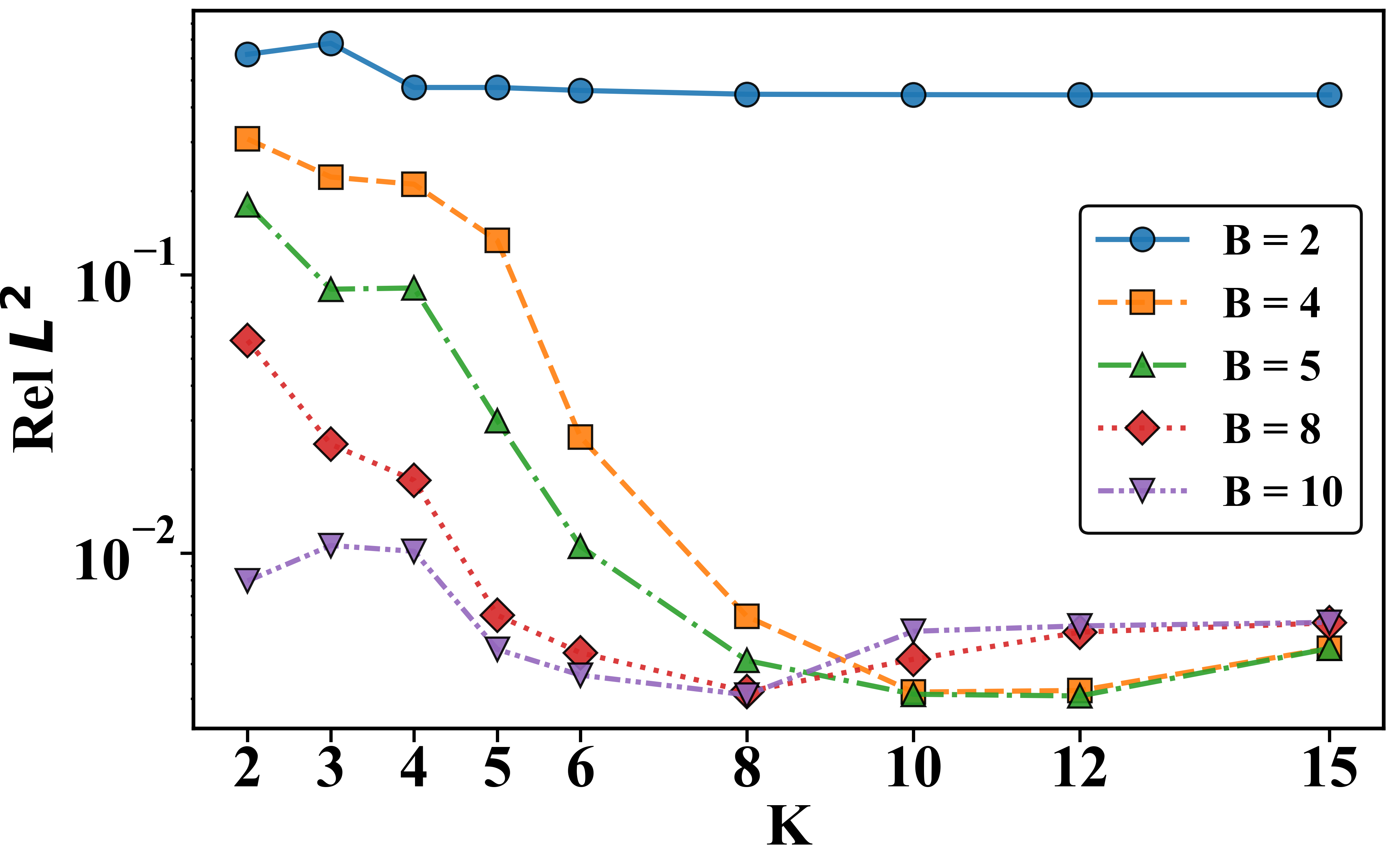}
        \caption{}
        \label{fig:multi_b_k_rel_l2}
    \end{subfigure}

    \vspace{0.5em}

    \begin{subfigure}{\linewidth}
        \centering
        \includegraphics[width=1.0\linewidth]{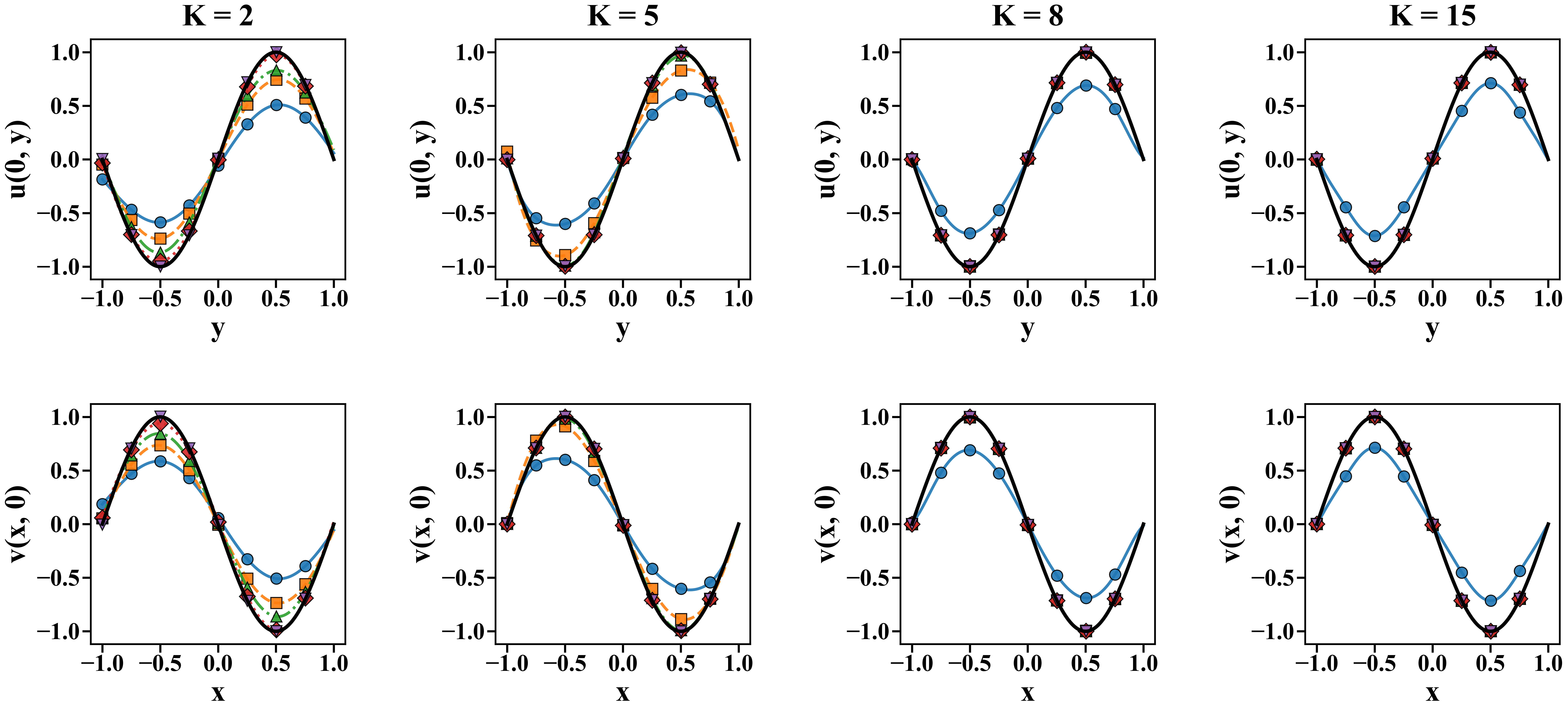}
        \caption{}
        \label{fig:multi_b_k_centerline}
    \end{subfigure}

    \caption{Spectral resolution and bottleneck width study of JetKAN-F on the 2D Taylor--Green vortex ($Re = 100$), varying the polynomial order $K \in \{2, 3, 4, 5, 6, 8, 10, 12, 15\}$ for five bottleneck widths $B \in \{2, 4, 5, 8, 10\}$. All runs use $P = 5$, a $51 \times 51$ collocation grid, seed $42$, and Adam at $10^{-3}$ for $50{,}000$ epochs.
    \textbf{(a)}~$\mathrm{Rel}~L^{2}$ error versus polynomial order $K$. The $B = 2$ curve remains flat above $4.4 \times 10^{-1}$ at every $K$, indicating that a two-dimensional bottleneck is insufficient for this problem. For $B \geq 4$, the error decreases with increasing $K$, reaches a minimum near $3.1 \times 10^{-3}$ in the range $K = 8$ to $12$, and increases again at $K = 15$ due to ill-conditioning of the high-order differentiation matrices.
    \textbf{(b)}~Centerline velocity profiles at selected $(B, K)$ combinations compared against the analytical solution (solid black line). Top row: horizontal velocity $u$ along the vertical centerline, $u(0, y)$. Bottom row: vertical velocity $v$ along the horizontal centerline, $v(x, 0)$. Columns correspond to $K = 2, 5, 8,$ and $15$. For $B = 2$, the predicted profiles deviate visibly from the analytical solution at every $K$ tested. For $B \geq 4$ and $K \geq 8$, the profiles are indistinguishable from the reference.}
    \label{fig:multi_b_k}
\end{figure}

We vary the polynomial order $K \in \{2,3,4,5,6,8,10,12,15\}$ and the bottleneck width $B \in \{2,4,5,8,10\}$ simultaneously to map the accuracy of \textit{JetKAN-F} across these two architectural parameters. All runs use the same $51 \times 51$ collocation grid, 50{,}000 epochs with Adam at $\text{lr} = 10^{-3}$, degree $P = 5$, and seed 42 on the Taylor--Green vortex at $Re = 100$. The $\mathrm{Rel}~L^{2}$ errors for all 45 configurations are reported in Table~\ref{tab:multi_b_k}, and the $\mathrm{Rel}~L^{2}$ error as a function of $K$ for each $B$ is plotted in Fig.~\ref{fig:multi_b_k_rel_l2}. Table~\ref{tab:best_b_k} lists the polynomial order that gives the lowest error for each $B$, along with the error at $K = 15$ to show how much accuracy is lost when the spectral order increases. The centerline velocity profiles for selected $(B, K)$ combinations are shown in Fig.~\ref{fig:multi_b_k_centerline}.

The bottleneck width $B$ sets a lower limit on accuracy. At the low value $B = 2$, the $\mathrm{Rel}~L^{2}$ error remains above $4.43 \times 10^{-1}$ at every $K$ tested (Table~\ref{tab:multi_b_k}, first data column). Even at $K = 15$ with $n_f = 225$ spectral features, all of them are compressed into just two latent variables before reaching the KAN layer. Two latent variables cannot carry enough information to reconstruct three independent fields, and the error stays flat regardless of $K$. The centerline velocity profiles for $B = 2$ in Fig.~\ref{fig:multi_b_k_centerline} confirm this: the predicted $u$ and $v$ profiles deviate visibly from the reference solution at every tested $K$. Raising $B$ to $4$ produces a qualitative change. The bottleneck now passes four independent linear combinations of the spectral features to the \textit{KAN} layer, which is sufficient for three output fields. At $B = 4$ and $K = 6$, the error drops to $2.61 \times 10^{-2}$, and at $K = 10$ it reaches as low as $3.16 \times 10^{-3}$ (Table~\ref{tab:multi_b_k}). Further increases to $B = 5$, $8$, and $10$ yield lower errors of $3.06 \times 10^{-3}$ ($K = 12$), $3.18 \times 10^{-3}$ ($K = 8$), and $3.09 \times 10^{-3}$ ($K = 8$), respectively (Table~\ref{tab:best_b_k}). Beyond $B = 4$, the minimum $\mathrm{Rel}~L^{2}$ error is effectively independent of $B$, settling near $3.1 \times 10^{-3}$ for all tested widths. This plateau is visible in Fig.~\ref{fig:multi_b_k_rel_l2}, where the curves for $B = 4$, $5$, $8$, and $10$ overlap in the range $K = 8$ to $12$. The centerline profiles in Fig.~\ref{fig:multi_b_k_centerline} for these configurations are indistinguishable from the reference solution. In brief, there is a critical value of $B$ at which the error is minimized. 

For a $B \geq 4$, the error first decreases with increasing $K$ and then increases again beyond an optimal value. From Table~\ref{tab:best_b_k}, this optimal order is $K = 10$ and   $K = 12$ are $B = 4$ and $B = 5$, respectively. However, it is $K = 8$ for both $B = 8$ and $B = 10$. The shift of the optimal $K$ toward lower values as $B$ increases may reflect the effect of compression. When $B$ is small relative to $n_f$, the bottleneck is a tight compression, and more spectral modes ($K$) are needed to ensure that the $B$ projected variables still contain enough spectral content to resolve the solution. When $B$ is large, the bottleneck preserves most of the spectral information even at moderate $K$, so additional modes contribute little and instead expose the optimizer to the ill-conditioning of the high-order differentiation matrices. At $K = 15$, the $\mathrm{Rel}~L^{2}$ error rises to $5.15 \times 10^{-3}$ for $B = 4$, $5.57 \times 10^{-3}$ for $B = 8$, and $5.63 \times 10^{-3}$ for $B = 10$, which is $1.6\times$, $1.8\times$, and $1.8\times$ above the respective minima (Table~\ref{tab:best_b_k}). This increase at high $K$ is consistent with the well-known ill-conditioning of polynomial differentiation matrices at high order~\cite{trefethen2000spectral}. The entries of the second-derivative Chebyshev matrix grow as $O(K^4)$ \cite{weideman1988eigenvalues}, and when these large values multiply small coefficient differences during the PDE residual evaluation, floating-point rounding errors accumulate and the loss landscape becomes harder for the optimizer to navigate. The error increase is more pronounced at larger $B$ ($1.8\times$ for $B = 8$ and $B = 10$ versus $1.6\times$ for $B = 4$), because a wider bottleneck passes more of these ill-conditioned high-order components through to the \textit{KAN} layer.

The training time is effectively constant across all 45 configurations, ranging from 519 to 544 seconds (approximately 10.4 to 10.9 ms/epoch). Neither $K$ nor $B$ introduces a measurable change in per-epoch cost within the tested range, because the parameter count remains small enough that fixed-size matrix operations in the spectral-basis evaluation and the \textit{KAN} layer dominate wall-clock time.


\subsubsection{The role of the \textit{KAN} layer}
\label{sec:linear_vs_jetkan}
To highlight the importance of the \textit{KAN} layer, we compare \textit{JetKAN-F} against a Chebyshev spectral solver that uses the same basis $\boldsymbol{\Phi}(x,y)$ but represents the solution as a direct linear expansion $\hat{\mathbf{U}} = \boldsymbol{\Phi}\mathbf{W}$, without the learnable nonlinear \textit{KAN} layer. Note that this spectral solver (Chebyshev expansion) is a fixed linear combination of $K_x \times K_y$ modes; its accuracy is bounded by how well such a combination can represent the nonlinear term $\mathbf{u}\cdot\nabla\mathbf{u}$.

\begin{figure}[htbp]
    \centering
    \includegraphics[width=0.8\linewidth]{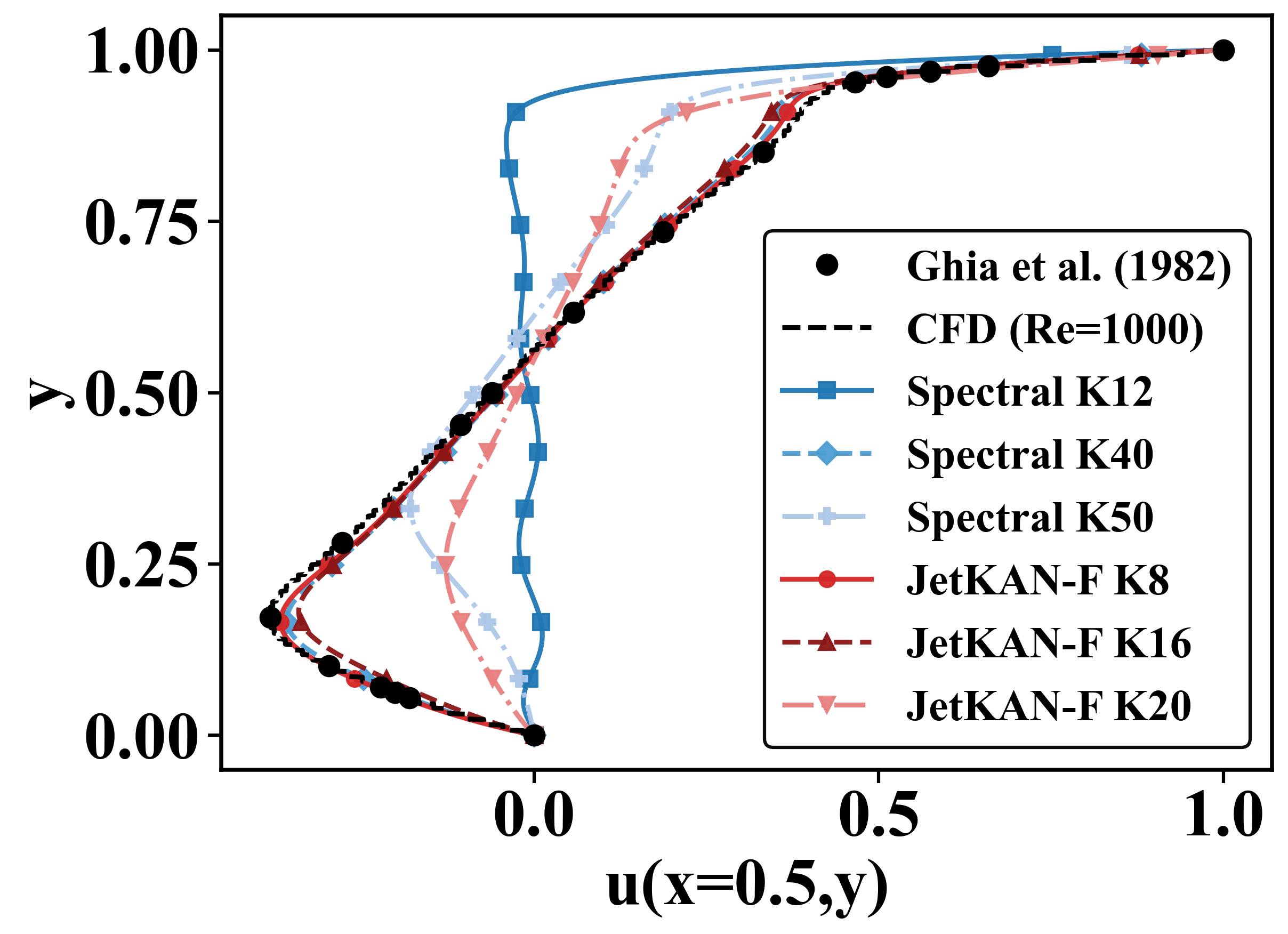}
    \par\noindent\hspace*{0.1\linewidth}\textbf{(a)}
    \vspace{0.5em}

    \includegraphics[width=0.8\linewidth]{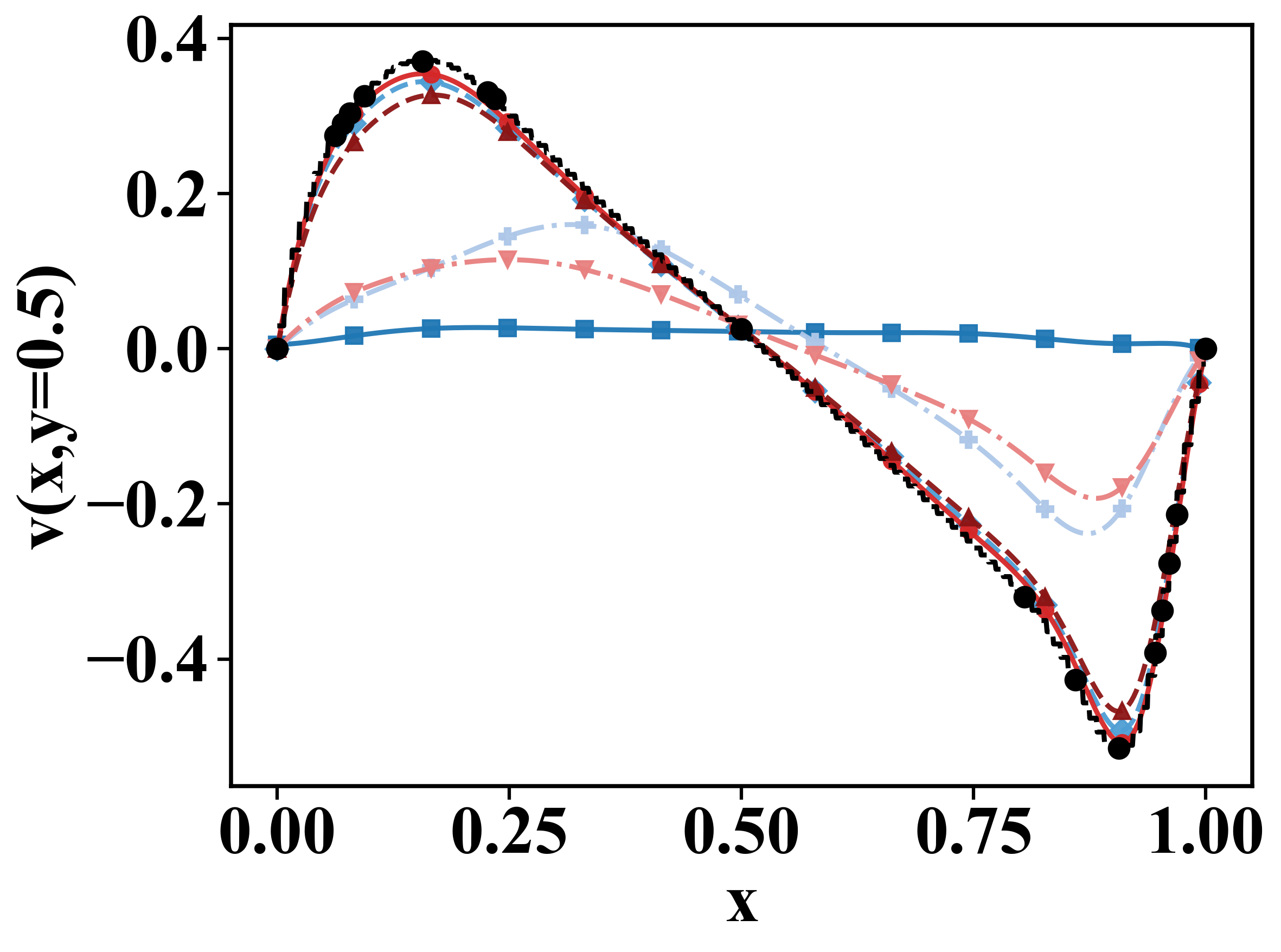}
    \par\noindent\hspace*{0.1\linewidth}\textbf{(b)}
    \vspace{0.5em}

    \caption{Centerline velocity profile comparison between the pure Chebyshev
spectral solver and JetKAN-F for the 2D lid-driven cavity flow at $Re = 1000$,
against the benchmark reference data of Ghia et al.~\cite{ghia1982highre}
(filled circles) and the in-house CFD solution (dashed line).
\textbf{(a)}~Horizontal velocity component $u$ along the vertical centerline
$u(0.5,y)$, comparing Spectral $K = 12$, $40$, $50$ and JetKAN-F $K = 8$,
$16$, $20$.
\textbf{(b)}~Vertical velocity component $v$ along the horizontal centerline
$v(x, 0.5)$, comparing the same set of models.
Spectral $K{=}12$ fails to reproduce the correct profile shape in both
components; Spectral $K{=}40$ matches the reference closely but degrades at
$K{=}50$. JetKAN-F $K{=}8$ tracks both reference curves most accurately
among all cases.}
    \label{fig:linear_jet_centerline}
\end{figure}

\begin{figure}
    \centering
    \includegraphics[width=1.0\linewidth]{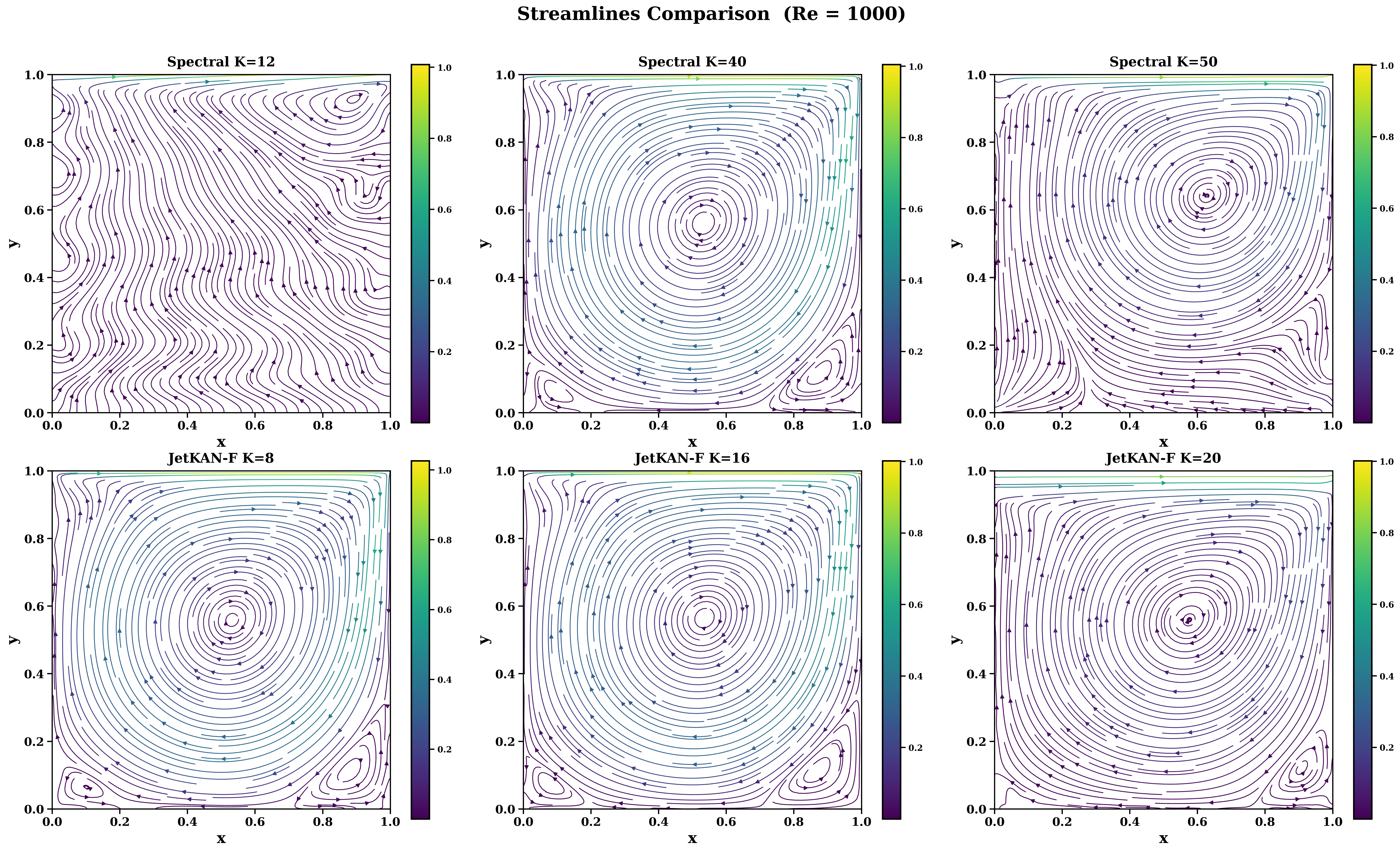}
    \caption{Streamlines of the predicted velocity field for the 2D lid-driven cavity flow at $Re=1000$, comparing the pure Chebyshev spectral solver (top row: $K=12$, $40$, $50$) and JetKAN-F (bottom row: $K=8$, $16$, $20$). Streamlines are colored by the velocity magnitude $|\boldsymbol{u}|$.}
    \label{fig:lin_jet_stream_1000}
\end{figure}

We evaluate the \textit{JetKAN-F} and Chebyshev expansions on the $2D$ lid-driven cavity at $Re = 1000$ over a range of mode counts $K$, using the same BRDR adaptive weighting and L-BFGS optimizer with pseudo-time stepping. Results are reported in Table~\ref{tab:spectral_comparison}, centerline velocity profiles are shown in Fig.~\ref{fig:linear_jet_centerline} and streamlines in Fig.~\ref{fig:lin_jet_stream_1000}. 

The Chebyshev spectral expansion model is tested with three different values of $ K=12, 40, $ and $50$. At $K{=}12$ ($n_f = 144$, 432 parameters) gives Rel\,$L^2(u) = 8.40\times10^{-1}$, which is effectively non-convergent. The centerline profile in Fig.~\ref{fig:linear_jet_centerline}(a) shows that the predicted $u(0.5,y)$ is nearly flat and departs substantially from both the Ghia et al.~\cite{ghia1982highre} reference and the CFD solution. Increasing $K{=}40$ ($n_f = 1{,}600$, 4{,}800 parameters) reduces the error to Rel\,$L^2(u) = 6.67\times10^{-2}$. The centerline profiles in Fig.~\ref{fig:linear_jet_centerline} closely follow the reference data. As $K{=}50$ ($n_f = 2{,}500$, 7{,}500 parameters), the solution degrades to Rel\,$L^2(u) = 5.05\times10^{-1}$. This is remarkable given that it uses $1.6\times$ more parameters than the $K=40$ model. The profile in Fig.~\ref{fig:linear_jet_centerline}(a) is visibly worse than $K{=}40$. The best linear model is thus $K=40$. The streamlines in Fig.~\ref{fig:lin_jet_stream_1000} demonstrate the distinction. At $K{=}12$ (top-left), it produces a chaotic flow field with no coherent vortex structure. At $K{=}40$ (top-center), the streamlines show the correct primary vortex but do not resolve the bottom-corner vortices. At $K{=}50$ (top-right), the streamlines show a distorted primary vortex center and visible spurious oscillations. The spectral solver also struggles to capture corner vortices for all $K$. This effect is attributed to the ill-conditioning of the second-derivative Chebyshev differentiation matrix (condition number $\sim O(K^4)$) \cite{weideman1988eigenvalues, trefethen2000spectral}, which causes the L-BFGS optimizer to stall. 

\textit{JetKAN-F} also needs a larger value of $K$ to resolve accurately the vortices. However, the required $K$ grows much more slowly because the nonlinear KAN layer compensates for the missing representational capacity. Three values of $K$ are tested: i) $K=8,16$ and 20. At $K{=}8$ ($n_f = 64$, 984 parameters), \textit{JetKAN-F}  achieves Rel\,$L^2(u) = 3.41\times10^{-2}$ and $MSE(u) = 5.38\times10^{-5}$. Note that these errors are lower than those of the spectral method ($K{=}40$) while using approximately one-fifth the parameters (984 vs.\ 4{,}800). The centerline profiles in Fig.~\ref{fig:linear_jet_centerline}(a,b) are the closest match to the Ghia et al.\ data among all cases. At $K{=}16$ and $K{=}20$, \textit{JetKAN-F} gives Rel\,$L^2(u) = 7.36\times10^{-2}$ and $5.59\times10^{-1}$, respectively. This is surprising, since accuracy degrades once $K$ is greater than 8. As seen in Fig.~\ref{fig:lin_jet_stream_1000}, \textit{JetKAN-F} at $K{=}8$ can resolve both the primary vortex and the bottom corner vortices. At $K{=}16$ and $K{=}20$, \textit{JetKAN-F} shows a slight distortion of the vortex center, which is consistent with the degraded Rel\,$L^2$ error. In brief, the $KAN$ layer plays a key role in simulating both the primary and secondary vortices.

\begin{figure}[htbp]
    \centering
    \includegraphics[width=0.8\linewidth]{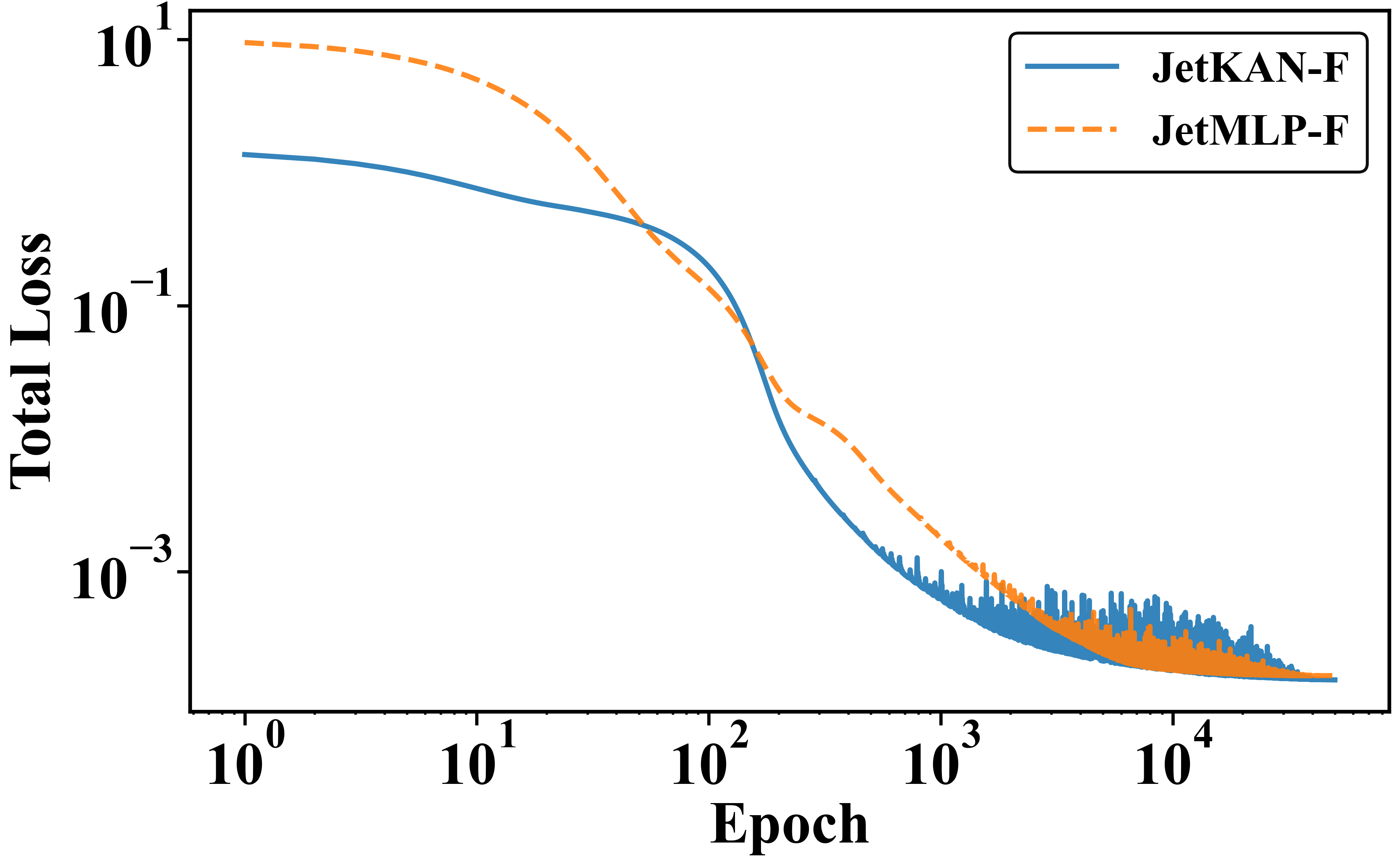}
    \par\noindent\hspace*{0.1\linewidth}\textbf{(a)}
    \vspace{0.5em}

    \includegraphics[width=0.8\linewidth]{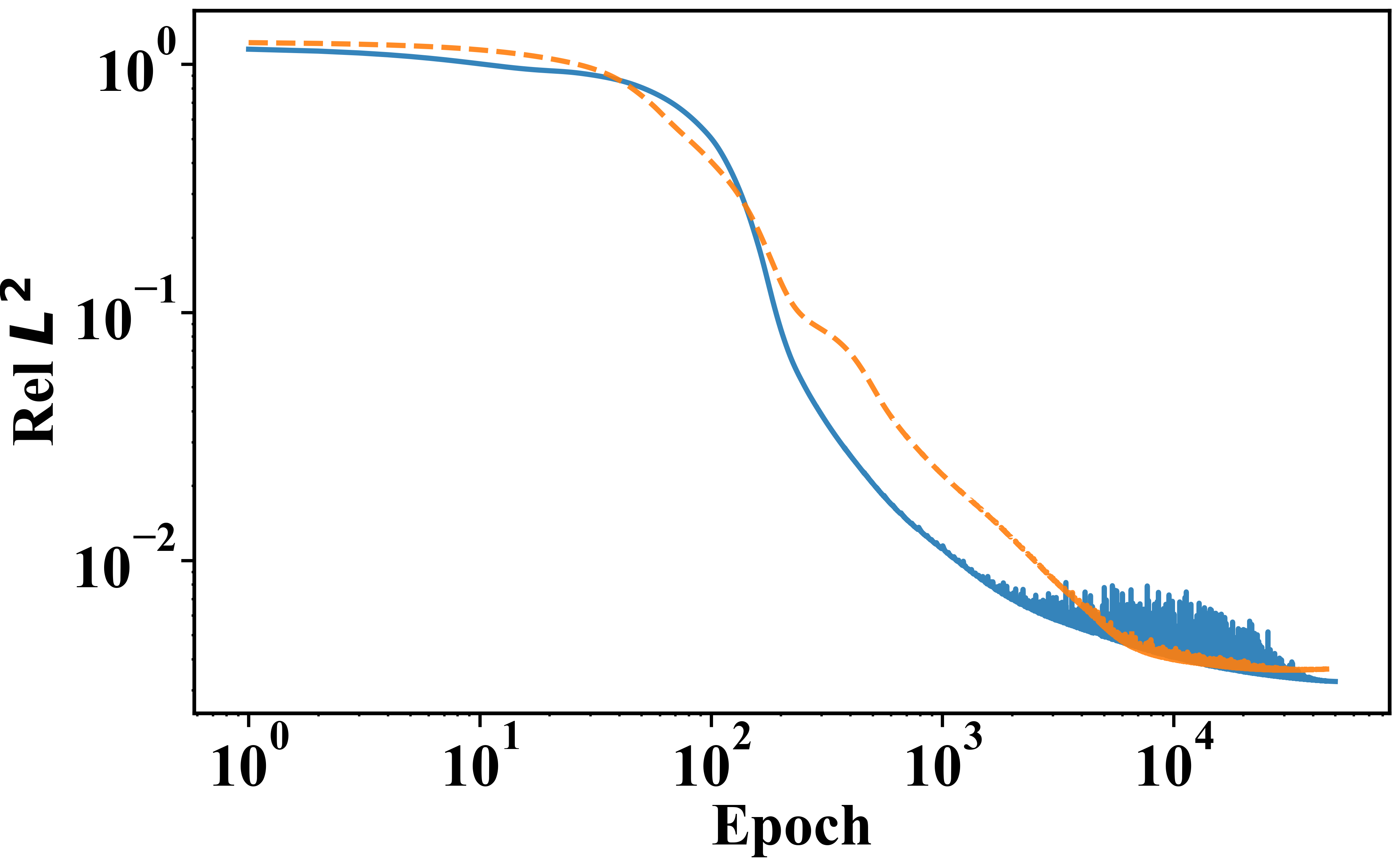}
    \par\noindent\hspace*{0.1\linewidth}\textbf{(b)}
    \vspace{0.5em}

    \caption{MLP head versus KAN head ablation on the 2D Taylor--Green vortex
($Re = 100$), comparing JetKAN-F (single Chebyshev KAN layer, 656 parameters)
and JetMLP-F (single linear layer with $\tanh$, 539 parameters), with both
models sharing the same spectral basis ($K_x = K_y = 8$), $B = 8$, analytic
jet propagation, a $51 \times 51$ collocation grid, and Adam at $10^{-3}$
for $50{,}000$ epochs.
\textbf{(a)}~Total loss versus epoch. JetMLP-F descends faster in the early
epochs but both curves converge to $\sim\!10^{-4}$ and are nearly
indistinguishable by the end of training.
\textbf{(b)}~$\mathrm{Rel}~L^{2}$ error versus epoch. Both curves track
closely throughout training, with JetKAN-F settling marginally lower than
JetMLP-F at the final epoch.}
    \label{fig:mlp_kan_loss_l2}
\end{figure}

To further examine the role of the $KAN$ layer, we replace the $KAN$ layer with a $MLP$ tanh network (\textit{JetMLP-F}) to test whether learnable activation functions provide any accuracy benefit over a standard MLP head. Both models use the same spectral basis ($K_x = K_y = 8$), bottleneck width $B = 8$, analytic jet propagation, a $51 \times 51$ grid, 50{,}000 epochs with Adam at $\text{lr} = 10^{-3}$. This \textit{JetMLP-F} has 539 parameters (27 in the head). As reported in Table~\ref{tab:mlp_kan_table} and in Figs.~\ref{fig:mlp_kan_loss_l2}, both \textit{JetKAN-F} and \textit{JetMLP-F} follow nearly identical loss trajectories. \textit{JetMLP-F} descends slightly faster in the early epochs, but converges to the same region as \textit{JetKAN-F} does near $10^{-4}$ by the end of training. The $\mathrm{Rel}~L^{2}$ error histories in Fig.~\ref{fig:mlp_kan_loss_l2}(b) show a similar picture. Both respective losses of \textit{JetKAN-F} and \textit{JetMLP-F} overlap for most of the training and settle at comparable final values. \textit{JetKAN-F} achieves an $MSE =5.96 \times 10^{-6}$ with $\mathrm{Rel}~L^{2}(u,v,p)$ errors of $(3.10 \times 10^{-3}, 3.07 \times 10^{-3}, 4.33 \times 10^{-3}$ respectively. \textit{JetMLP-F} achieves an $MSE=7.53 \times 10^{-6}$ with $\mathrm{Rel}~L^{2}(u,v,p)$ errors of $(3.52 \times 10^{-3}, 3.45 \times 10^{-3}, 4.78 \times 10^{-3})$ as reported in Table~\ref{tab:mlp_kan_table}. The insignificant improvement factors of the $KAN$ head over the $MLP$ head are $1.14\times$, $1.12\times$, and $1.10\times$ for the $u$, $v$, and $p$ components respectively. The $PDE$ loss values are also close, $1.29 \times 10^{-4}$ for JetKAN-F versus $1.39 \times 10^{-4}$ for JetMLP-F. However, the training of \textit{JetMLP-F} is faster at $6.1$~ms/epoch compared to $9.2$~ms/epoch for \textit{JetKAN-F} ($1.51\times$). This is expected since the $MLP$ head has fewer parameters and a simpler forward pass than the Chebyshev polynomial evaluation in the $KAN$ head. 
The KAN head provides a marginal improvement of $1.26\times$ in MSE and up to $1.14\times$ in $\mathrm{Rel}~L^{2} (u)$ error at $1.51\times$ longer training time. In conclusion,  the choice between an MLP head and a KAN head has a minimal effect on accuracy for the Taylor--Green vortex at this resolution. However, the $KAN$ head retains one practical advantage that is not reflected in these metrics. Because its activation functions are explicit Chebyshev series with learned coefficients, the trained model admits a closed-form symbolic representation of the solution, which is not available with the fixed tanh nonlinearity in the $MLP$ head.

\subsubsection{Collocation points}
\label{sec:collocation}
\begin{figure}[htbp]
    \centering
    \includegraphics[width=0.8\linewidth]{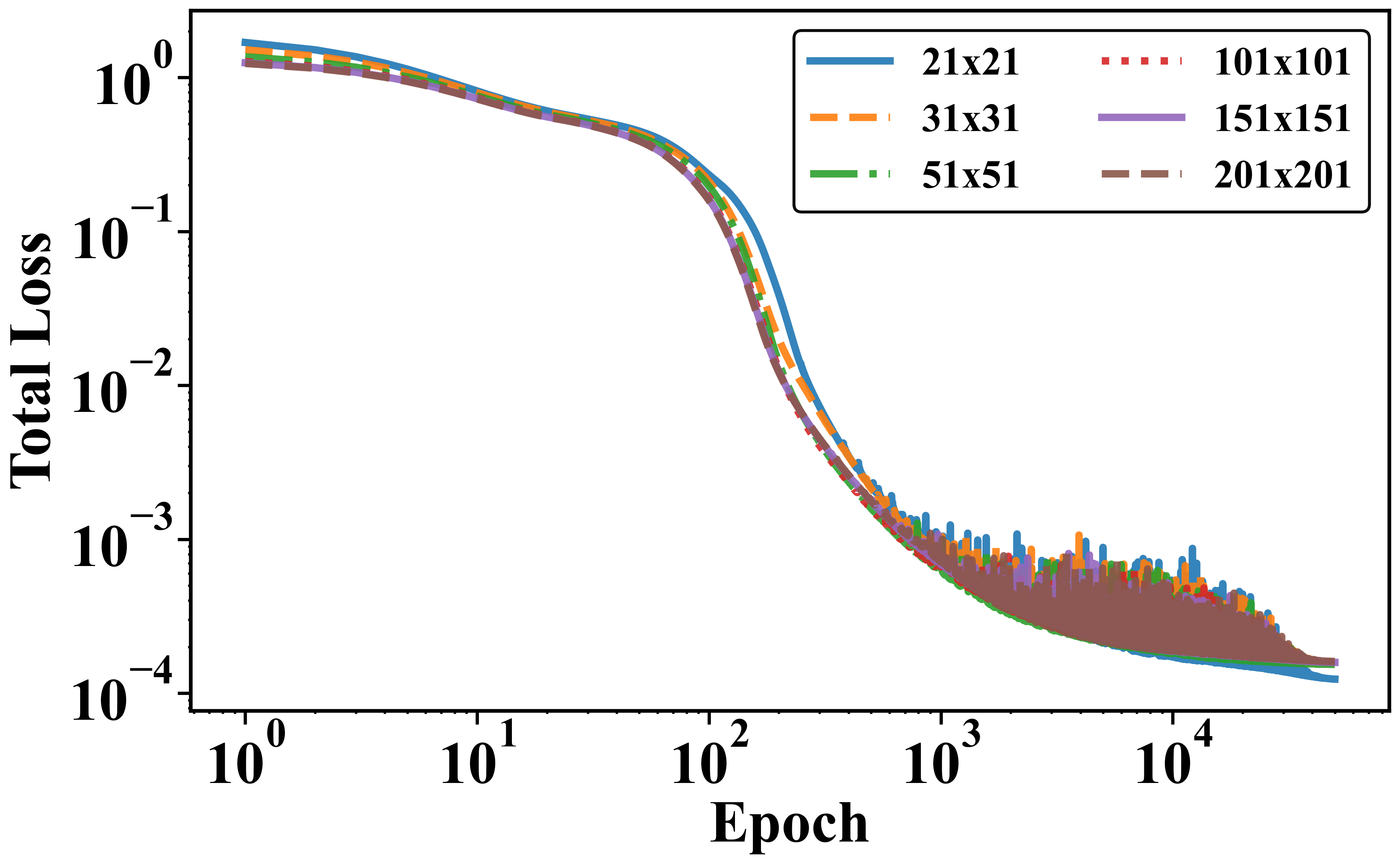}
    \par\noindent\hspace*{0.1\linewidth}\textbf{(a)}
    \vspace{0.5em}

    \includegraphics[width=0.8\linewidth]{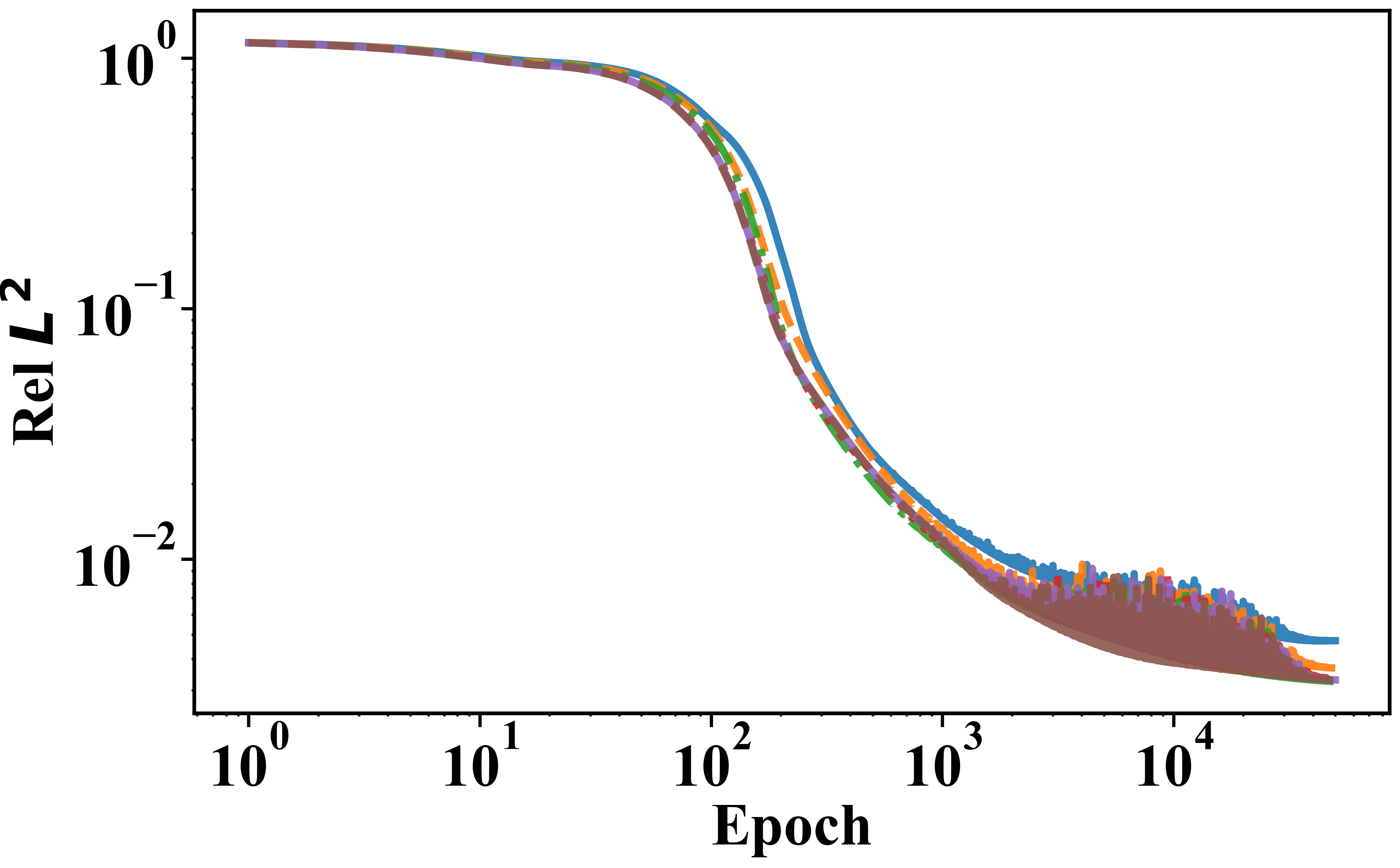}
    \par\noindent\hspace*{0.1\linewidth}\textbf{(b)}
    \vspace{0.5em}

    \caption{Grid sensitivity study of JetKAN-F on the 2D
Taylor--Green vortex ($Re = 100$), using six collocation grids
($21 \times 21$, $31 \times 31$, $51 \times 51$, $101 \times 101$,
$151 \times 151$, $201 \times 201$) with $K_x = K_y = 8$, $B = 8$,
$P = 5$, seed $42$, and Adam at $10^{-3}$ for $50{,}000$ epochs.
\textbf{(a)}~Total loss versus epoch. All six curves converge to
$\sim\!10^{-4}$ with no visible separation between grids by the end
of training.
\textbf{(b)}~$\mathrm{Rel}~L^{2}$ error versus epoch. The $21 \times 21$
grid settles at a distinctly higher error compared to the remaining
grids, which cluster together and converge to the same level, indicating
that accuracy saturates beyond $51 \times 51$.}
    \label{fig:collocation_loss_l2}
\end{figure}

We test the sensitivity of \textit{JetKAN-F} to collocation grid density by varying the collocation grid from $21 \times 21$ (441~points) to $201 \times 201$ (40{,}401~points). All experiments use \textit{JetKAN-F} on the steady Taylor--Green vortex at $Re=100$  with the same 656~parameters ($K_x = K_y = 8$, $B = 8$, $P = 5$), 50{,}000~epochs with Adam at $\text{lr} = 10^{-3}$, and seed~42. the training histories are shown in Figs.~\ref{fig:collocation_loss_l2}(a) and (b).

As shown in Table~\ref{tab:collocation_table}, at $21 \times 21$ grid, $MSE=1.26 \times 10^{-5}$ and  $\mathrm{Rel}~L^{2}(u, v, p)$ are respectively $(4.01 \times 10^{-3}, 3.88 \times 10^{-3}, 8.62 \times 10^{-3})$. At $31 \times 31$, the $MSE=7.63 \times 10^{-6}$ and the $\mathrm{Rel}~L^{2}$ errors drop noticeably, particularly for the pressure field ($\mathrm{Rel}~L^{2}(p)= 5.06 \times 10^{-3}$. In particular, the $51 \times 51$ grid gives the lowest $MSE=5.96 \times 10^{-6}$ with $\mathrm{Rel}~L^{2}(u,v,p)$ errors of ($3.10 \times 10^{-3}, 3.07 \times 10^{-3}, 4.33 \times 10^{-3}$), respectively. However, neither the $MSE$ nor the $\mathrm{Rel}~L^{2}$ errors continue to improve further at higher grid densities of $101 \times 101$, $151 \times 151$, and $201 \times 201$. Both $MSE$, $\mathrm{Rel}~L^{2}$ errors, and the PDE loss remain comparable to those on the $51 \times 51$ grid. 

The $\mathrm{Rel}~L^{2}$ error histories in Fig.~\ref{fig:collocation_loss_l2}(b) further elaborate the limited effect of collocation points. The error of the $21 \times 21$ grid is clearly separated from the rest, settling at a higher error around $10^{-3}$. The errors of the other grids are nearly indistinguishable by the end of training. The total loss histories in Fig.~\ref{fig:collocation_loss_l2}(a) show a similar behavior with a close overlap of all curves and a final value of approximately $10^{-4}$. 

The per-epoch training time remains consistently at $8.5-8.7$~ms across all grids. This is because the collocation points in \textit{JetKAN-F} do not add trainable parameters. The model has a fixed $656$) parameters across all grids, and the bottleneck projection compresses features into a fixed $B = 8$ representation before the KAN layer. Hence, the per-epoch cost remains the same regardless of grid size. The grid density can therefore be chosen based solely on accuracy, without affecting training time.

\subsubsection{Network depth}
\label{sec:ablation_depth}
The effect of network depth on accuracy and training cost was evaluated by varying the number of \textit{KAN} layers from $1$ layer ($1L$) to $6$ ($6L$) on the $2D$ Taylor--Green
vortex at $Re = 100$. All cases use $K_x = K_y = 8$, $B = 5$, degree $P = 5$, and are trained for 50{,}000 epochs with Adam at lr $= 10^{-3}$. 

\begin{figure}[htbp]
    \centering
    \includegraphics[width=0.8\linewidth]{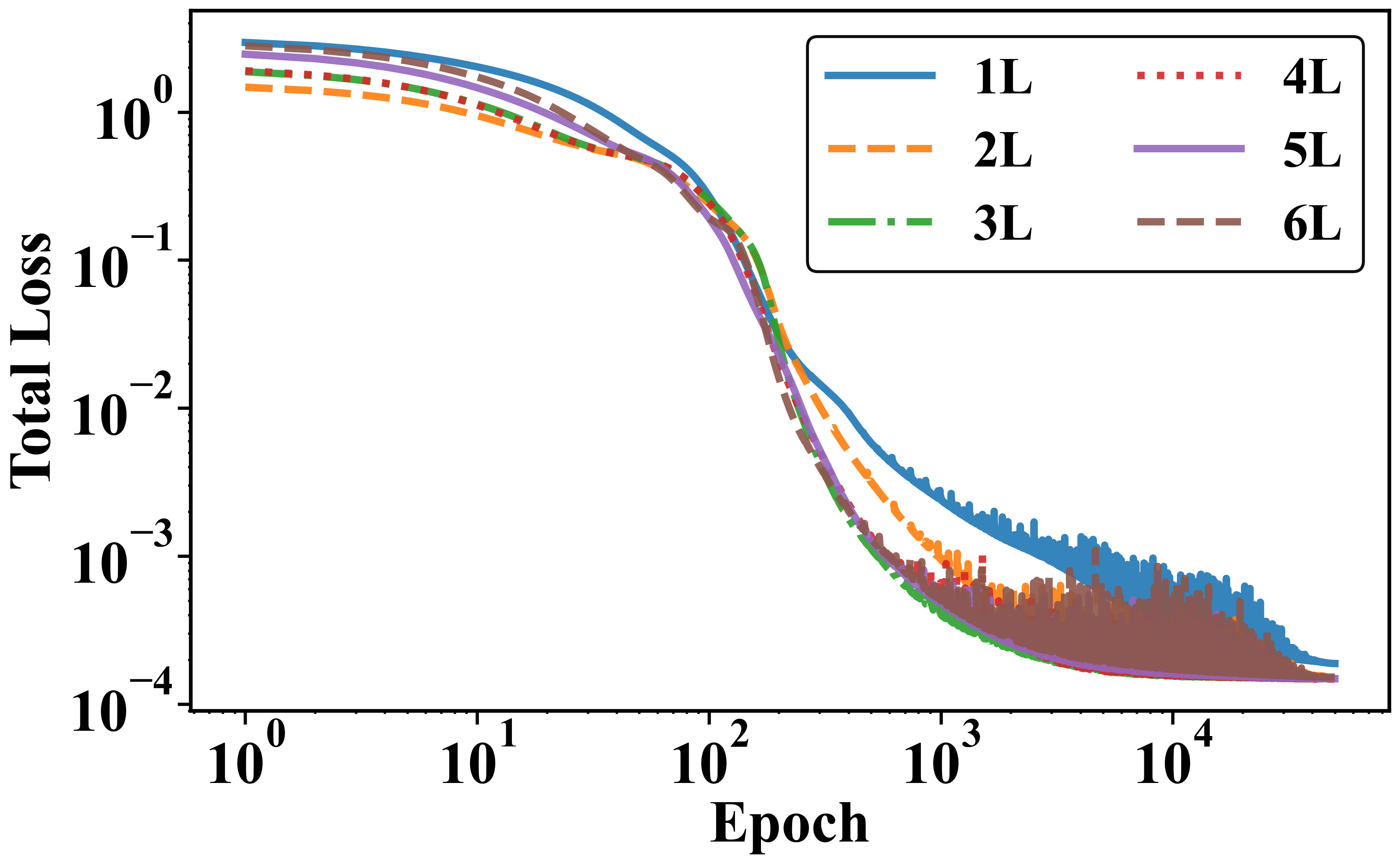}
    \par\noindent\hspace*{0.1\linewidth}\textbf{(a)}
    \vspace{0.5em}

    \includegraphics[width=0.8\linewidth]{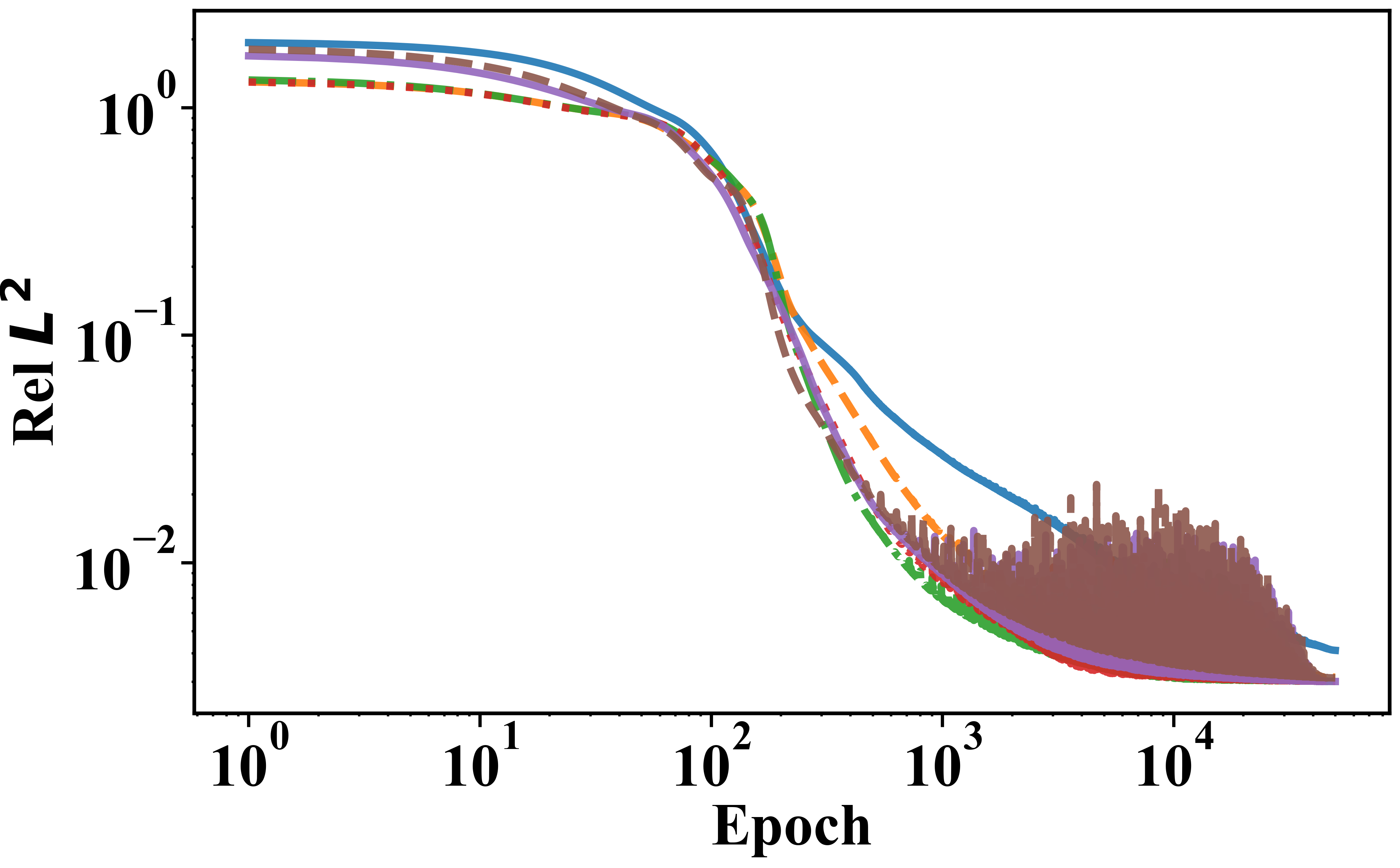}
    \par\noindent\hspace*{0.1\linewidth}\textbf{(b)}
    \vspace{0.5em}

    \caption{Network depth study of JetKAN-F on the 2D Taylor--Green vortex
($Re = 100$) for six configurations ranging from 1 to 6 KAN layers
($K_x = K_y = 8$, $B = 5$, $P = 5$, Adam at $10^{-3}$, 50{,}000 epochs).
Hidden dims are $[5]$, $[5,5]$, $[5,5,5]$, $[5,5,5,5]$, $[5,5,5,5,5]$,
$[5,5,5,5,5,5]$ for 1L--6L respectively.
\textbf{(a)}~Total loss versus epoch.
The 1L curve converges to a higher final floor ($\sim\!1.5\times10^{-4}$)
than the deeper networks; 2L--6L reach $\sim\!10^{-4}$, with 6L showing
progressively larger oscillations in the tail phase.
\textbf{(b)}~Rel\,$L^{2}$ error versus epoch.
1L settles noticeably above the rest; 2L--6L cluster tightly, but 6L
exhibits the most rapid fluctuations and fails to settle to a clean
minimum by epoch $5\times10^{4}$.}
    \label{fig:depth_loss_l2}
\end{figure}

As summarized in Table~\ref{tab:depth_study} and Fig.~\ref{fig:depth_loss_l2}, the network depth increases from $1L$ to $2L$ KAN layers, the \textit{MSE} drops from $9.54\times10^{-6}$ to $5.54\times10^{-6}$ (a reduction of 42\%) and Rel\,$L^2$ from $3.69\times10^{-3}$ to $2.98\times10^{-3}$ ($u$). Adding a third layer gives a further moderate reduction to $MSE = 5.10\times10^{-6}$ and Rel\,$L^2 = 2.84\times10^{-3}$ ($u$). As the network depth further increases to $4L$ and $5L$, the value of \textit{MSE} reduces by less than 0.5\% relative to one at $3L$ while the Rel\,$L^2$ values of $3L, 4L$ and $5L$ are indistinguishable (approximately $2.83\times10^{-3}$ ($u$)). At $6L$ the final Rel\,$L^2$ rises back to $2.95\times10^{-3}$ ($u$), slightly above the one at $3L$. As seen in Fig.~\ref{fig:depth_loss_l2}, as the depth increases, the loss curves develop progressively larger oscillations in the tail phase (epochs $10^3$--$5\times10^4$). Especially, the one at $6L$ fails to settle into a clean minimum. These high-frequency fluctuations prevent the optimizer from reaching the accuracy achieved by shallower networks. Note that the per-epoch training time increases roughly linearly with depth. It increases from $10.0\,ms$ at $1L$ to $43.5\,ms$ at $6L$ (an increase of approximately $4.3\times$). At the same time, the gained accuracy (Rel\,$L^2$ ($u$)) beyond $2L$ is less than $6\%$. In this particular problem, the $3L$ configuration (712 parameters, 23.8\,ms/epoch) offers the best balance between accuracy and training cost.

\subsubsection{Reproducibility}
\label{subsec: 6.3 seed-sensitivity}
Cui et al.~\cite{cui2025pikan} report that the \textit{PDE} loss of the Navier--Stokes equations varied by an order of magnitude across seeds when using \textit{PIKANs}. Thus, it is critical to examine \textit{JetKAN-F} 's sensitivity to random seeds. We select eight values ($3$, $16$, $32$, $64$, $128$, $256$, $512$, $1024$) to examine whether the trained model depends on initialization. To validate the transient effect over time, we use the unsteady Taylor--Green vortex (Eqs.~(\ref{eq:tgv_u})--(\ref{eq:tgv_p})) at $Re = 100$ for $t \in [0, 2]$. The loss function is Eq.~(\ref{eqn:total_loss}) with $\lambda_R = \lambda_{BC} = \lambda_{IC} = 1$. The  $\mathcal{L}_R$ contains the momentum and continuity residuals, $\mathcal{L}_{BC}$ enforces periodicity of $(\hat{u}, \hat{v}, \hat{p})$ on all four sides for at all time $t$, and $\mathcal{L}_{IC}$ prescribes the analytical velocity at $t = 0$. The basis functions $\phi_{ij}$ of Eq.~(\ref{eq:7. ckan univariate}) are extended with a Chebyshev factor in time, $\phi_{ijk} = T_i(\xi_x)\,T_j(\xi_y)\,T_k(\xi_t)$, so that $n_f = K_x K_y K_t$. All other settings are kept fixed: $K_x = K_y = 8$, $K_t = 4$, $B = 8$, $P = 5$, a $51 \times 51 \times 51$ space--time collocation grid, $50{,}000$ epochs with Adam at a learning rate of $10^{-3}$, and $2{,}192$ trainable parameters. 

\begin{figure}[htbp]
    \centering
    \includegraphics[width=0.8\linewidth]{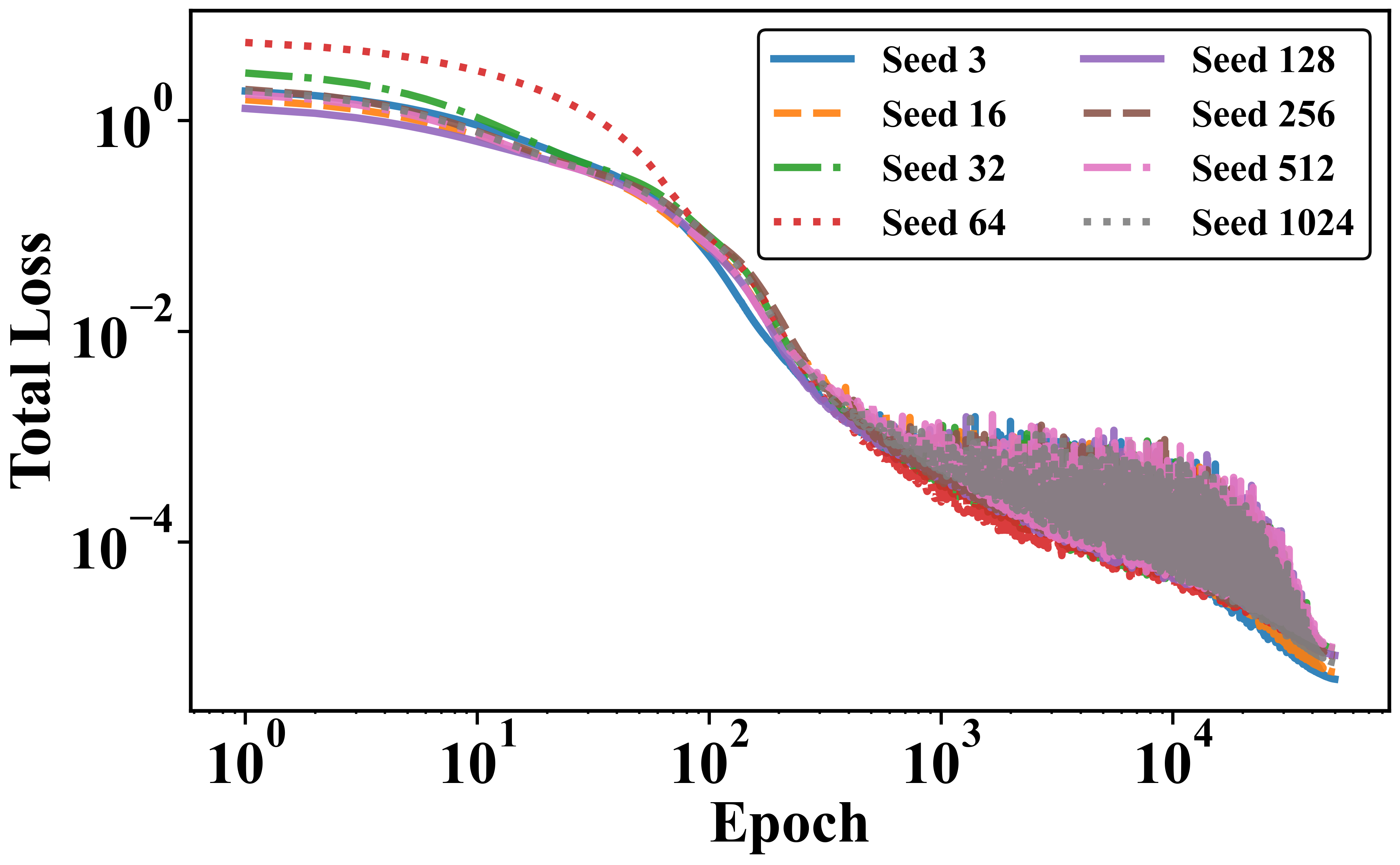}
    \par\noindent\hspace*{0.1\linewidth}\textbf{(a)}
    \vspace{0.5em}

    \includegraphics[width=0.8\linewidth]{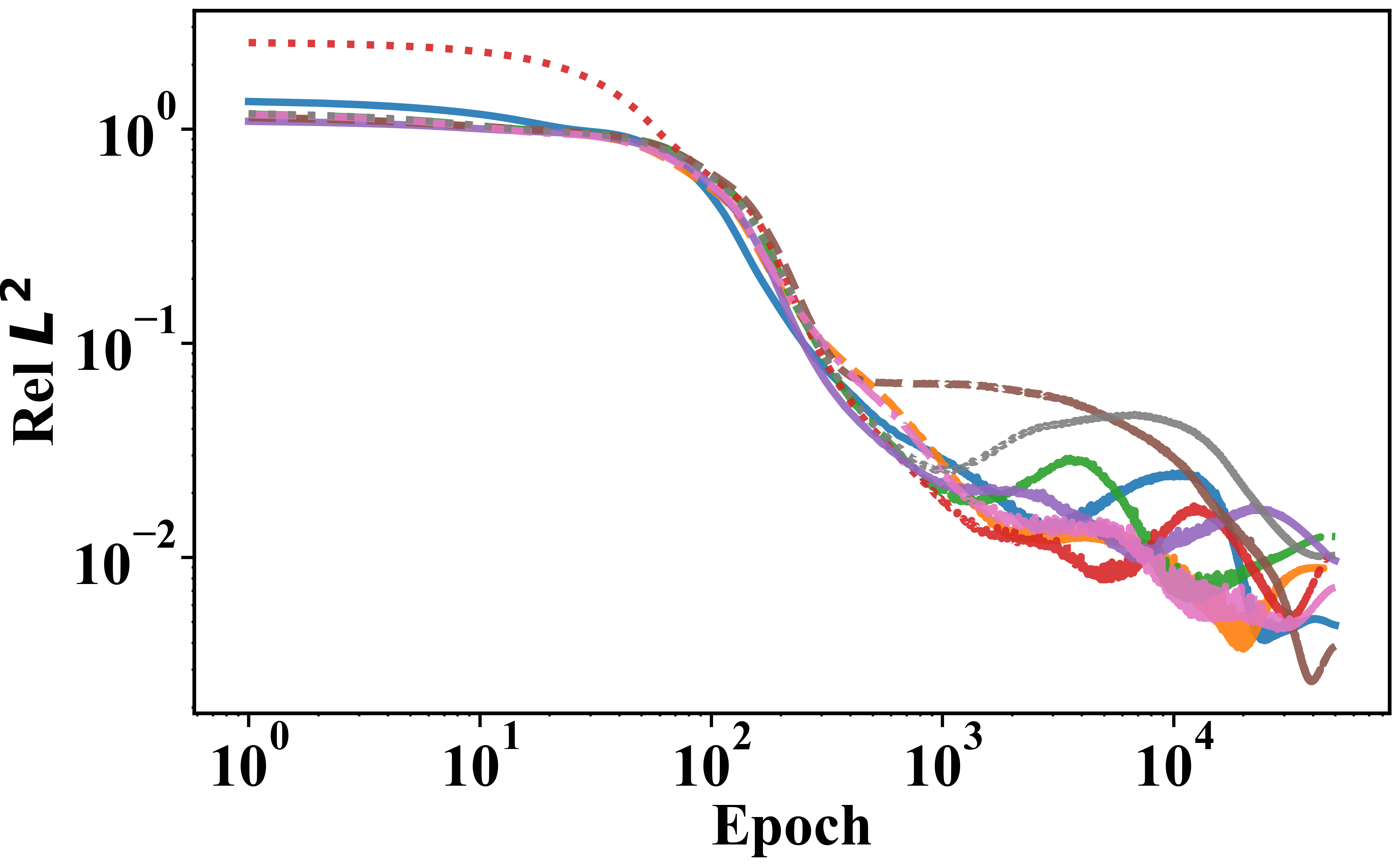}
    \par\noindent\hspace*{0.1\linewidth}\textbf{(b)}
    \vspace{0.5em}

    \caption{Seed-sensitivity study of \textit{JetKAN-F} on the 2D unsteady Taylor--Green vortex ($Re = 100$) over eight random seeds (3, 16, 32, 64, 128, 256, 512, 1024) with $K_x = K_y = 8$, $B = 8$, $P = 5$, a $51 \times 51 \times 51$ collocation grid, and Adam at $10^{-3}$ for 50{,}000 epochs. \textbf{(a)} Total loss versus epoch. All eight curves follow nearly identical trajectories and converge to $\sim 10^{-5}$ by the end of training. \textbf{(b)} Rel\,$L^2$ error versus epoch. The eight curves remain tightly clustered throughout and settle between $\sim 10^{-3}$ and $\sim 10^{-2}$, with no seed producing an outlier or failing to converge.}
    \label{fig:seed_loss_l2}
\end{figure}

The total loss histories for all seeds are shown in Figure~\ref{fig:seed_loss_l2}(a). All curves follow the same path during training, and every run converges. The final value of $\mathcal{L}_R$ ranges from $4.45 - 8.65 \times 10^{-6}$, and that of $\mathcal{L}_{BC}$ from $2.04 - 4.18 \times 10^{-7}$, a factor of about 2 between the largest and the smallest value in both cases. As shown in Fig.~\ref{fig:seed_loss_l2}(b), all losses decrease together within a narrow band. The $\mathrm{Rel}~L^{2}$ error ranges from $1.26 -1.99 \times 10^{-3}$ ($u$) and $1.16 -2.25 \times 10^{-3}$ ($v$), so the velocity error changes by less than a factor of 2 between the best and the worst seed. The pressure error ranges from $1.05 -3.75 \times 10^{-2}$; the spread is larger because the pressure enters the loss only through its gradient and one mean-value constraint. In \textit{JetKAN-F}, the seed sets only the initial values of the trainable parameters, $W$ and $c_{b,o,n}$. The feature matrix $\Phi$ and its derivative matrices are computed once before training and do not depend on the seed, so all cases represent the solution in the same fixed basis and differ only in the optimizer's starting point. Our result shows that a single training run of \textit{JetKAN-F} is sufficient to provide reliable simulation results. There is no need to repeat the training with several seeds and report the best or the average result. 

\subsubsection{Model complexity}
\begin{figure}
    \centering
    \includegraphics[width=1.1\linewidth]{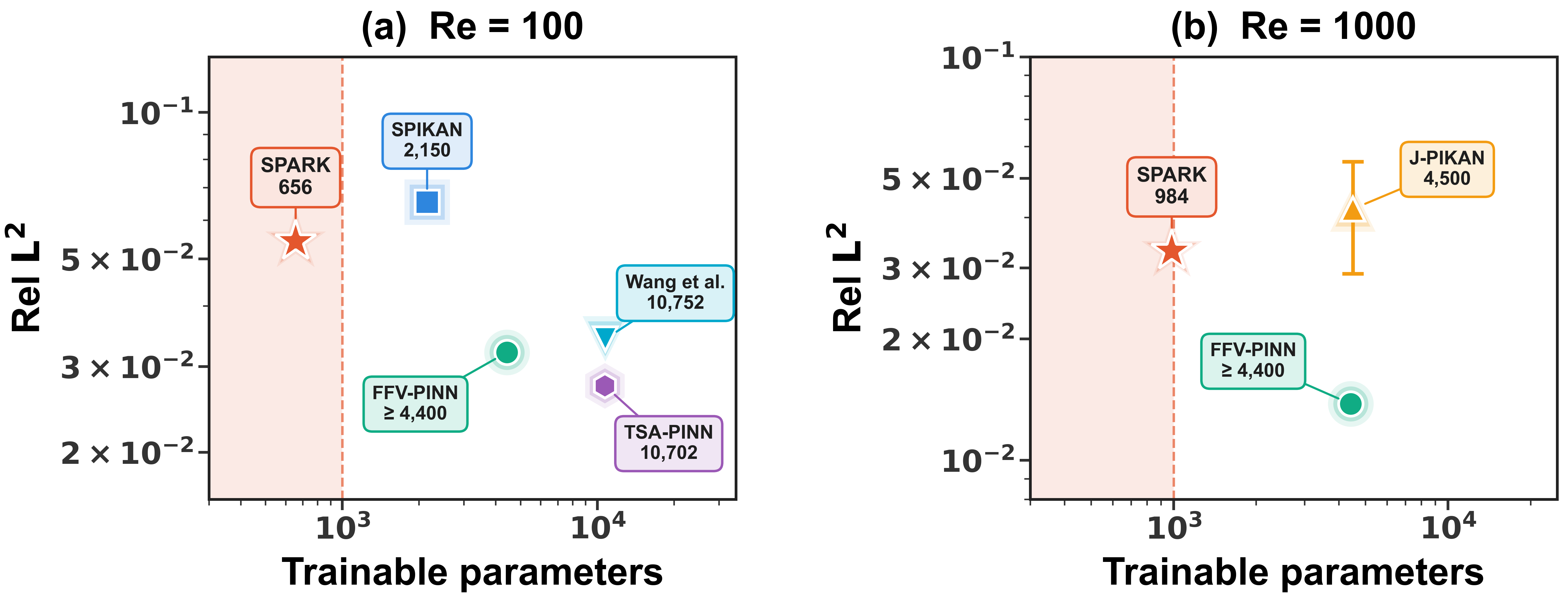}
    \caption{Relative $L^2$ error versus the number of trainable parameters
for SPARK (JetKAN-F) and existing methods in the literature on the 2D
lid-driven cavity at (a) $Re = 100$ and (b) $Re = 1000$. The shaded
region marks models with fewer than $10^3$ trainable parameters, and
SPARK is the only method in this region at both Reynolds numbers.
SPARK uses 656 parameters at $Re = 100$ and 984 parameters at
$Re = 1000$. The other methods use about 3 to 16 times more parameters
at $Re = 100$ and about 4.5 times more at $Re = 1000$, while their
errors are of the same order as that of SPARK.}
    \label{fig:literature_comparision}
\end{figure}

The performance of \textit{JetKAN-F} is compared with existing methods  (\textit{SPIKAN} \cite{jacob2025spikans}, \textit{FFV-PINN} \cite{wei2025ffvpinn}, \textit{TSA-PINN} \cite{khademi2025physics}, \textit{J-PIKAN} \cite{xiong2025jpikan}) for the lid-driven cavity flow at $Re = 100$ and $Re = 1000$, which are the most reported  Reynolds numbers. At both Reynolds numbers, \textit{JetKAN-F} uses the fewest parameters. At $Re=100$ \textit{JetKAN-F} reaches an error $\mathrm{Rel}~L^2$ error of $5.43 \times 10^{-2}$ with only 656 parameters as seen in Fig.~\ref{fig:literature_comparision}(a). At $Re = 1000$ (Fig.~\ref{fig:literature_comparision}(b)), \textit{JetKAN-F} reaches $3.31 \times 10^{-2}$ with 984 parameters. These errors are comparable to those of other methods. However, \textit{JetKAN-F} uses the fewest parameters. In \textit{SPIKAN, J-PIKAN, FFV-PINN, TSA-PINN}, the network learns the spatial structure of the solution directly from the input coordinates, and this requires a large number of trainable weights. In \textit{JetKAN-F}, the spatial structure is carried by the fixed Chebyshev basis, the linear bottleneck compresses it into a few latent channels, and a single KAN layer maps these channels to the solution. The trainable parameters are therefore limited to the bottleneck matrix $W$ and the coefficients $c_{b,o,n}$, which keeps the parameter count below $10^3$ for both Reynolds numbers.

\section{Limitation and future work}
The current \textit{SJ-KAN} architecture has three practical challenges that should be acknowledged. First, the spectral basis is defined on a rectangular domain, which restricts direct application to simple geometries. Extending to curved boundaries or bodies with internal features would require domain decomposition. Second, the optimal choices of spectral order $K$ and bottleneck width $B$ are problem-specific (Section~\ref{sec:ablation}). Third, the architecture does not address the weighting of the loss terms. For example, the cavity flow at high $Re$ still requires adaptive \textit{BRDR} weighting. Zhang et al.~\cite{zhang2025alpkan} incorporated an augmented Lagrangian formulation into a hybrid \textit{KAN} architecture to enforce multiple physical constraints simultaneously, reducing the sensitivity to loss-weight tuning that typically requires manual adjustment. This constraint treatment is independent of how the derivatives are computed and can be combined with the spectral jet.

\section{Conclusion}
The main challenge for Physics-Informed Neural solvers of the incompressible Navier--Stokes equations is the pressure-velocity $(u,v,p)$ coupling. The simultaneous constraints through the momentum and continuity equations create imbalances between these residual terms and lead to stalled convergence. In this work, the Spectral--Jet KAN (\textit{SJ--KAN}) is introduced, which embeds the input coordinates in a tensor-product Chebyshev basis, compresses the basis through a linear bottleneck, and maps the latent features to the solution through a single nonlinear Chebyshev KAN (\textit{CKAN}) layer. Because the Chebyshev derivatives are known in closed form and the bottleneck is linear, all coordinate derivatives required by the \textit{PDE} residual propagate analytically through the network, removing automatic differentiation from the residual evaluation entirely. In \textit{SJ-KAN}, the three fields are produced simultaneously from the same bottleneck channels by one KAN layer, so the momentum and continuity residuals act on a shared set of latent features. Finally, three inference variants of $SJ-KAN$ (\textit{JetKAN}, \textit{JetKAN-F}, \textit{JetKAN-H}) are validated against $AD$-based methods (\textit{CKAN} and \textit{PINN} on the Taylor--Green vortex and the two-dimensional lid-driven cavity. The results show that \textit{JetKAN-F} performs excellently at Reynolds numbers up to 10\,000, the highest Reynolds number reported to date for KAN-based physics-informed solvers for the lid-driven cavity problem. 

The principal findings are summarized below.

\begin{enumerate}

  \item \textit{Seed robustness.}In \textit{SJ-KAN}, the spectral basis and all differentiation matrices are fixed before training, so the seed changes only the starting values of the bottleneck and KAN-layer weights, and a single run gives a representative result.

  \item \textit{Accuracy and cost of the analytic jet.} The \textit{SJ-KAN} network and $AD$-based methods provide the same solution. However, \textit{SJ-KAN} requires much lower computational cost. 
 
  \item \textit{Parameter efficiency.} \textit{SJ-KAN} network achieves the lowest \textit{MSE} with a small number of parameters, in comparison to \textit{CKAN} and \textit{PINN} methods. 
 
  \item \textit{Necessity of the nonlinear KAN layer.} Removing the nonlinear KAN output layer and retaining only the spectral basis with the linear bottleneck limits both accuracy and scalability. Replacing the Chebyshev KAN layer with a single linear-plus-tanh MLP layer (\textit{JetMLP-F}) yields higher $MSE$ error, albeit at lower per-epoch cost. Both types of layers (MLP or KAN) receive the same spectral basis and bottleneck projection, and the comparable accuracy confirms that the spectral jet structure, not the choice of nonlinear layer, drives the solution. 
\end{enumerate}

\section{Acknowledgement}
 The authors express their gratitude to the funding provided to support this study from National Science Foundation (NSF) EPSCoR RII Track-2 Program under the grant number OIA-2119691 and MRI-Track 2 2510020. The findings and opinions expressed in this article are those of the authors only and do not necessarily reflect the views of the sponsors. 

\clearpage
{\small
\renewcommand{\arraystretch}{1.1}
\begin{longtable*}{ll}
\caption*{List of symbols and acronyms}\label{tab:symbols}\\
\toprule
\textbf{Symbol} & \textbf{Description} \\
\midrule
\endfirsthead
\toprule
\textbf{Symbol} & \textbf{Description} \\
\midrule
\endhead
\midrule
\multicolumn{2}{r}{\textit{Continued on next page}} \\
\endfoot
\bottomrule
\endlastfoot
$B$ & Bottleneck width \\
$b$ & Bottleneck channel index \\
$c$ & Chebyshev coefficients \\
$\mathbf{D}_1$, $\mathbf{D}_2$ & First- and second-derivative Chebyshev matrices \\
$d$ & Number of input variables \\
$d_l$ & Number of nodes in layer $l$ ($N_n$) \\
$g$ & Chebyshev derivative sums of the activation \\
$H_x$, $H_y$ & Domain lengths \\
$h_x$, $h_y$ & Scale factors (half-lengths of the domain) \\
$i$, $j$ & Chebyshev degrees in $x$ and $y$ \\
$K$, $K_x$, $K_y$ & Number of Chebyshev modes (per direction) \\
$k$ & Chebyshev degree index (row of $\mathbf{D}_1$) \\
$L$ & Number of KAN layers ($N_l$) \\
$l$ & Layer index \\
$\mathcal{L}$ & Training loss \\
$m$ & Collocation-point index \\
$N$ & Number of collocation points \\
$\mathcal{N}$ & Nonlinear differential operator \\
$n$ & Chebyshev degree index \\
$n_f$ & Number of basis functions \\
$O$ & Number of output channels \\
$o$ & Output channel index \\
$P$ & Highest degree of the KAN activation \\
$p$ & Pressure \\
$q$ & Outer-function index \\
$Re$ & Reynolds number \\
$\mathcal{R}$ & Steady residual \\
$r$ & Pseudo-step index \\
$s$ & $\tanh$-scaled latent variable \\
$T_n$ & Chebyshev polynomial of degree $n$ \\
$t$, $t_f$ & Time and final time \\
$\mathbf{U}$ & Solution vector $(u,v,p)$ \\
$\hat{\mathbf{U}}$ & Network prediction \\
$u$, $v$ & Velocity components \\
$\mathbf{V}$ & Velocity field \\
$\mathbf{W}$ & Bottleneck weight matrix \\
$\mathbf{W}_{\mathrm{lin}}$ & Linear-head weight matrix (JetKAN-H) \\
$x$, $y$ & Cartesian coordinates \\
$\mathbf{Z}$ & Latent matrix \\
$z$ & Latent variable \\
\midrule
$\alpha$ & Node index \\
$\beta$ & Base function of a CKAN edge \\
$\gamma$ & Normalization constant of $\mathbf{D}_1$ \\
$\varepsilon$ & Pseudo-time residual \\
$\kappa$ & Input-variable index \\
$\lambda$ & Loss weight \\
$\nu$ & Kinematic viscosity \\
$\xi$ & Normalized coordinate \\
$\sigma$ & Derivative of $\tanh$ \\
$\tau$ & Pseudo-time \\
$\phi$ & Univariate (Chebyshev) function \\
$\Phi$ & Outer function; KAN function matrix \\
$\boldsymbol{\Phi}$ & Feature matrix \\
$\Omega$ & Computational domain \\
$\omega$ & Weight of the base function \\
\midrule
\textbf{Acronym} & \textbf{Expansion} \\
\midrule
AD & Automatic differentiation \\
BC, IC & Boundary condition, initial condition \\
BRDR & Balanced residual decay rate \\
CKAN & Chebyshev Kolmogorov--Arnold network \\
JetKAN & Spectral--Jet KAN (einsum variant) \\
JetKAN-F & Spectral--Jet KAN (fused variant) \\
JetKAN-H & Spectral--Jet KAN (hybrid variant) \\
KAN & Kolmogorov--Arnold network \\
L-BFGS & Limited-memory Broyden--Fletcher--Goldfarb--Shanno \\
MLP & Multilayer perceptron \\
MSE & Mean squared error \\
PIKAN & Physics-informed Kolmogorov--Arnold network \\
PINN & Physics-informed neural network \\
RAR & Residual-based adaptive refinement \\
SJ-KAN & Spectral--Jet Kolmogorov--Arnold network \\
SPARK & Spectral Preconditioned Analytic Residual KAN \\
\end{longtable*}
}
\addtocounter{table}{-1} 

\clearpage
\bibliographystyle{plain}
\bibliography{pikan}

\clearpage
\section{Table}

\clearpage
\begin{table}[htbp]
\centering
\caption{Models compared in this work. ``Jet'' denotes analytic
derivative propagation and ``AD'' reverse-mode automatic
differentiation. Parameter counts are given for the Taylor--Green
vortex (TGV) and for the lid-driven cavity at $Re=100$ (LDC).}
\label{tab:arch-summary}
\small
\setlength{\tabcolsep}{5pt}
\begin{tabular}{llllrr}
\toprule
Model & Input features & Trainable part & Derivatives & TGV & LDC \\
\midrule
JetKAN   & Chebyshev TP basis & $\mathbf{W}$ + one CKAN layer (einsum) & Jet & 656 & 656 \\
JetKAN-F & Chebyshev TP basis & $\mathbf{W}$ + one CKAN layer (fused)  & Jet & 656 & 656 \\
JetKAN-H & Chebyshev TP basis & $\mathbf{W}$ + one CKAN layer + $\mathbf{W}_{lin}$ & Jet & 848 & 848 \\
CKAN     & $(x,y)$            & stacked CKAN layers                    & AD  & 1{,}148 & 11{,}880 \\
PINN     & $(x,y)$            & MLP with $\tanh$                       & AD  & 12{,}867 & 30{,}903 \\
\bottomrule
\end{tabular}
\end{table}

\clearpage
\begin{sidewaystable}[htbp]
\centering
\caption{Performance comparison of neural network architectures for the 2D Taylor--Green vortex (steady Navier--Stokes).}
\label{tab:tgv_results}
\resizebox{\textwidth}{!}{%
\large
\begin{tabular}{lcccccccccc}
\hline
\textbf{Model} & \textbf{Parameters} & \textbf{Epochs} & \textbf{Optimizer} & \textbf{PDE Loss} & \textbf{BC Loss} & \textbf{MSE} & \textbf{Rel $L^2$ ($u$)} & \textbf{Rel $L^2$ ($v$)} & \textbf{Rel $L^2$ ($p$)} & \textbf{ms/epoch} \\
\hline
JetKAN   & 656    & 50,000 & Adam & $1.27 \times 10^{-4}$ & $1.85 \times 10^{-5}$ & $5.69 \times 10^{-6}$          & $2.99 \times 10^{-3}$          & $3.05 \times 10^{-3}$          & $4.23 \times 10^{-3}$          & 9.0 \\
JetKAN-F & 656    & 50,000 & Adam & $1.29 \times 10^{-4}$ & $1.87 \times 10^{-5}$ & $5.96 \times 10^{-6}$          & $3.10 \times 10^{-3}$          & $3.07 \times 10^{-3}$          & $4.33 \times 10^{-3}$          & $\mathbf{7.8}$ \\
JetKAN-H & 848    & 50,000 & Adam & $1.22 \times 10^{-4}$ & $1.80 \times 10^{-5}$ & $\mathbf{5.17 \times 10^{-6}}$ & $\mathbf{2.86 \times 10^{-3}}$ & $\mathbf{2.83 \times 10^{-3}}$ & $\mathbf{4.18 \times 10^{-3}}$ & 9.5 \\
CKAN  & 1,148  & 50,000 & Adam & $2.33 \times 10^{-4}$ & $1.70 \times 10^{-5}$ & $1.84 \times 10^{-5}$          & $4.95 \times 10^{-3}$          & $4.88 \times 10^{-3}$          & $6.81 \times 10^{-3}$          & 144.8 \\
PINN & 12,867 & 50,000 & Adam & $1.40 \times 10^{-4}$ & $2.05 \times 10^{-5}$ & $7.77 \times 10^{-6}$          & $3.53 \times 10^{-3}$          & $3.55 \times 10^{-3}$          & $5.27 \times 10^{-3}$          & 15.7 \\
\hline
\end{tabular}%
}
\end{sidewaystable}

\clearpage
\begin{sidewaystable}[htbp]
\centering
\caption{Performance comparison of neural network architectures for the 2D lid-driven cavity
flow (steady Navier--Stokes) across Reynolds numbers $Re \in \{100,\,1000,\,5000,\,10{,}000\}$.}
\label{tab:ldc_all_re}
\setlength{\tabcolsep}{4pt}
\renewcommand{\arraystretch}{1.1}
\begin{tabular}{cccclrrrccccr}
\hline
$Re$ & \textbf{Model} & $K_x$ & $K_y$ & $B$ & \textbf{Params} & \textbf{Epochs} & \textbf{Opt.}
  & \textbf{MSE}$(u)$ & \textbf{MSE}$(v)$
  & \textbf{Rel}\,$L^{2}(u)$ & \textbf{Rel}\,$L^{2}(v)$
  & \textbf{ms/ep} \\
\hline
\multirow{5}{*}{100}
  & JetKAN   & 8 & 8 &  8 & 656 & 100{,}000 & Adam
    & $7.78\times10^{-5}$ & $9.67\times10^{-5}$
    & $4.13\times10^{-2}$ & $6.48\times10^{-2}$ &   8.8 \\
  & JetKAN-F & 8 & 8 & 8  &  656 & 100{,}000 & Adam
    & $9.61\times10^{-5}$ & $8.70\times10^{-5}$
    & $4.59\times10^{-2}$ & $6.15\times10^{-2}$ &   8.2 \\
  & JetKAN-H & 8 & 8 & 8 &  848 & 100{,}000 & Adam
    & $5.51\times10^{-5}$ & $5.09\times10^{-5}$
    & $3.47\times10^{-2}$ & $4.71\times10^{-2}$ &   9.1 \\
  & CKAN & 8 & 8 & 8  & 11{,}880 & 100{,}000 & Adam
    & $1.96\times10^{-4}$ & $3.63\times10^{-4}$
    & $6.54\times10^{-2}$ & $1.26\times10^{-1}$ & 267.6 \\
  & PINN & 8 & 8 & 8 & 30{,}903 & 100{,}000 & Adam
    & $4.00\times10^{-5}$ & $4.32\times10^{-5}$
    & $2.96\times10^{-2}$ & $4.33\times10^{-2}$ &  17.5 \\
\hline
\multirow{3}{*}{1{,}000}
  & JetKAN   & 8 & 8 & 12 &  984 & 10{,}000 & L-BFGS
    & $2.75\times10^{-4}$ & $3.60\times10^{-4}$
    & $7.71\times10^{-2}$ & $9.35\times10^{-2}$ & 196.0 \\
  & JetKAN-F & 8 & 8 & 12 &  984 & 10{,}000 & L-BFGS
    & $5.38\times10^{-5}$ & $4.22\times10^{-5}$
    & $3.41\times10^{-2}$ & $3.20\times10^{-2}$ & 150.0 \\
  & JetKAN-H & 8 & 8 & 12 &  1{,}176 & 10{,}000 & L-BFGS
    & $5.58\times10^{-5}$ & $7.12\times10^{-5}$
    & $3.47\times10^{-2}$ & $4.16\times10^{-2}$ & 217.6 \\
\hline
\multirow{3}{*}{5{,}000}
  & JetKAN & 12 & 12 & 24  & 3{,}888 & 20{,}000 & L-BFGS
    & $3.35\times10^{-4}$ & $3.49\times10^{-4}$
    & $8.70\times10^{-2}$ & $9.02\times10^{-2}$ & 539.3 \\
  & JetKAN-F & 12 & 12 & 24 & 3{,}888 & 20{,}000 & L-BFGS
    & $2.45\times10^{-4}$ & $2.73\times10^{-4}$
    & $7.45\times10^{-2}$ & $7.98\times10^{-2}$ & 189.0 \\
  & JetKAN-H & 12 & 12 & 24 &  4{,}320 & 20{,}000 & L-BFGS
    & $1.44\times10^{-3}$ & $1.62\times10^{-3}$
    & $1.81\times10^{-1}$ & $1.95\times10^{-1}$ & 621.6 \\
\hline
\multirow{3}{*}{10{,}000}
  & JetKAN   & 16 & 16 & 32 & 8{,}768 & 40{,}000 & L-BFGS
    & $1.59\times10^{-4}$ & $2.23\times10^{-4}$
    & $6.36\times10^{-2}$ & $7.60\times10^{-2}$ &  910.8 \\
  & JetKAN-F & 16 & 16 & 32 & 8{,}768 & 40{,}000 & L-BFGS
    & $2.30\times10^{-4}$ & $2.78\times10^{-4}$
    & $7.66\times10^{-2}$ & $8.47\times10^{-2}$ &  287.4 \\
  & JetKAN-H & 16 & 16 & 32 & 9{,}536 & 40{,}000 & L-BFGS
    & $2.42\times10^{-4}$ & $2.97\times10^{-4}$
    & $7.85\times10^{-2}$ & $8.77\times10^{-2}$ & 1093.3 \\
\hline
\end{tabular}
\end{sidewaystable}


\clearpage
\begin{sidewaystable}[htbp]
\centering
\caption{Jet vs.\ autograd (AD) ablation: analytic jet propagation vs.\ PyTorch autograd for derivative computation on the identical JetKAN-F architecture for the 2D Taylor--Green vortex (steady Navier--Stokes, $Re = 100$). Both models share 656 parameters, $51\times51$ grid, $K_x = K_y = 8$, $B = 8$, degree $= 5$, 50{,}000 epochs, Adam optimizer, lr $= 0.001$.}
\label{tab:jet_ad_table}
\resizebox{\textwidth}{!}{%
\large
\begin{tabular}{cccccccccccc}
\hline
\textbf{Derivative Strategy} & \textbf{Parameters} & \textbf{Epochs} & \textbf{Optimizer} & \textbf{PDE Loss} & \textbf{BC Loss} & \textbf{MSE} & \textbf{Rel $L^2$ ($u$)} & \textbf{Rel $L^2$ ($v$)} & \textbf{Rel $L^2$ ($p$)} & \textbf{ms/epoch} & \textbf{Total time (s)} \\
\hline
Jet (analytic) & 656 & 50,000 & Adam & $1.29 \times 10^{-4}$ & $1.87 \times 10^{-5}$ & $5.95 \times 10^{-6}$ & $3.10 \times 10^{-3}$ & $3.06 \times 10^{-3}$ & $4.33 \times 10^{-3}$ & $\mathbf{7.4}$ & $\mathbf{371.3}$ \\
AD (autograd)  & 656 & 50,000 & Adam & $1.29 \times 10^{-4}$ & $1.87 \times 10^{-5}$ & $5.94 \times 10^{-6}$ & $3.10 \times 10^{-3}$ & $3.06 \times 10^{-3}$ & $4.33 \times 10^{-3}$ & 60.7 & 3035.4 \\
\hline
\multicolumn{4}{c}{\textbf{Ratio (AD / Jet)}} & $1.00$ & $1.00$ & $1.00$ & $1.00$ & $1.00$ & $1.00$ & $\mathbf{8.2\times}$ & $\mathbf{8.2\times}$ \\
\hline
\end{tabular}%
}
\end{sidewaystable}

\clearpage
\begin{table}[htbp]
\centering
\caption{$\mathrm{Rel}~L^{2}$ error for JetKAN-F on the Taylor--Green vortex ($Re = 100$) as a function of polynomial order $K$ and bottleneck width $B$. All runs use a $51 \times 51$ grid, 50{,}000 epochs, Adam optimizer, $\text{lr} = 10^{-3}$, degree $P = 5$, seed 42.}
\label{tab:multi_b_k}
\resizebox{\textwidth}{!}{%
\large
\begin{tabular}{rrrrrrr}
\hline
\textbf{$K$} & \textbf{$n_f$} & \textbf{$B = 2$} & \textbf{$B = 4$} & \textbf{$B = 5$} & \textbf{$B = 8$} & \textbf{$B = 10$} \\
\hline
 2 &   4 & $6.19 \times 10^{-1}$ & $3.07 \times 10^{-1}$ & $1.78 \times 10^{-1}$ & $5.81 \times 10^{-2}$ & $7.94 \times 10^{-3}$ \\
 3 &   9 & $6.79 \times 10^{-1}$ & $2.24 \times 10^{-1}$ & $8.87 \times 10^{-2}$ & $2.46 \times 10^{-2}$ & $1.07 \times 10^{-2}$ \\
 4 &  16 & $4.71 \times 10^{-1}$ & $2.11 \times 10^{-1}$ & $8.96 \times 10^{-2}$ & $1.82 \times 10^{-2}$ & $1.02 \times 10^{-2}$ \\
 5 &  25 & $4.71 \times 10^{-1}$ & $1.33 \times 10^{-1}$ & $2.99 \times 10^{-2}$ & $5.98 \times 10^{-3}$ & $4.51 \times 10^{-3}$ \\
 6 &  36 & $4.59 \times 10^{-1}$ & $2.61 \times 10^{-2}$ & $1.06 \times 10^{-2}$ & $4.38 \times 10^{-3}$ & $3.64 \times 10^{-3}$ \\
 8 &  64 & $4.45 \times 10^{-1}$ & $5.92 \times 10^{-3}$ & $4.11 \times 10^{-3}$ & $3.18 \times 10^{-3}$ & $3.09 \times 10^{-3}$ \\
10 & 100 & $4.44 \times 10^{-1}$ & $3.16 \times 10^{-3}$ & $3.11 \times 10^{-3}$ & $4.17 \times 10^{-3}$ & $5.23 \times 10^{-3}$ \\
12 & 144 & $4.43 \times 10^{-1}$ & $3.22 \times 10^{-3}$ & $3.06 \times 10^{-3}$ & $5.32 \times 10^{-3}$ & $5.46 \times 10^{-3}$ \\
15 & 225 & $4.43 \times 10^{-1}$ & $5.15 \times 10^{-3}$ & $4.50 \times 10^{-3}$ & $5.57 \times 10^{-3}$ & $5.63 \times 10^{-3}$ \\
\hline
\end{tabular}%
}
\end{table}

\begin{table}[htbp]
\centering
\caption{Summary of the $B \times K$ parametric study (Table~\ref{tab:multi_b_k}). For each bottleneck width $B$, the table reports the polynomial order $K$ that gives the lowest $\mathrm{Rel}~L^{2}$ error and the $\mathrm{Rel}~L^{2}$ error at the highest tested order $K = 15$. The ratio of these two errors quantifies the accuracy loss when the spectral order is increased beyond the optimum.}
\label{tab:best_b_k}
\resizebox{0.75\textwidth}{!}{%
\large
\begin{tabular}{rrrrc}
\hline
\textbf{$B$} & \textbf{Best $K$} & \textbf{Best Rel $L^2$} & \textbf{Rel $L^2$ at $K{=}15$} & \textbf{Ratio} \\
\hline
 2 & 12 & $4.43 \times 10^{-1}$ & $4.43 \times 10^{-1}$ & $1.0\times$ \\
 4 & 10 & $3.16 \times 10^{-3}$ & $5.15 \times 10^{-3}$ & $1.6\times$ \\
 5 & 12 & $3.06 \times 10^{-3}$ & $4.50 \times 10^{-3}$ & $1.5\times$ \\
 8 &  8 & $3.18 \times 10^{-3}$ & $5.57 \times 10^{-3}$ & $1.8\times$ \\
10 &  8 & $3.09 \times 10^{-3}$ & $5.63 \times 10^{-3}$ & $1.8\times$ \\
\hline
\end{tabular}%
}
\end{table}

\clearpage
\begin{table}[htbp]
\centering
\caption{Comparison of pure spectral and JetKAN-F models on the lid-driven
cavity flow ($Re = 1000$). $K$ is the polynomial order per direction,
$n_f = K^2$ is the total number of spectral features, and Params counts
all trainable parameters.}
\label{tab:spectral_comparison}
\resizebox{\textwidth}{!}{%
\begin{tabular}{lrrrcccc}
\toprule
\textbf{Model}
  & \textbf{$K$}
  & \textbf{$n_f$}
  & \textbf{Params}
  & \textbf{MSE$(u)$}
  & \textbf{MSE$(v)$}
  & \textbf{Rel\,$L^{2}(u)$}
  & \textbf{Rel\,$L^{2}(v)$} \\
\midrule
Spectral $K{=}12$ & 12 &  144 &    432
  & $3.27\times10^{-2}$ & $3.81\times10^{-2}$
  & $8.40\times10^{-1}$ & $9.62\times10^{-1}$ \\
Spectral $K{=}40$ & 40 & 1{,}600 & 4{,}800
  & $2.06\times10^{-4}$ & $2.30\times10^{-4}$
  & $6.67\times10^{-2}$ & $7.46\times10^{-2}$ \\
Spectral $K{=}50$ & 50 & 2{,}500 & 7{,}500
  & $1.18\times10^{-2}$ & $1.55\times10^{-2}$
  & $5.05\times10^{-1}$ & $6.13\times10^{-1}$ \\
\midrule
JetKAN-F $K{=}8$  &  8 &  64 &    984
  & $\mathbf{5.38\times10^{-5}}$ & $\mathbf{4.22\times10^{-5}}$
  & $\mathbf{3.41\times10^{-2}}$ & $\mathbf{3.20\times10^{-2}}$ \\
JetKAN-F $K{=}16$ & 16 & 256 & 3{,}288
  & $4.06\times10^{-4}$ & $5.85\times10^{-4}$
  & $7.36\times10^{-2}$ & $8.19\times10^{-2}$ \\
JetKAN-F $K{=}20$ & 20 & 400 & 5{,}016
  & $1.45\times10^{-2}$ & $1.90\times10^{-2}$
  & $5.59\times10^{-1}$ & $6.79\times10^{-1}$ \\
\bottomrule
\end{tabular}%
}
\end{table}


\clearpage
\begin{sidewaystable}[htbp]
\centering
\caption{MLP vs.\ KAN head ablation: JetMLP-F (single linear layer on $\tanh$) vs.\ JetKAN-F (single Chebyshev KAN layer) for the 2D Taylor--Green vortex (steady Navier--Stokes, $Re = 100$). Both models share the same $K_x = K_y = 8$ spectral basis, $B{=}8$, $51\times51$ grid, 50{,}000 epochs, Adam optimizer, lr $= 0.001$.}
\label{tab:mlp_kan_table}
\resizebox{\textwidth}{!}{%
\begin{tabular}{ccccccccccc}
\hline
\textbf{Model} & \textbf{Total Params} & \textbf{Epochs} & \textbf{Optimizer} & \textbf{PDE Loss} & \textbf{BC Loss} & \textbf{MSE} & \textbf{Rel $L^2$ ($u$)} & \textbf{Rel $L^2$ ($v$)} & \textbf{Rel $L^2$ ($p$)} & \textbf{ms/epoch} \\
\hline
JetMLP-F & 539 & 50,000 & Adam & $1.39 \times 10^{-4}$ & $1.91 \times 10^{-5}$ & $7.53 \times 10^{-6}$ & $3.52 \times 10^{-3}$ & $3.45 \times 10^{-3}$ & $4.78 \times 10^{-3}$ & $\mathbf{6.1}$ \\
JetKAN-F & 656 & 50,000 & Adam & $\mathbf{1.29 \times 10^{-4}}$ & $\mathbf{1.87 \times 10^{-5}}$ & $\mathbf{5.96 \times 10^{-6}}$ & $\mathbf{3.10 \times 10^{-3}}$ & $\mathbf{3.07 \times 10^{-3}}$ & $\mathbf{4.33 \times 10^{-3}}$ & 9.2 \\
\hline
\multicolumn{4}{c}{\textbf{Ratio (KAN / MLP)}} & $1.08\times$ better & $1.02\times$ better & $\mathbf{1.26\times}$ \textbf{better} & $\mathbf{1.14\times}$ \textbf{better} & $\mathbf{1.12\times}$ \textbf{better} & $\mathbf{1.10\times}$ \textbf{better} & $1.51\times$ slower \\
\hline
\end{tabular}%
}
\end{sidewaystable}

\clearpage
\begin{sidewaystable}[htbp]
\centering
\caption{Grid resolution study of JetKAN-F for the 2D Taylor--Green vortex (steady Navier--Stokes, $Re = 100$). All cases use 656 parameters, 50{,}000 epochs, Adam optimizer, $K_x = K_y = 8$, $B = 8$, degree $= 5$, lr $= 0.001$, seed $= 42$.}
\label{tab:collocation_table}
\resizebox{\textwidth}{!}{%
\large
\begin{tabular}{cccccccccccc}
\hline
\textbf{Grid} & \textbf{Colloc.\ pts} & \textbf{Parameters} & \textbf{Epochs} & \textbf{Optimizer} & \textbf{PDE Loss} & \textbf{BC Loss} & \textbf{MSE} & \textbf{Rel $L^2$ ($u$)} & \textbf{Rel $L^2$ ($v$)} & \textbf{Rel $L^2$ ($p$)} & \textbf{ms/epoch} \\
\hline
$21\times21$   & 441    & 656 & 50,000 & Adam & $1.05 \times 10^{-4}$ & $5.75 \times 10^{-6}$ & $1.26 \times 10^{-5}$          & $4.01 \times 10^{-3}$          & $3.88 \times 10^{-3}$          & $8.62 \times 10^{-3}$          & 8.7 \\
$31\times31$   & 961    & 656 & 50,000 & Adam & $1.31 \times 10^{-4}$ & $1.76 \times 10^{-5}$ & $7.63 \times 10^{-6}$          & $3.44 \times 10^{-3}$          & $3.48 \times 10^{-3}$          & $5.06 \times 10^{-3}$          & 8.5 \\
$51\times51$   & 2,601  & 656 & 50,000 & Adam & $1.29 \times 10^{-4}$ & $1.87 \times 10^{-5}$ & $\mathbf{5.96 \times 10^{-6}}$ & $\mathbf{3.10 \times 10^{-3}}$ & $\mathbf{3.07 \times 10^{-3}}$ & $4.33 \times 10^{-3}$          & 8.6 \\
$101\times101$ & 10,201 & 656 & 50,000 & Adam & $1.35 \times 10^{-4}$ & $1.94 \times 10^{-5}$ & $6.27 \times 10^{-6}$          & $3.22 \times 10^{-3}$          & $3.16 \times 10^{-3}$          & $4.31 \times 10^{-3}$          & 8.5 \\
$151\times151$ & 22,801 & 656 & 50,000 & Adam & $1.33 \times 10^{-4}$ & $1.97 \times 10^{-5}$ & $6.13 \times 10^{-6}$          & $3.22 \times 10^{-3}$          & $3.09 \times 10^{-3}$          & $4.26 \times 10^{-3}$          & 8.6 \\
$201\times201$ & 40,401 & 656 & 50,000 & Adam & $1.35 \times 10^{-4}$ & $1.97 \times 10^{-5}$ & $6.11 \times 10^{-6}$          & $3.16 \times 10^{-3}$          & $3.13 \times 10^{-3}$          & $\mathbf{4.26 \times 10^{-3}}$ & 8.6 \\
\hline
\end{tabular}%
}
\end{sidewaystable}


\clearpage
\begin{sidewaystable}[htbp]
\centering
\caption{Network depth study of JetKAN-F for the 2D Taylor--Green vortex
(steady Navier--Stokes, $Re = 100$). All cases use $K_x = K_y = 8$,
$B = 5$, degree $= 5$, 50{,}000 epochs, Adam optimizer, lr $= 10^{-3}$.}
\label{tab:depth_study}
\resizebox{\textwidth}{!}{%
\large
\begin{tabular}{clccccccccccc}
\hline
\textbf{Depth} & \textbf{Hidden dims} & \textbf{Parameters} & \textbf{Epochs} & \textbf{Optimizer}
  & \textbf{PDE Loss} & \textbf{BC Loss} & \textbf{MSE}
  & \textbf{Rel $L^2$ ($u$)} & \textbf{Rel $L^2$ ($v$)} & \textbf{Rel $L^2$ ($p$)}
  & \textbf{ms/epoch} \\
\hline
1L & $[5]$ & 410  & 50{,}000 & Adam
   & $1.59\times10^{-4}$ & $1.96\times10^{-5}$ & $9.54\times10^{-6}$
   & $3.69\times10^{-3}$ & $3.94\times10^{-3}$ & $5.94\times10^{-3}$ & 10.0 \\
2L & $[5,5]$ & 561  & 50{,}000 & Adam
   & $1.30\times10^{-4}$ & $1.83\times10^{-5}$ & $5.54\times10^{-6}$
   & $2.98\times10^{-3}$ & $2.94\times10^{-3}$ & $4.25\times10^{-3}$ & 17.2 \\
3L & $[5,\,5, \,5]$ & 712  & 50{,}000 & Adam
   & $1.25\times10^{-4}$ & $\mathbf{1.82\times10^{-5}}$ & $5.10\times10^{-6}$
   & $2.84\times10^{-3}$ & $\mathbf{2.83\times10^{-3}}$ & $4.11\times10^{-3}$ & 23.8 \\
4L & $[5,\,5,\,5, \,5]$ & 863  & 50{,}000 & Adam
   & $\mathbf{1.24\times10^{-4}}$ & $1.83\times10^{-5}$ & $5.09\times10^{-6}$
   & $2.83\times10^{-3}$ & $2.86\times10^{-3}$ & $\mathbf{4.05\times10^{-3}}$ & 30.8 \\
5L & $[5,\,5,\,5,\,5, \,5]$ & 1{,}014 & 50{,}000 & Adam
   & $1.25\times10^{-4}$ & $1.84\times10^{-5}$ & $\mathbf{5.08\times10^{-6}}$
   & $\mathbf{2.82\times10^{-3}}$ & $2.83\times10^{-3}$ & $4.13\times10^{-3}$ & 36.9 \\
6L & $[5,\,5,\,5,\,5,\,5, \,5]$ & 1{,}165 & 50{,}000 & Adam
   & $1.28\times10^{-4}$ & $1.88\times10^{-5}$ & $5.46\times10^{-6}$
   & $2.95\times10^{-3}$ & $2.94\times10^{-3}$ & $4.18\times10^{-3}$ & 43.5 \\
\hline
\end{tabular}%
}
\end{sidewaystable}

\clearpage
\begin{sidewaystable}[htbp]
\centering
\caption{Seed sensitivity study of JetKAN-F for the 2D unsteady Taylor--Green vortex (unsteady Navier--Stokes, $Re = 100$, $t \in [0, 2]$), trained with physics-only loss (PDE + BC + IC at $t = 0$). All cases use $51\times51\times51$ grid, $K_x = K_y = 8$, $K_t = 4$, $B = 8$, degree $= 5$, lr $= 0.001$.}
\label{tab:seed_study_table}
\resizebox{\textwidth}{!}{%
\large
\begin{tabular}{ccccccccccc}
\hline
\textbf{Seed} & \textbf{Parameters} & \textbf{Epochs} & \textbf{Optimizer} & \textbf{PDE Loss} & \textbf{BC Loss} & \textbf{MSE} & \textbf{Rel $L^2$ ($u$)} & \textbf{Rel $L^2$ ($v$)} & \textbf{Rel $L^2$ ($p$)} & \textbf{ms/epoch} \\
\hline
3    & 2,192 & 50,000 & Adam & $4.45 \times 10^{-6}$ & $2.04 \times 10^{-7}$ & $1.26 \times 10^{-5}$          & $1.45 \times 10^{-3}$          & $\mathbf{1.16 \times 10^{-3}}$ & $1.41 \times 10^{-2}$          & 14.8 \\
16   & 2,192 & 50,000 & Adam & $5.15 \times 10^{-6}$ & $2.14 \times 10^{-7}$ & $4.26 \times 10^{-5}$          & $\mathbf{1.26 \times 10^{-3}}$ & $1.38 \times 10^{-3}$          & $2.66 \times 10^{-2}$          & 14.8 \\
32   & 2,192 & 50,000 & Adam & $8.65 \times 10^{-6}$ & $4.03 \times 10^{-7}$ & $8.44 \times 10^{-5}$          & $1.80 \times 10^{-3}$          & $1.91 \times 10^{-3}$          & $3.75 \times 10^{-2}$          & 14.8 \\
64   & 2,192 & 50,000 & Adam & $7.28 \times 10^{-6}$ & $2.22 \times 10^{-7}$ & $6.46 \times 10^{-5}$          & $1.65 \times 10^{-3}$          & $1.70 \times 10^{-3}$          & $3.28 \times 10^{-2}$          & 14.8 \\
128  & 2,192 & 50,000 & Adam & $7.29 \times 10^{-6}$ & $4.18 \times 10^{-7}$ & $5.06 \times 10^{-5}$          & $1.74 \times 10^{-3}$          & $1.93 \times 10^{-3}$          & $2.88 \times 10^{-2}$          & 14.8 \\
256  & 2,192 & 50,000 & Adam & $7.44 \times 10^{-6}$ & $2.99 \times 10^{-7}$ & $\mathbf{8.02 \times 10^{-6}}$ & $1.85 \times 10^{-3}$          & $1.73 \times 10^{-3}$          & $\mathbf{1.05 \times 10^{-2}}$ & 14.8 \\
512  & 2,192 & 50,000 & Adam & $8.64 \times 10^{-6}$ & $2.74 \times 10^{-7}$ & $2.83 \times 10^{-5}$          & $1.99 \times 10^{-3}$          & $2.25 \times 10^{-3}$          & $2.11 \times 10^{-2}$          & 14.8 \\
1024 & 2,192 & 50,000 & Adam & $6.17 \times 10^{-6}$ & $2.26 \times 10^{-7}$ & $5.65 \times 10^{-5}$          & $1.79 \times 10^{-3}$          & $1.75 \times 10^{-3}$          & $3.05 \times 10^{-2}$          & 14.8 \\
\hline
\end{tabular}%
}
\end{sidewaystable}

\clearpage
\begin{table}[htbp]
\centering
\caption{Comparison of \textit{JetKAN-F} with existing methods for the \textit{2D} lid-driven cavity flow at $Re = 100$ and $Re = 1000$.}
\label{tab:literature_comparison}
\begin{tabular}{clccccc}
\toprule
$Re$ & Method & Params & Epochs & Rel $L^2(u)$ & Rel $L^2(v)$ & Rel $L^2$ (total) \\
\midrule
\multirow{5}{*}{100}
 & \textit{JetKAN-F} (present)            & \textbf{656}             & 100{,}000 & $4.59\times10^{-2}$ & $6.15\times10^{-2}$ & $5.43\times10^{-2}$ \\
 & SPIKAN~\cite{jacob2025spikans}       & 2{,}150                  & 50{,}000  & $5.83\times10^{-2}$ & $7.18\times10^{-2}$ & $6.54\times10^{-2}$ \\
 & FFV-PINN~\cite{wei2025ffvpinn}        & $\geq$4{,}400$^\ddagger$ & --        & $3.51\times10^{-2}$ & $2.88\times10^{-2}$ & $3.21\times10^{-2}$ \\
 & TSA-PINN~\cite{khademi2025physics}        & 10{,}702$^\dagger$       & 8{,}000   & --                  & --                  & $2.74\times10^{-2}$ \\
 & Wang et al.~\cite{wang2021understanding}    & 10{,}752$^\dagger$       & 40{,}000  & --                  & --                  & $3.42\times10^{-2}$ \\
\midrule
\multirow{3}{*}{1000}
 & \textit{JetKAN-F} (present)            & \textbf{984}             & 10{,}000  & $3.41\times10^{-2}$ & $3.20\times10^{-2}$ & $3.31\times10^{-2}$ \\
 & J-PIKAN~\cite{xiong2025jpikan}      & 4{,}500                  & --        & --                  & --                  & $(4.2\pm1.3)\times10^{-2}$ \\
 & FFV-PINN~\cite{wei2025ffvpinn}        & $\geq$4{,}400$^\ddagger$ & --        & $1.36\times10^{-2}$ & $1.40\times10^{-2}$ & $1.38\times10^{-2}$ \\
\bottomrule
\end{tabular}

\vspace{2pt}
\begin{minipage}{\linewidth}
\footnotesize
$^\dagger$Computed from the architecture stated in the original paper.\\
$^\ddagger$Not reported in the original paper; minimum count consistent with the stated architecture.
\end{minipage}
\end{table}


\clearpage
\section{Figures}
\clearpage

\clearpage

\clearpage

\clearpage

\clearpage

\clearpage


\clearpage


\clearpage


\clearpage

\clearpage

\clearpage

\clearpage


\clearpage

\clearpage

\begin{figure}
    \centering
    \includegraphics[width=1.0\linewidth]{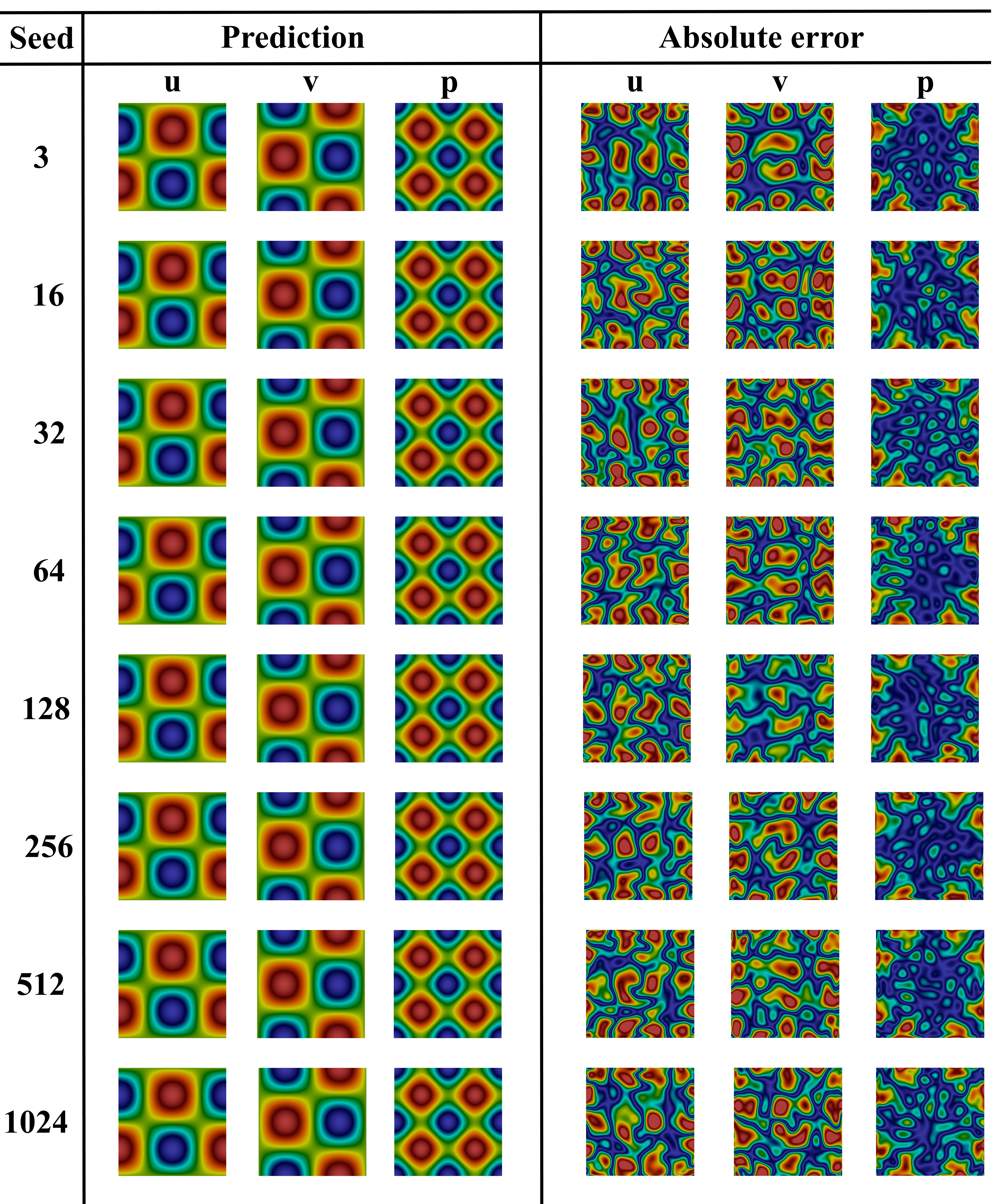}
    \caption{Predicted velocity and pressure fields and corresponding absolute error for JetKAN-F on the 2D Taylor--Green vortex, across the eight random seeds ($3$, $16$, $32$, $64$, $128$, $256$, $512$, $1024$) of the seed-sensitivity study. For each seed, the left three columns show the predicted $u$, $v$, and $p$ fields, and the right three columns show the corresponding absolute error $|u-u_{\text{exact}}|$, $|v-v_{\text{exact}}|$, and $|p-p_{\text{exact}}|$.}
    \label{fig:placeholder}
\end{figure}

\clearpage






\clearpage
\appendix
\section*{Appendix}
\label{appendix}

This Appendix provides the derivations supporting \Cref{subsec: 3.1 sjkan}. \ref{sec:A. chebyshev Differentiation Matrices} 
derives the Chebyshev differentiation matrices $\mathbf{D}_1$ and 
$\mathbf{D}_2$. Appendix~B derives the two-dimensional derivative feature 
matrices $\boldsymbol{\Phi}_x$, $\boldsymbol{\Phi}_{xx}$, 
$\boldsymbol{\Phi}_y$, $\boldsymbol{\Phi}_{yy}$ from the one-dimensional 
matrices of Appendix~A. Appendix~C derives the bottleneck derivative 
equations $\mathbf{Z}_x = \boldsymbol{\Phi}_x\,\mathbf{W}$, 
$\mathbf{Z}_{xx} = \boldsymbol{\Phi}_{xx}\,\mathbf{W}$, 
$\mathbf{Z}_y = \boldsymbol{\Phi}_y\,\mathbf{W}$, 
$\mathbf{Z}_{yy} = \boldsymbol{\Phi}_{yy}\,\mathbf{W}$ from the linearity 
of the bottleneck projection. Appendix~D derives the first and second 
spatial derivatives of the network output analytically.


\section{Chebyshev Differentiation Matrices}
\label{sec:A. chebyshev Differentiation Matrices}

This appendix derives the differentiation matrices $\mathbf{D}_1$ and 
$\mathbf{D}_2$ introduced in~\Cref{eq:22. precomputed matrices} of 
Section~\ref{subsubsec:3.1.2 Precomputed differentiation matrices}. The first-kind Chebyshev polynomials, used to construct the basis 
functions $\phi_{ij}$ in~\Cref{eq:20. tensor product}, are defined by the 
three-term recurrence
\begin{equation}
  T_0(\xi)=1, \quad T_1(\xi)=\xi, \quad
  T_{n+1}(\xi)=2\xi\,T_n(\xi)-T_{n-1}(\xi), \quad n\ge 1.
  \label{eq:A_recurrence}
\end{equation}
They satisfy $|T_n(\xi)|\le 1$ on $[-1,1]$. The derivative of 
$T_n$ can be written as a Chebyshev series,
\begin{equation}
  T_n'(\xi)=\sum_{k=0}^{n-1}(\mathbf{D}_1)_{kn}\,T_k(\xi),
  \label{eq:A2}
\end{equation}
where the entries of the matrix $\mathbf{D}_1$ are
\begin{equation}
  (\mathbf{D}_1)_{kn}=
  \begin{cases}
    2n/\gamma_k, & k<n \text{ and } (n+k) \text{ odd},\\
    0,               & \text{otherwise},
  \end{cases}
  \qquad \gamma_0=2,\quad \gamma_{k\ge 1}=1.
  \label{eq:A3}
\end{equation}
In other words, the $n$-th column of $\mathbf{D}_1$ contains the 
Chebyshev coefficients of $T_n'$. If a function is represented as 
$f(\xi)=\sum_{n=0}^{K-1}a_n T_n(\xi)$ with coefficient vector 
$\mathbf{a} \in \mathbb{R}^K$, then 
$f'(\xi)=\sum_{n=0}^{K-1}a'_n T_n(\xi)$ with 
$\mathbf{a}'=\mathbf{D}_1\,\mathbf{a}$, where $\mathbf{D}_1$ is 
of size $K \times K$. The second-derivative matrix is 
$\mathbf{D}_2=\mathbf{D}_1\,\mathbf{D}_1$, as stated in \Cref{subsubsec:3.1.2 Precomputed differentiation matrices}. For $K=3$, direct evaluation of~\Cref{eq:A3} gives
\begin{equation}
  \mathbf{D}_1=
  \begin{pmatrix}0 & 1 & 0\\0 & 0 & 4\\0 & 0 & 0\end{pmatrix},
  \qquad
  \mathbf{D}_2=\mathbf{D}_1^2=
  \begin{pmatrix}0 & 0 & 4\\0 & 0 & 0\\0 & 0 & 0\end{pmatrix},
  \label{eq:A_D1K3}
\end{equation}
which is consistent with $T_0'=0$, $T_1'=1$, $T_2'(\xi)=4\xi$ 
and $T_0''=T_1''=0$, $T_2''=4$. In the SJ-KAN forward pass, the polynomial values 
$\mathbf{T}(s_b) = [T_0(s_b),\ldots,T_P(s_b)]^{\top} 
\in \mathbb{R}^{P+1}$ are computed at each collocation point 
using the recurrence~\eqref{eq:A_recurrence}. The derivative 
values $[T_0'(s_b),\ldots,T_P'(s_b)]$ are then obtained by the 
matrix product $\mathbf{D}_1^{\top}\,\mathbf{T}(s_b)$, which 
follows from~\Cref{eq:A2}: writing 
$T_n'(s_b) = \sum_k (\mathbf{D}_1)_{kn}\,T_k(s_b)$ and 
collecting all $n$ gives entry $n$ of 
$\mathbf{D}_1^{\top}\,\mathbf{T}(s_b)$. The second derivative 
values follow by replacing $\mathbf{D}_1$ with $\mathbf{D}_2$. 
Here $\mathbf{D}_1$ and $\mathbf{D}_2$ are of size 
$(P+1) \times (P+1)$, as used in~\Cref{eq:gcoeffs} of 
Section~\ref{subsubsec:3.1.4 KAN activation layer and analytic jet}. 
For the spatial derivative feature matrices 
in~\Cref{eqn:23 . precomputed matrices}, the same construction applies with 
$\mathbf{D}_1$ of size $K_x \times K_x$ or $K_y \times K_y$, 
as described in 
Section~\ref{subsubsec:3.1.2 Precomputed differentiation matrices}. 
In both cases, the matrices depend only on their size and are 
computed once before training.


\section{Tensor-Product Derivative Feature Matrices}
\label{sec:B. ensor-Product Derivative Feature Matrices}

This appendix derives the derivative feature matrices
$\boldsymbol{\Phi}_x$, $\boldsymbol{\Phi}_{xx}$,
$\boldsymbol{\Phi}_y$, $\boldsymbol{\Phi}_{yy}$ introduced
in~\eqref{eqn:23 . precomputed matrices} of
Section~\ref{subsubsec:3.1.2 Precomputed differentiation matrices}.
The derivation uses the differentiation matrices $\mathbf{D}_1$ and
$\mathbf{D}_2$ from Appendix~\ref{sec:A. chebyshev Differentiation Matrices} and follows
from the chain rule applied to the tensor-product basis
functions $\phi_{ij}$ defined in~\eqref{eq:20. tensor product}. Recall from~\eqref{eq:20. tensor product} that each basis function is a product
of one-dimensional Chebyshev polynomials,
\begin{equation}
  \phi_{ij}(x,y) = T_i(\xi_x)\,T_j(\xi_y),
  \qquad i = 0,\ldots,K_x{-}1, \quad j = 0,\ldots,K_y{-}1,
  \label{eq:B1}
\end{equation}
where $\xi_x$ and $\xi_y$ are the normalized coordinates 
from~\eqref{eq:18. normalization}, mapping the physical coordinates $x$ 
and $y$ to $[-1,1]$. Differentiating it gives
\begin{equation}
  \frac{\partial \xi_x}{\partial x}
    = \frac{2}{x_{\max} - x_{\min}},
  \qquad
  \frac{\partial \xi_y}{\partial y}
    = \frac{2}{y_{\max} - y_{\min}},
  \label{eq:B_jacobian}
\end{equation}
and all cross-derivatives vanish:
$\partial \xi_x / \partial y = 0$ and
$\partial \xi_y / \partial x = 0$. The feature matrix $\boldsymbol{\Phi} \in \mathbb{R}^{N \times n_f}$
stores the basis function values at all
$N$ collocation points. Each row corresponds to collocation point
$m$ with coordinates $(x_m, y_m)$ and normalized coordinates
$\xi_{x,m}$, $\xi_{y,m}$. Each column corresponds to one
basis function $\phi_{ij}$, with the index pair $(i,j)$ flattened
into a single column index. The entry in row $m$ and column $(i,j)$
is
\begin{equation}
  \boldsymbol{\Phi}_{m,(i,j)}
    = \phi_{ij}(x_m, y_m)
    = T_i(\xi_{x,m})\,T_j(\xi_{y,m}).
  \label{eq:B3}
\end{equation}

To obtain the derivative feature matrices, we apply the chain rule
to~\eqref{eq:B1}. Since $\phi_{ij}$ is a product of two
functions of separate variables, the derivative with respect to $x$
acts only on the $T_i(\xi_x)$ factor,
\begin{equation}
  \frac{\partial \phi_{ij}}{\partial x}
    = T_i'(\xi_x)\,
      \frac{\partial \xi_x}{\partial x}\,T_j(\xi_y)
    = T_i'(\xi_x)\,
      \frac{2}{x_{\max} - x_{\min}}\,T_j(\xi_y),
  \label{eq:B_dphidx}
\end{equation}
and the second derivative with respect to $x$ is
\begin{equation}
  \frac{\partial^2 \phi_{ij}}{\partial x^2}
    = T_i''(\xi_x)\,
      \left(\frac{\partial \xi_x}{\partial x}\right)^{\!2}
      T_j(\xi_y)
    = T_i''(\xi_x)\,
      \left(\frac{2}{x_{\max} - x_{\min}}\right)^{\!2}\,T_j(\xi_y),
  \label{eq:B_d2phidx2}
\end{equation}
where the squared Jacobian appears because
$\partial\xi_x/\partial x$ is a constant. The derivatives with
respect to $y$ follow the same structure but act on the
$T_j(\xi_y)$ factor,
\begin{equation}
  \frac{\partial \phi_{ij}}{\partial y}
    = T_i(\xi_x)\,T_j'(\xi_y)\,
      \frac{2}{y_{\max} - y_{\min}},
  \label{eq:B_dphidy}
\end{equation}
\begin{equation}
  \frac{\partial^2 \phi_{ij}}{\partial y^2}
    = T_i(\xi_x)\,T_j''(\xi_y)\,
      \left(\frac{2}{y_{\max} - y_{\min}}\right)^{\!2}.
  \label{eq:B_d2phidy2}
\end{equation}

The derivative feature matrices are constructed by evaluating these
expressions at every collocation point $m = 1,\ldots,N$. The matrix
$\boldsymbol{\Phi}_x \in \mathbb{R}^{N \times n_f}$ has the entry
in row $m$ and column $(i,j)$ given by
\begin{equation}
  (\boldsymbol{\Phi}_x)_{m,(i,j)}
    = T_i'(\xi_{x,m})\,
      \frac{2}{x_{\max} - x_{\min}}\,T_j(\xi_{y,m}),
  \label{eq:B8}
\end{equation}
and the remaining three matrices are
\begin{align}
  (\boldsymbol{\Phi}_{xx})_{m,(i,j)}
    &= T_i''(\xi_{x,m})\,
       \left(\frac{2}{x_{\max} - x_{\min}}\right)^{\!2}\,
       T_j(\xi_{y,m}),
  \label{eq:B_Phixx_entry} \\[6pt]
  (\boldsymbol{\Phi}_y)_{m,(i,j)}
    &= T_i(\xi_{x,m})\,T_j'(\xi_{y,m})\,
       \frac{2}{y_{\max} - y_{\min}},
  \label{eq:B10} \\[6pt]
  (\boldsymbol{\Phi}_{yy})_{m,(i,j)}
    &= T_i(\xi_{x,m})\,T_j''(\xi_{y,m})\,
       \left(\frac{2}{y_{\max} - y_{\min}}\right)^{\!2}.
  \label{eq:B11}
\end{align}

The Chebyshev derivatives $T_i'$ and $T_i''$ in these entries are
computed using the differentiation matrices from
Appendix~\ref{sec:A. chebyshev Differentiation Matrices}. The values
$T_0'(\xi_{x,m}), \ldots, T_{K_x-1}'(\xi_{x,m})$ are obtained by
applying $\mathbf{D}_1^{\top}$ (of size $K_x \times K_x$) to the
vector $[T_0(\xi_{x,m}), \ldots, T_{K_x-1}(\xi_{x,m})]^{\top}$,
as described in Appendix~\ref{sec:A. chebyshev Differentiation Matrices}. The second derivatives
use $\mathbf{D}_2^{\top}$ in the same way. The $y$-direction
derivatives use $\mathbf{D}_1$ and $\mathbf{D}_2$ of size
$K_y \times K_y$ applied to the Chebyshev values in $\xi_y$. To see how these matrices look in practice, consider $K_x = K_y = 2$,
which gives $n_f = 4$ basis functions with index pairs $(0,0)$,
$(1,0)$, $(0,1)$, $(1,1)$. The Chebyshev polynomials are
$T_0(\xi) = 1$ and $T_1(\xi) = \xi$, with derivatives $T_0' = 0$,
$T_1' = 1$, $T_0'' = 0$, $T_1'' = 0$. Substituting
into~\eqref{eq:B3}--\eqref{eq:B11} gives the
following matrices for $N$ collocation points:
\begin{equation}
  \boldsymbol{\Phi} =
  \begin{pmatrix}
    1 & \xi_{x,1} & \xi_{y,1} & \xi_{x,1}\,\xi_{y,1} \\
    1 & \xi_{x,2} & \xi_{y,2} & \xi_{x,2}\,\xi_{y,2} \\
    \vdots & \vdots & \vdots & \vdots \\
    1 & \xi_{x,N} & \xi_{y,N} & \xi_{x,N}\,\xi_{y,N}
  \end{pmatrix},
  \label{eq:B_Phi_K2}
\end{equation}
\begin{equation}
  \boldsymbol{\Phi}_x = \frac{2}{x_{\max} - x_{\min}}
  \begin{pmatrix}
    0 & 1 & 0 & \xi_{y,1} \\
    0 & 1 & 0 & \xi_{y,2} \\
    \vdots & \vdots & \vdots & \vdots \\
    0 & 1 & 0 & \xi_{y,N}
  \end{pmatrix},
  \qquad
  \boldsymbol{\Phi}_y = \frac{2}{y_{\max} - y_{\min}}
  \begin{pmatrix}
    0 & 0 & 1 & \xi_{x,1} \\
    0 & 0 & 1 & \xi_{x,2} \\
    \vdots & \vdots & \vdots & \vdots \\
    0 & 0 & 1 & \xi_{x,N}
  \end{pmatrix},
  \label{eq:B_Phixy_K2}
\end{equation}
\begin{equation}
  \boldsymbol{\Phi}_{xx} = \mathbf{0}_{N \times 4},
  \qquad
  \boldsymbol{\Phi}_{yy} = \mathbf{0}_{N \times 4}.
  \label{eq:B_Phixx_K2}
\end{equation}
The second-derivative matrices are zero because the $2 \times 2$
matrix $\mathbf{D}_2$ from Appendix~\ref{sec:A. chebyshev Differentiation Matrices} is the
zero matrix when $K = 2$. Nonzero second derivatives require at
least $K_x \geq 3$ (or $K_y \geq 3$), since $T_2''(\xi) = 4$
is the first Chebyshev polynomial with a nonzero second derivative. All five matrices have the same size $N \times n_f$ and are computed
once before training, since the collocation points and the domain
bounds are fixed. These precomputed matrices are then used in the
bottleneck projection (Appendix~\ref{sec:C. Linearity of the Bottleneck Derivatives}) and the
analytic jet computation (~\ref{sec:D. Derivation of the Analytic Jet}).


\section{Linearity of the Bottleneck Derivatives}
\label{sec:C. Linearity of the Bottleneck Derivatives}

This \Cref{sec:C. Linearity of the Bottleneck Derivatives} derives the spatial derivatives of the bottleneck
output $\mathbf{Z}$ introduced in
Section~\ref{subsubsec:3.1.3 Bottleneck projection}. Because the
weight matrix $\mathbf{W}$ does not depend on the spatial
coordinates, each derivative of $\mathbf{Z}$ reduces to a single
matrix product with the corresponding derivative feature matrix
from Appendix~\ref{sec:B. ensor-Product Derivative Feature Matrices}. The bottleneck weight matrix
$\mathbf{W} \in \mathbb{R}^{n_f \times B}$ has one row per basis
function and one column per bottleneck channel,
\begin{equation}
  \mathbf{W} =
  \begin{pmatrix}
    W_{1,1} & W_{1,2} & \cdots & W_{1,B} \\
    W_{2,1} & W_{2,2} & \cdots & W_{2,B} \\
    \vdots  & \vdots  & \ddots & \vdots  \\
    W_{n_f,1} & W_{n_f,2} & \cdots & W_{n_f,B}
  \end{pmatrix}.
  \label{eq:C_W}
\end{equation}
These entries are learnable parameters that do not depend on the
spatial coordinates $(x,y)$. The bottleneck output is defined as
\begin{equation}
  \mathbf{Z} = \boldsymbol{\Phi}\,\mathbf{W}
  \in \mathbb{R}^{N \times B},
  \label{eq:C_Zdef}
\end{equation}
where $\boldsymbol{\Phi} \in \mathbb{R}^{N \times n_f}$ is the
feature matrix from~\eqref{eqn:23 . precomputed matrices}. Each row of
$\mathbf{Z}$ corresponds to one collocation point $m$ and each
column to one bottleneck channel $b$. Writing out all rows,
\begin{equation}
  \mathbf{Z} =
  \begin{pmatrix}
    z_{1,1} & z_{1,2} & \cdots & z_{1,B} \\
    z_{2,1} & z_{2,2} & \cdots & z_{2,B} \\
    \vdots  & \vdots  & \ddots & \vdots  \\
    z_{N,1} & z_{N,2} & \cdots & z_{N,B}
  \end{pmatrix},
  \label{eq:C_Zmatrix}
\end{equation}
where the entry in row $m$ and column $b$ is
\begin{equation}
  z_{m,b}
    = \sum_{f=1}^{n_f} \boldsymbol{\Phi}_{m,f}\,W_{f,b},
  \qquad m = 1,\ldots,N, \quad b = 1,\ldots,B.
  \label{eq:C_zmb}
\end{equation}

To differentiate $z_{m,b}$ with respect to $x$, we note that
$W_{f,b}$ is independent of $x$, so only $\boldsymbol{\Phi}_{m,f}$
carries spatial dependence. Differentiating each term in the sum
gives
\begin{equation}
  \frac{\partial z_{m,b}}{\partial x}
  = \sum_{f=1}^{n_f}
    \frac{\partial \boldsymbol{\Phi}_{m,f}}{\partial x}\,W_{f,b}
  = \sum_{f=1}^{n_f}
    (\boldsymbol{\Phi}_x)_{m,f}\,W_{f,b},
  \label{eq:C5. C_dzdx}
\end{equation}
where $(\boldsymbol{\Phi}_x)_{m,f}$ is the entry of the
first-derivative feature matrix defined
in~\eqref{eq:B8}. The right-hand side
of~\eqref{eq:C5. C_dzdx} is exactly the $(m,b)$ entry of the matrix
product $\boldsymbol{\Phi}_x\,\mathbf{W}$. Collecting all $N$
rows and $B$ columns into a matrix,
\begin{equation}
  \mathbf{Z}_x =
  \begin{pmatrix}
    \dfrac{\partial z_{1,1}}{\partial x}
      & \dfrac{\partial z_{1,2}}{\partial x}
      & \cdots
      & \dfrac{\partial z_{1,B}}{\partial x} \\[8pt]
    \dfrac{\partial z_{2,1}}{\partial x}
      & \dfrac{\partial z_{2,2}}{\partial x}
      & \cdots
      & \dfrac{\partial z_{2,B}}{\partial x} \\[8pt]
    \vdots & \vdots & \ddots & \vdots \\[4pt]
    \dfrac{\partial z_{N,1}}{\partial x}
      & \dfrac{\partial z_{N,2}}{\partial x}
      & \cdots
      & \dfrac{\partial z_{N,B}}{\partial x}
  \end{pmatrix}
  = \boldsymbol{\Phi}_x\,\mathbf{W}
  \;\in\; \mathbb{R}^{N \times B}.
  \label{eq:C6}
\end{equation}

The same argument applies to the $y$-derivative. Since $W_{f,b}$
is also independent of $y$,
\begin{equation}
  \frac{\partial z_{m,b}}{\partial y}
  = \sum_{f=1}^{n_f}
    \frac{\partial \boldsymbol{\Phi}_{m,f}}{\partial y}\,W_{f,b}
  = \sum_{f=1}^{n_f}
    (\boldsymbol{\Phi}_y)_{m,f}\,W_{f,b},
  \label{eq:C7}
\end{equation}
where $(\boldsymbol{\Phi}_y)_{m,f}$ is defined
in~\eqref{eq:B10}. Collecting into a matrix,
\begin{equation}
  \mathbf{Z}_y =
  \begin{pmatrix}
    \dfrac{\partial z_{1,1}}{\partial y}
      & \dfrac{\partial z_{1,2}}{\partial y}
      & \cdots
      & \dfrac{\partial z_{1,B}}{\partial y} \\[8pt]
    \dfrac{\partial z_{2,1}}{\partial y}
      & \dfrac{\partial z_{2,2}}{\partial y}
      & \cdots
      & \dfrac{\partial z_{2,B}}{\partial y} \\[8pt]
    \vdots & \vdots & \ddots & \vdots \\[4pt]
    \dfrac{\partial z_{N,1}}{\partial y}
      & \dfrac{\partial z_{N,2}}{\partial y}
      & \cdots
      & \dfrac{\partial z_{N,B}}{\partial y}
  \end{pmatrix}
  = \boldsymbol{\Phi}_y\,\mathbf{W}
  \;\in\; \mathbb{R}^{N \times B}.
  \label{eq:C8}
\end{equation}

For the second derivatives, the same reasoning holds because
$W_{f,b}$ remains constant regardless of the order of
differentiation. Differentiating~\eqref{eq:C5. C_dzdx} once more with
respect to $x$ gives
\begin{equation}
  \frac{\partial^2 z_{m,b}}{\partial x^2}
  = \sum_{f=1}^{n_f}
    \frac{\partial^2 \boldsymbol{\Phi}_{m,f}}{\partial x^2}\,W_{f,b}
  = \sum_{f=1}^{n_f}
    (\boldsymbol{\Phi}_{xx})_{m,f}\,W_{f,b},
  \label{eq:C_d2zdx2}
\end{equation}
and differentiating~\eqref{eq:C7} with respect to $y$ gives
\begin{equation}
  \frac{\partial^2 z_{m,b}}{\partial y^2}
  = \sum_{f=1}^{n_f}
    \frac{\partial^2 \boldsymbol{\Phi}_{m,f}}{\partial y^2}\,W_{f,b}
  = \sum_{f=1}^{n_f}
    (\boldsymbol{\Phi}_{yy})_{m,f}\,W_{f,b}.
  \label{eq:C_d2zdy2}
\end{equation}
The corresponding matrices are
\begin{equation}
  \mathbf{Z}_{xx}
  = \boldsymbol{\Phi}_{xx}\,\mathbf{W}
  \;\in\; \mathbb{R}^{N \times B},
  \qquad
  \mathbf{Z}_{yy}
  = \boldsymbol{\Phi}_{yy}\,\mathbf{W}
  \;\in\; \mathbb{R}^{N \times B},
  \label{eq:C_Zxxyy}
\end{equation}

To summarize, the complete set of bottleneck matrices is
\begin{align}
  \mathbf{Z}     &= \boldsymbol{\Phi}\,\mathbf{W},
  \label{eq:C_Zall} \\
  \mathbf{Z}_x   &= \boldsymbol{\Phi}_x\,\mathbf{W},
  \label{eq:C_Zx_final} \\
  \mathbf{Z}_{xx}&= \boldsymbol{\Phi}_{xx}\,\mathbf{W},
  \label{eq:C_Zxx_final} \\
  \mathbf{Z}_y   &= \boldsymbol{\Phi}_y\,\mathbf{W},
  \label{eq:C_Zy_final} \\
  \mathbf{Z}_{yy}&= \boldsymbol{\Phi}_{yy}\,\mathbf{W},
  \label{eq:C_Zyy_final}
\end{align}
where all five matrices have size $N \times B$. Each derivative
matrix is obtained by replacing $\boldsymbol{\Phi}$ with the
corresponding precomputed derivative feature matrix from
Appendix~\ref{sec:B. ensor-Product Derivative Feature Matrices} and multiplying by the same weight
matrix $\mathbf{W}$. No nonlinear chain rule is required at this
stage because the bottleneck is a linear map. These five matrices
form the input to the analytic jet computation in
Appendix~\ref{sec:D. Derivation of the Analytic Jet}.


\section{Derivation of the Analytic Jet}
\label{sec:D. Derivation of the Analytic Jet}

This \ref{sec:D. Derivation of the Analytic Jet} derives the first and second spatial derivatives of
the network output $\hat{U}_o$ with respect to the physical
coordinates $x$ and $y$. These derivatives, together with
$\hat{U}_o$ itself, form the spectral jet introduced in
Section~\ref{subsubsec:3.1.4 KAN activation layer and analytic jet}. The
derivation builds on the bottleneck matrices $\mathbf{Z}$,
$\mathbf{Z}_x$, $\mathbf{Z}_{xx}$, $\mathbf{Z}_y$,
$\mathbf{Z}_{yy}$ from Appendix~\ref{sec:C. Linearity of the Bottleneck Derivatives} and the
differentiation matrices $\mathbf{D}_1$, $\mathbf{D}_2$ from
~\ref{sec:A. chebyshev Differentiation Matrices}. After the bottleneck projection, the $b$-th latent channel at
collocation point $m$ is $z_b \equiv z_{m,b}$, with spatial
derivatives $\partial z_b / \partial x$,
$\partial^2 z_b / \partial x^2$, $\partial z_b / \partial y$, and
$\partial^2 z_b / \partial y^2$ given by the $(m,b)$ entries of
$\mathbf{Z}_x$, $\mathbf{Z}_{xx}$, $\mathbf{Z}_y$, and
$\mathbf{Z}_{yy}$ from \ref{sec:C. Linearity of the Bottleneck Derivatives}. The
KAN activation maps each $z_b$ through a Chebyshev expansion in a
transformed variable $s_b$,
\begin{equation}
  \phi_{b,o}(z_b) = \sum_{n=0}^{P} c_{b,o,n}\,T_n(s_b),
  \qquad s_b = \tanh(z_b),
  \label{eq:D_phi}
\end{equation}
where $c_{b,o,n}$ are the trainable Chebyshev coefficients, $o =
1,\ldots,O$ indexes the output channel, and $P$ is the polynomial
degree. The $\tanh$ maps $z_b \in \mathbb{R}$ to $s_b \in (-1,1)$,
which is the natural domain of the Chebyshev polynomials. The
predicted field for output channel $o$ at collocation point $m$ is
\begin{equation}
  \hat{U}_o = \sum_{b=1}^{B} \phi_{b,o}(z_b).
  \label{eq:D2}
\end{equation}

To compute the spatial derivatives of $\hat{U}_o$, we first need
the derivatives of each activation $\phi_{b,o}$ with respect to
$z_b$. Differentiating the Chebyshev series~\eqref{eq:D_phi} with
respect to $s_b$ gives
\begin{align}
  g_{b,o}^{(1)}
    &= \frac{\mathrm{d}\phi_{b,o}}{\mathrm{d}s_b}
     = \sum_{n=0}^{P} c_{b,o,n}\,T_n'(s_b),
  \label{eq:D_g1} \\[6pt]
  g_{b,o}^{(2)}
    &= \frac{\mathrm{d}^2\phi_{b,o}}{\mathrm{d}s_b^2}
     = \sum_{n=0}^{P} c_{b,o,n}\,T_n''(s_b),
  \label{eq:D_g2}
\end{align}
where $T_n'(s_b)$ and $T_n''(s_b)$ are computed by applying
$\mathbf{D}_1^{\top}$ and $\mathbf{D}_2^{\top}$ from
~\ref{sec:A. chebyshev Differentiation Matrices}, here of size
$(P{+}1) \times (P{+}1)$, to the vector
$[T_0(s_b), \ldots, T_P(s_b)]^{\top}$. Let $\sigma_b$ denote the derivative of $\tanh$. Since
$s_b = \tanh(z_b)$,
\begin{equation}
  \sigma_b
    = \frac{\mathrm{d}s_b}{\mathrm{d}z_b}
    = 1 - s_b^2.
  \label{eq:D_sigma}
\end{equation}
The first derivative of $\phi_{b,o}$ with respect to $z_b$ then
follows from the chain rule,
\begin{equation}
  \frac{\mathrm{d}\phi_{b,o}}{\mathrm{d}z_b}
    = \frac{\mathrm{d}\phi_{b,o}}{\mathrm{d}s_b}\,
      \frac{\mathrm{d}s_b}{\mathrm{d}z_b}
    = g_{b,o}^{(1)}\,\sigma_b.
  \label{eq:D6}
\end{equation}

For the second derivative of $\phi_{b,o}$ with respect to $z_b$,
we apply the product rule to~\eqref{eq:D6},
\begin{equation}
  \frac{\mathrm{d}^2\phi_{b,o}}{\mathrm{d}z_b^2}
    = \frac{\mathrm{d}}{\mathrm{d}z_b}
      \!\left[g_{b,o}^{(1)}\,\sigma_b\right]
    = \frac{\mathrm{d}g_{b,o}^{(1)}}{\mathrm{d}z_b}\,\sigma_b
      + g_{b,o}^{(1)}\,
        \frac{\mathrm{d}\sigma_b}{\mathrm{d}z_b}.
  \label{eq:D7}
\end{equation}
The two terms on the right are expanded separately.
Differentiating $g_{b,o}^{(1)}$ through $s_b = \tanh(z_b)$,
\begin{equation}
  \frac{\mathrm{d}g_{b,o}^{(1)}}{\mathrm{d}z_b}
    = \frac{\mathrm{d}g_{b,o}^{(1)}}{\mathrm{d}s_b}\,
      \frac{\mathrm{d}s_b}{\mathrm{d}z_b}
    = g_{b,o}^{(2)}\,\sigma_b.
  \label{eq:D8}
\end{equation}
Differentiating $\sigma_b = 1 - s_b^2$ through
$s_b = \tanh(z_b)$,
\begin{equation}
  \frac{\mathrm{d}\sigma_b}{\mathrm{d}z_b}
    = -2\,s_b\,\frac{\mathrm{d}s_b}{\mathrm{d}z_b}
    = -2\,s_b\,\sigma_b.
  \label{eq:D9}
\end{equation}
Substituting~\eqref{eq:D8} and~\eqref{eq:D9}
into~\eqref{eq:D7},
\begin{equation}
  \frac{\mathrm{d}^2\phi_{b,o}}{\mathrm{d}z_b^2}
    = g_{b,o}^{(2)}\,\sigma_b^2
      - 2\,s_b\,\sigma_b\,g_{b,o}^{(1)}.
  \label{eq:D10}
\end{equation}

With~\Cref{eq:D6} and~\Cref{eq:D10} in hand, the
spatial derivatives of $\hat{U}_o$ follow from the chain rule
applied to~\Cref{eq:D2}. The first derivative with respect to
$x$ is
\begin{equation}
  \frac{\partial \hat{U}_o}{\partial x}
    = \sum_{b=1}^{B}
      \frac{\mathrm{d}\phi_{b,o}}{\mathrm{d}z_b}\,
      \frac{\partial z_b}{\partial x}
    = \sum_{b=1}^{B}
      g_{b,o}^{(1)}\,\sigma_b\,
      \frac{\partial z_b}{\partial x},
  \label{eq:D11}
\end{equation}
where $\partial z_b / \partial x$ is the $(m,b)$ entry of
$\mathbf{Z}_x$ from~\Cref{eq:C6}. The first derivative with
respect to $y$ is
\begin{equation}
  \frac{\partial \hat{U}_o}{\partial y}
    = \sum_{b=1}^{B}
      \frac{\mathrm{d}\phi_{b,o}}{\mathrm{d}z_b}\,
      \frac{\partial z_b}{\partial y}
    = \sum_{b=1}^{B}
      g_{b,o}^{(1)}\,\sigma_b\,
      \frac{\partial z_b}{\partial y},
  \label{eq:D_dUdy}
\end{equation}
where $\partial z_b / \partial y$ is the $(m,b)$ entry of
$\mathbf{Z}_y$ from~\eqref{eq:C8}. For the second derivative with respect to $x$, we
differentiate~\eqref{eq:D11} and apply the product rule to each
term in the sum,
\begin{equation}
  \frac{\partial^2 \hat{U}_o}{\partial x^2}
    = \sum_{b=1}^{B}
      \frac{\partial}{\partial x}
      \!\left[
        \frac{\mathrm{d}\phi_{b,o}}{\mathrm{d}z_b}\,
        \frac{\partial z_b}{\partial x}
      \right]
    = \sum_{b=1}^{B}
      \left[
        \frac{\mathrm{d}^2\phi_{b,o}}{\mathrm{d}z_b^2}\,
        \left(\frac{\partial z_b}{\partial x}\right)^{\!2}
        +\,
        \frac{\mathrm{d}\phi_{b,o}}{\mathrm{d}z_b}\,
        \frac{\partial^2 z_b}{\partial x^2}
      \right],
  \label{eq:D_d2Udx2}
\end{equation}
where $\mathrm{d}^2\phi_{b,o}/\mathrm{d}z_b^2$ is given
by~\Cref{eq:D10},
$\mathrm{d}\phi_{b,o}/\mathrm{d}z_b = g_{b,o}^{(1)}\,\sigma_b$
from~\Cref{eq:D6}, and $\partial z_b / \partial x$,
$\partial^2 z_b / \partial x^2$ are the $(m,b)$ entries of
$\mathbf{Z}_x$ and $\mathbf{Z}_{xx}$. The second derivative with
respect to $y$ follows the same product rule,
\begin{equation}
  \frac{\partial^2 \hat{U}_o}{\partial y^2}
    = \sum_{b=1}^{B}
      \left[
        \frac{\mathrm{d}^2\phi_{b,o}}{\mathrm{d}z_b^2}\,
        \left(\frac{\partial z_b}{\partial y}\right)^{\!2}
        +\,
        \frac{\mathrm{d}\phi_{b,o}}{\mathrm{d}z_b}\,
        \frac{\partial^2 z_b}{\partial y^2}
      \right],
  \label{eq:D_d2Udy2}
\end{equation}
with $\partial z_b / \partial y$ and
$\partial^2 z_b / \partial y^2$ from the $(m,b)$ entries of
$\mathbf{Z}_y$ and $\mathbf{Z}_{yy}$. The Laplacian required by
the Navier--Stokes equations is
\begin{equation}
  \Delta \hat{U}_o
    = \frac{\partial^2 \hat{U}_o}{\partial x^2}
      + \frac{\partial^2 \hat{U}_o}{\partial y^2}.
  \label{eq:D_laplacian}
\end{equation}

Collecting the output and all its required spatial derivatives at
collocation point $m$ for output channel $o$ defines the spectral
jet from~\Cref{eq:32. jet},
\begin{equation}
  \mathrm{Jet}(\hat{U}_o)_m
    = \left(
        \hat{U}_o,\;
        \frac{\partial \hat{U}_o}{\partial x},\;
        \frac{\partial^2 \hat{U}_o}{\partial x^2},\;
        \frac{\partial \hat{U}_o}{\partial y},\;
        \frac{\partial^2 \hat{U}_o}{\partial y^2}
      \right)_{\!m},
  \label{eq:D_jet}
\end{equation}
which provides all the derivative information needed to evaluate
the PDE residuals without automatic differentiation through the
network.

\end{document}